\documentclass[a4paper,11pt]{article}
\pdfoutput=1
\usepackage{jheppub}

\usepackage{amssymb,amsmath,amsfonts}
\usepackage{mathcomp}
\usepackage[normalem]{ulem}
\usepackage[utf8x]{inputenc}
\usepackage{slashed}
\usepackage{graphicx}
\usepackage{tabularx}
\usepackage{here}
\usepackage{color}
\usepackage{csquotes} 
\usepackage{comment}
\usepackage{mathrsfs}
\usepackage{float}
\usepackage{ascmac}
\usepackage{multirow}
\usepackage{longtable}
\usepackage{bm}
\usepackage{ulem}
\usepackage{subcaption}
\usepackage{booktabs}
\usepackage{makecell}

\usepackage[italicdiff]{physics}

\newcounter{c1}
\newcounter{c}
\newcounter{c2}
\newcounter{c3}
\newcounter{c4}
\newcommand{\meV}{\mbox{meV}}
\newcommand{\eV}{\mbox{eV}}

\makeatletter
\newcommand*\rel@kern[1]{\kern#1\dimexpr\macc@kerna}
\newcommand*\widebar[1]{%
  \begingroup
  \def\mathaccent##1##2{%
    \rel@kern{0.8}%
    \overline{\rel@kern{-0.8}\macc@nucleus\rel@kern{0.2}}%
    \rel@kern{-0.2}%
  }%
  \macc@depth\@ne
  \let\math@bgroup\@empty \let\math@egroup\macc@set@skewchar
  \mathsurround\z@ \frozen@everymath{\mathgroup\macc@group\relax}%
  \macc@set@skewchar\relax
  \let\mathaccentV\macc@nested@a
  \macc@nested@a\relax111{#1}%
  \endgroup
}
\makeatother

\numberwithin{equation}{section}

\preprint{
\begin{minipage}{5cm}
\small
\flushright
KYUSHU-HET-366\\
KUNS-3112
\end{minipage}}

\title{
Revisiting One-Zero and Two-Zero Neutrino Mass Textures in Light of Recent Oscillation and Cosmological Data
}

\author{Haruto Kitagawa$^{1}$,}
\author{Coh Miyao$^{2}$,}
\author{Satsuki Nishimura$^{3}$, and}
\author{Hajime Otsuka$^{1,4}$}
\affiliation{
$^1$Department of Physics, Kyushu University, \\ 744 Motooka, Nishi-ku, Fukuoka 819-0395, Japan}
\affiliation{
$^2$Department of Physics and Astronomy, Faculty of Science and Technology, \\ Tokyo University of Science, Yamazaki, Noda, Chiba 278-8510, Japan}
\affiliation{
$^3$Department of Physics, Kyoto University, \\ Kitashirakawa-Oiwakecho, Sakyo-ku, Kyoto 606-8502, Japan
}
\affiliation{
$^4$Quantum and Spacetime Research Institute (QuaSR), Kyushu University, \\ 744 Motooka, Nishi-ku, Fukuoka 819-0395, Japan}
\emailAdd{kitagawa.haruto.691@s.kyushu-u.ac.jp}
\emailAdd{miyao.coh@rs.tus.ac.jp}
\emailAdd{satsuki@gauge.scphys.kyoto-u.ac.jp}
\emailAdd{otsuka.hajime@phys.kyushu-u.ac.jp}

\abstract{
We revisit one-zero and two-zero textures of the neutrino mass matrix under current experimental and cosmological constraints. 
We identify the phenomenologically viable texture structures using the latest results on neutrino oscillation parameters, the cosmological bound on the sum of neutrino masses, the kinematic bound on the effective electron-neutrino mass, and limits from neutrinoless double-beta decay. 
For two-zero textures, several structures are still allowed if only the CMB bound on the neutrino mass sum is imposed. 
Among them, the $B$-series textures show a characteristic prediction for the Dirac CP phase, with $\delta_{\rm CP}$ lying around $\pi/2$ and $3\pi/2$, and are within the reach of future neutrinoless double-beta decay searches. 
When the stronger CMB+BAO constraint is included, however, only the $A$-series textures remain viable. 
Therefore, we also analyze one-zero textures by using machine learning techniques, particularly flow matching. 
It turns out that some of the texture structures are already excluded by current data, while the allowed ones give distinct predictions for $\sum_i m_i$, $m_{\nu_e}^{\rm eff}$, $\langle m_{ee}\rangle$, and $\delta_{\rm CP}$. We further discuss how the one-zero texture structures can arise from non-invertible selection rules.
}

\makeatletter
\gdef\@fpheader{}
\makeatother

\begin{document}

\maketitle

\section{Introduction}
\label{sec:Intro}

The Standard Model (SM) of particle physics, completed by the discovery of the Higgs boson in 2012 \cite{ATLAS:2012yve,CMS:2012qbp}, is a remarkably successful theory that describes most current experimental results. 
However, the SM cannot explain neutrino masses. 
Neutrino oscillations, which imply nonzero neutrino masses, therefore provide one of the clearest motivations for exploring physics beyond the Standard Model (BSM).

\medskip

Experimentally, neutrino oscillation parameters have been measured with increasing precision by experiments such as SK \cite{Super-Kamiokande:1998kpq}, T2K \cite{T2K:2011ypd}, KamLAND \cite{PhysRevLett.90.021802,PhysRevLett.94.081801,PhysRevD.88.033001}, and JUNO \cite{JUNO:2025gmd}. 
These and other experimental results are combined in the NuFIT global analysis~\cite{Esteban:2024eli}, which provides the experimentally favored values of the mixing angles $\theta_{ij}$ ($ij = 12,13,23$), the CP phase $\delta_{\mathrm{CP}}$, and the neutrino mass-squared differences $\Delta m^2_{21}$ and $\Delta m^2_{3\ell}$ ($\ell = 1$ for normal ordering and $\ell = 2$ for inverted ordering), as summarized in Table \ref{tab:NuFIT}. 
\renewcommand{\arraystretch}{1.25}
\begin{table}[b]
\caption{Results of the NuFIT~6.0 global fit~\cite{Esteban:2024eli} for the neutrino oscillation parameters (IC24 with SK atmospheric data).}
\label{tab:NuFIT}
\centering
\small
   \begin{tabular}{|c||c|c||c|c|} \hline
     \multirow{2}{*}{Observables} & \multicolumn{2}{c||}{Normal Ordering (NO)} & \multicolumn{2}{c|}{Inverted Ordering (IO)}  \\
     \cline{2-5}
       & best fit $\pm 1\sigma$ & $3\sigma$ range & best fit $\pm 1\sigma$ & $3\sigma$ range  \\
     \hline
     $\sin^{2}\theta_{12}$ & $0.308_{-0.011}^{+0.012}$ & $0.275\rightarrow 0.345$ & $0.308_{-0.011}^{+0.012}$ & $0.275\rightarrow 0.345$ \\
     $\theta_{12}/\tcdegree$ & $33.68_{-0.70}^{+0.73}$ & $31.63\rightarrow 35.95$ & $33.68_{-0.70}^{+0.73}$ & $31.63\rightarrow 35.95$ \\
     \hline
     $\sin^{2}\theta_{13}$ & $0.02215_{-0.00058}^{+0.00056}$ & $0.02030\rightarrow 0.02388$ & $0.02231_{-0.00056}^{+0.00056}$ & $0.02060\rightarrow 0.02409$ \\
     $\theta_{13}/\tcdegree$ & $8.56_{-0.11}^{+0.11}$ & $8.19\rightarrow 8.89$ & $8.59_{-0.11}^{+0.11}$ & $8.25\rightarrow 8.93$ \\
     \hline
      $\sin^{2}\theta_{23}$ & $0.470_{-0.013}^{+0.017}$ & $0.435\rightarrow 0.585$ & $0.550_{-0.015}^{+0.012}$ & $0.440\rightarrow 0.584$ \\
      $\theta_{23}/\tcdegree$ & $43.3_{-0.8}^{+1.0}$ & $41.3\rightarrow 49.9$ & $47.9_{-0.9}^{+0.7}$ & $41.5\rightarrow 49.8$ \\
     \hline
      $\delta_{\text{CP}}/\pi$ & $1.18_{-0.23}^{+0.14}$ & $0.69\rightarrow 2.02$ & $1.52_{-0.14}^{+0.12}$ & $1.12\rightarrow 1.86$ \\
      $\delta_{\text{CP}}/\tcdegree$ & $212_{-41}^{+26}$ & $124\rightarrow 364$ & $274_{-25}^{+22}$ & $201\rightarrow 335$ \\
     \hline
     $\dfrac{\Delta m_{21}^{2}}{10^{-5}\,\mathrm{eV}^{2}}$ & $7.49_{-0.19}^{+0.19}$ & $6.92\rightarrow 8.05$ & $7.49_{-0.19}^{+0.19}$ & $6.92\rightarrow 8.05$ \\
     \hline
     $\dfrac{\Delta m_{3l}^{2}}{10^{-3}\,\mathrm{eV}^{2}}$ & $+2.513_{-0.019}^{+0.021}$ & $+2.451\rightarrow +2.578$ & $-2.484_{-0.020}^{+0.020}$ & $-2.547\rightarrow -2.421$ \\ 
     \hline
 \end{tabular}
 \renewcommand{\arraystretch}{1}
\end{table}
In the future, next-generation experiments such as HK \cite{Hyper-Kamiokande:2018ofw} and DUNE \cite{DUNE:2015lol} are expected to further improve the precision of neutrino oscillation measurements.

\medskip

From a theoretical perspective, these experimental results can be used to probe BSM physics. In the simplest low-energy effective theory, neutrino masses are described by the dimension-five Weinberg operator~\cite{Weinberg:1979sa}, 
\begin{align}
\frac{c_{ij}}{\Lambda}(L_i\tilde{H})(L_j\tilde{H}),
\end{align}
with $\tilde{H}=i\sigma^2H^\ast$. 
After the Higgs field $H$ acquires a vacuum expectation value (VEV), this operator generates neutrino masses. 
In many cases, the structure originating from high-energy physics is encoded in the coefficients $c_{ij}$. 
Therefore, the structure of the neutrino mass matrix can provide insight into the underlying BSM physics. 

\medskip

A well-known example is the study of two-zero textures and two-zero minors of the neutrino mass matrix~\cite{Frampton:2002yf}. 
A two-zero texture has two vanishing elements in the symmetric neutrino mass matrix, whereas a two-zero minor corresponds to two vanishing elements in its inverse. 
The two zeros yield two complex equations that lead to predictions for neutrino oscillation parameters and neutrino masses~\cite{Fritzsch:1999ee,Frampton:2002yf,Xing:2002ta,Xing:2002ap,Guo:2002ei,Dev:2006qe,Lashin:2007dm,Fritzsch:2011qv,Meloni:2012ci,Ludl:2014axa,Liao:2013saa,Zhou:2015qua}. 
This framework enables a model-independent classification of viable structures consistent with current neutrino oscillation data. 
Indeed, such two-zero structures have been shown to arise in models with a U(1)$_{L_\mu-L_\tau}$ gauge symmetry \cite{Asai:2017ryy,Asai:2018ocx,Asai:2019ciz}, indicating that they are well-motivated from the viewpoint of model building.

\medskip

The precision of neutrino oscillation data has now reached a level at which some of these structures are severely constrained. 
Indeed, Refs.~\cite{Asai:2024pzx,Ibe:2025rwk} reported that certain two-zero textures and minors are now strongly constrained by the experimental results. 
Following the analysis of Ref.~\cite{Asai:2024pzx}, we revisit all two-zero textures in light of recent experimental data, including cosmological measurements from Planck, the Atacama Cosmology Telescope (ACT), and the Dark Energy Spectroscopic Instrument (DESI). 
Our comprehensive analysis shows that eight two-zero textures are already inconsistent with current experimental results, while the viability of the remaining textures depends on the neutrino mass ordering, normal (NO) or inverted (IO). 
This situation suggests that nature may favor less restrictive structures than two-zero textures. 
Motivated by this possibility, we also analyze one-zero textures and minors. 

\medskip

A one-zero texture has a single vanishing element in the neutrino mass matrix, while a one-zero minor has a single vanishing element in its inverse. 
These structures are of particular interest because they can be realized both in models based on non-invertible selection rules~\cite{Kobayashi:2024yqq,Kobayashi:2024cvp}, which have recently received much attention, and in models with a U(1)$_{L_\mu-L_\tau}$ gauge symmetry. 
Although several studies of these structures have appeared in recent years \cite{Lashin:2011dn,Deepthi:2011sk,Gautam:2015kya,Priya:2025khf}, a comprehensive investigation of the current constraints on all one-zero textures and minors has not yet been performed.

\medskip

In this work, we aim to derive comprehensive constraints on one-zero texture and minor structures using a new analysis method. 
We also use recently developed machine-learning techniques to explore the parameter space and identify regions favored by experimental data. 
The resulting distributions reveal characteristic features of each one-zero structure and allow us to derive predictions for the corresponding model parameters. 
We further discuss how one-zero textures can be realized by non-invertible selection rules. 

\medskip

The organization of this paper is as follows. 
In Sec.~\ref{sec:twozero}, we analyze all two-zero textures, following the approach of Refs.~\cite{Asai:2024pzx,Ibe:2025rwk}. 
The current constraints on two-zero textures are summarized in Table \ref{tab:two-zero-texture}. 
In Sec.~\ref{sec:onezero}, we investigate one-zero textures.
First we apply machine learning techniques to analyze the one-zero textures in Sec.~\ref{sec:flow_matching}. 
Specifically, we utilize flow matching, known as one of the generative artificial intelligence (generative AI) frameworks. 
We find that viable parameter regions exhibit distinct patterns in the neutrino observables for different one-zero textures, and it turns out that some one-zero textures are disfavored by current data, as summarized in Table \ref{tab:one-zero-texture}. 
In addition, we analyze the one-zero textures analytically and discuss the results from the machine learning techniques.
In Sec.~\ref{sec:UV}, we discuss a realization of one-zero neutrino mass textures based on non-invertible selection rules. 
Finally, Sec.~\ref{sec:con} is devoted to the conclusion. 
In Appendix~\ref{app:two-zero}, we present the distributions of $\delta_{\mathrm{CP}}/\pi$ and $\Sigma\,m_{i}$ as functions of $\theta_{23}$ for the viable two-zero textures other than $A_1$ and $A_2$. 
The results for two-zero and one-zero minors are summarized in Appendices~\ref{app:two-zero_minor} and~\ref{app:one-zero_minors}, respectively.

\section{Two-zero textures}
\label{sec:twozero}

In this section, we consider neutrino mass matrices with two zero entries. 
These structures are classified as shown in Table \ref{tab:texture-minor}.

\medskip

When the neutrino mass matrix exhibits one of these patterns, it is referred to as a two-zero texture structure. 
On the other hand, when the inverse neutrino mass matrix has one of these patterns, it is called a two-zero minor structure. 
Details of the two-zero minor case are provided in Appendix \ref{app:two-zero_minor}. 
By investigating such structures, we can obtain predictions for neutrino oscillation parameters that remain experimentally uncertain, following the methodology introduced in Ref.~\cite{Frampton:2002yf}. 
Therefore, we revisit this analytical approach and update the predictions using the latest global analysis results for neutrino oscillation data from NuFIT \cite{Esteban:2024eli}.

\begin{table}[H]
    \caption{Classification of two-zero textures.}
    \label{tab:texture-minor}
    \centering
    \begin{tabular}{|cccc|}\hline
         $A_1: \begin{pmatrix}  0 & 0 & * \\ 0 & * & * \\ * & * & * \end{pmatrix}$&
         $A_2: \begin{pmatrix}  0 & * & 0 \\ * & * & * \\ 0 & * & * \end{pmatrix}$& &\\ \hline
         $B_1: \begin{pmatrix}  * & * & 0 \\ * & 0 & * \\ 0 & * & * \end{pmatrix}$&
         $B_2: \begin{pmatrix}  * & 0 & * \\ 0 & * & * \\ * & * & 0 \end{pmatrix}$&
         $B_3: \begin{pmatrix}  * & 0 & * \\ 0 & 0 & * \\ * & * & * \end{pmatrix}$&
         $B_4: \begin{pmatrix}  * & * & 0 \\ * & * & * \\ 0 & * & 0 \end{pmatrix}$\\ \hline
         $C: \begin{pmatrix}  * & * & * \\ * & 0 & * \\ * & * & 0 \end{pmatrix}$&  &  & \\ \hline
         $D_1: \begin{pmatrix}  * & * & * \\ * & 0 & 0 \\ * & 0 & * \end{pmatrix}$&
         $D_2: \begin{pmatrix}  * & * & * \\ * & * & 0\\ * & 0 & 0 \end{pmatrix}$&  & \\ \hline
         $E_1: \begin{pmatrix}  0 & * & * \\ * & 0 & * \\ * & * & * \end{pmatrix}$&
         $E_2: \begin{pmatrix}  0 & * & * \\ * & * & * \\ * & * & 0 \end{pmatrix}$&
         $E_3: \begin{pmatrix}  0 & * & * \\ * & * & 0 \\ * & 0 & * \end{pmatrix}$& \\ \hline
         $F_1: \begin{pmatrix}  * & 0 & 0 \\ 0 & * & * \\ 0 & * & * \end{pmatrix}$&
         $F_2: \begin{pmatrix}  * & 0 & * \\ 0 & * & 0 \\ * & 0 & * \end{pmatrix}$&
         $F_3: \begin{pmatrix}  * & * & 0 \\ * & * & 0 \\ 0 & 0 & * \end{pmatrix}$& \\ \hline
    \end{tabular}
\end{table}

\subsection{Analytical method for two-zero textures}

First, we review the analysis method. 
Conventionally, the neutrino mass matrix in the flavor basis is diagonalized by using the PMNS matrix, which is parametrized as
\begin{align}
    &U_{\rm PMNS} \nonumber \\
    &= \begin{pmatrix}
    1 & 0 & 0 \\
    0 & c_{23} & s_{23} \\
    0 & -s_{23} & c_{23}
    \end{pmatrix}
    \begin{pmatrix}
    c_{13} & 0 & s_{13}e^{-i\delta_{\rm CP}} \\
    0 & 1 & 0 \\
    -s_{13}e^{i\delta_{\rm CP}} & 0 & c_{13}
    \end{pmatrix}
    \begin{pmatrix}
    c_{12} & s_{12} & 0 \\
    -s_{12} & c_{12} & 0 \\
    0 & 0 & 1
    \end{pmatrix}
    \begin{pmatrix}
    1 & 0 & 0 \\
    0 & e^{\frac{i\alpha_2}{2}} & 0 \\
    0 & 0 & e^{\frac{i\alpha_3}{2}}
    \end{pmatrix} \nonumber \\
    &=
    \begin{pmatrix}
    c_{12} c_{13} & c_{13} s_{12}e^{i\frac{\alpha_2}{2}} & s_{13}e^{i\left(\frac{\alpha_3}{2}-\delta_{\rm CP}\right)} \\
    -c_{23} s_{12} -c_{12} s_{13} s_{23} e^{i \delta_{\rm CP}} &
    (c_{12}c_{23}-s_{12}s_{13}s_{23}e^{i\delta_{\rm CP}})e^{i\frac{\alpha_2}{2}} &
    c_{13} s_{23}e^{i\frac{\alpha_3}{2}} \\
    s_{12} s_{23} -c_{12} c_{23} s_{13} e^{i \delta_{\rm CP}} &
    (-c_{12}s_{23}-c_{23} s_{12}s_{13}e^{i\delta_{\rm CP}})e^{i\frac{\alpha_2}{2}} &
    c_{13} c_{23} e^{i\frac{\alpha_3}{2}}
    \end{pmatrix} \\
    &\equiv
    \begin{pmatrix}
    V_{11} & V_{12}e^{i\frac{\alpha_2}{2}} & V_{13}e^{i\frac{\alpha_3}{2}}\\
    V_{21} & V_{22} e^{i\frac{\alpha_2}{2}} & V_{23} e^{i\frac{\alpha_3}{2}} \\
    V_{31} & V_{32} e^{i\frac{\alpha_2}{2}} & V_{33} e^{i\frac{\alpha_3}{2}}
    \end{pmatrix}.
    \label{eq:PMNS}
\end{align}
Here, we define $\cos \theta_{ij} \equiv c_{ij}$ and $\sin \theta_{ij} \equiv s_{ij}$. 
The parameters $\delta_{\rm CP},~\alpha_{2,3}$ and $\theta_{ij}$ denote the Dirac CP phase, the Majorana phases, and the mixing angles, respectively. 
The quantity $\Delta m^2_{ij}\equiv m_i^2 - m_j^2$ represents the neutrino mass-squared difference. 
Using this PMNS matrix, the neutrino mass matrix in the flavor basis, $\mathcal{M}_{\nu}^{\rm flavor}$, is related to the diagonal mass matrix,  $\mathcal{M}_{\nu}^{\rm diag}$, by following transformation:
\begin{align}
    \mathcal{M}_{\nu}^{\rm diag} = U_{\rm PMNS}^{T}\, \mathcal{M}_{\nu}^{\rm flavor} \,U_{\rm PMNS} \label{eq:nu-mass_diag-flavor}.
\end{align}

\subsection*{Two-zero texture}

Here, we investigate the two-zero texture case.
From Eqs.~\eqref{eq:PMNS} and \eqref{eq:nu-mass_diag-flavor}, the components of the mass matrix in the flavor basis can be written as
\begin{align}
        \left(\mathcal{M}_{\nu}^{\rm flavor}\right)^*_{ij}&= V_{i1}V_{j1}m_1+e^{i\alpha_2}V_{i2}V_{j2}m_2+e^{i\alpha_3}V_{i3}V_{j3}m_3. \label{tex_component}
\end{align}
Assuming a two-zero texture, we impose
\begin{align}
     \left(\mathcal{M}_{\nu}^{\rm flavor}\right)^*_{ij}&=0, \\
     \left(\mathcal{M}_{\nu}^{\rm flavor}\right)^*_{\rho \sigma}&=0,
\end{align}
with $ij \ne \rho \sigma$.
Solving these complex equations for the Majorana phases, we find
\begin{align}
e^{i\alpha_2} &= \frac{m_1}{m_2} \frac{-V_{i3}V_{j3}V_{\rho 1}V_{\sigma 1}+V_{i1}V_{j1}V_{\rho 3}V_{\sigma 3}}{V_{i3}V_{j3}V_{\rho 2}V_{\sigma 2}-V_{i 2}V_{j2}V_{\rho 3}V_{\sigma 3}} \equiv  \frac{m_1}{m_2} R_2 (\theta_{12},\theta_{13},\theta_{23},\delta_{\rm CP}), \label{eq:R2_texture}\\
e^{i\alpha_3} &= \frac{m_1}{m_3} \frac{V_{i2}V_{j2}V_{\rho1}V_{\sigma 1}-V_{i1}V_{j1}V_{\rho2}V_{\sigma2}}{V_{i3}V_{j3}V_{\rho2}V_{\sigma2}-V_{i2}V_{j2}V_{\rho3}V_{\sigma3}} \equiv  \frac{m_1}{m_3} R_3 (\theta_{12},\theta_{13},\theta_{23},\delta_{\rm CP})\label{eq:R3_texture}.
\end{align}
The quantities $R_{2,3}$ are functions of $\theta_{12},\theta_{13},\theta_{23}$ and $\delta_{\rm CP}$.
Since the magnitudes of $e^{i \alpha_{2,3}}$ are unity, we obtain the following relations:
\begin{align}
    \frac{m_2}{m_1} &= |R_2 (\theta_{12},\theta_{13},\theta_{23},\delta_{\rm CP})|,\\
    \frac{m_3}{m_1} &= |R_3 (\theta_{12},\theta_{13},\theta_{23},\delta_{\rm CP})|.
\end{align}
For the normal ordering (NO), using these relations, we can rewrite the neutrino mass-squared differences in terms of $|R_{2,3}|$ as
\begin{align}
    \Delta m^2_{21} &\equiv m_2^2 -m_1^2 = m_1^2(|R_2|^2-1),\\
    \Delta m^2_{31} &\equiv m_3^2 -m_1^2 = m_1^2(|R_3|^2-1).
\end{align}
Solving these equations for $m_1^2$ and combining them, we finally obtain the equation:
\begin{align}
    \frac{\Delta m^2_{21}}{|R_2|^2-1} = \frac{\Delta m^2_{31}}{|R_3|^2-1}. \label{eq:tex-NO_d-t23}
\end{align}
This equation includes six parameters: $\theta_{12},\theta_{13},\theta_{23},\delta_{\rm CP},\Delta m^2_{21},\Delta m^2_{31}$. 
By fixing the four most precisely measured parameters, $\theta_{12},\theta_{13},\Delta m^2_{21},\Delta m^2_{31}$, to the best-fit values in Table \ref{tab:NuFIT}, we can determine $\delta_{\mathrm{CP}}$ as a function of $\theta_{23}$ by solving Eq.~\eqref{eq:tex-NO_d-t23}.

\medskip

After obtaining the $\theta_{23}$ dependence of $\delta_{\rm CP}$, we can predict the individual neutrino masses as
\begin{align}
    m_1&=\sqrt{\frac{\Delta m_{21}^2}{|R_2|^2-1}} = \sqrt{\frac{\Delta m_{31}^2}{|R_3|^2-1}}, \label{eq:nu-texture_m1_NO}\\
    m_2&=\sqrt{\frac{\Delta m_{21}^2 |R_2|^2}{|R_2|^2-1}}, \label{eq:nu-texture_m2_NO}\\
    m_3&= \sqrt{\frac{\Delta m_{31}^2 |R_3|^2}{|R_3|^2-1}}. \label{eq:nu-texture_m3_NO}
\end{align}
The sum of these masses is constrained by cosmological limits, as discussed in Sec.~\ref{subsec:const_msum}. 
Hence, we can classify the viable texture structures. 
Moreover, using these results, we can express the Majorana phases $\alpha_{2,3}$ in terms of $\theta_{23}$.

\medskip

The same analysis can be applied to the inverted ordering (IO).
In this case, the mass-squared differences are written as
\begin{align}
    \Delta m^2_{21} &\equiv m_2^2 -m_1^2 = m_1^2(|R_2|^2-1),\\
    \Delta m^2_{32} &\equiv m_3^2 -m_2^2 = m_1^2(|R_3|^2-|R_2|^2),
\end{align}
and we obtain
\begin{align}
    \frac{\Delta m^2_{21}}{|R_2|^2-1} = \frac{\Delta m^2_{32}}{|R_3|^2-|R_2|^2}.
    \label{eq:tex-IO_d-t23}
\end{align}
By fixing $\theta_{12},\theta_{13},\Delta m^2_{21},\Delta m^2_{32}$ to the best-fit values in Table \ref{tab:NuFIT}, we can obtain the predictions.
Note that the mass formulas for IO are 
\begin{align}
    m_1&=\sqrt{\frac{\Delta m_{21}^2}{|R_2|^2-1}} = \sqrt{\frac{\Delta m_{32}^2}{|R_3|^2-|R_2|^2}}, \label{eq:nu-texture_m1_IO}\\
    m_2&=\sqrt{\frac{\Delta m_{21}^2 |R_2|^2}{|R_2|^2-1}}, \label{eq:nu-texture_m2_IO}\\
    m_3&= \sqrt{\frac{(\Delta m_{32}^2+\Delta m_{21}^2) |R_3|^2}{|R_3|^2-1}}. \label{eq:nu-texture_m3_IO}
\end{align}

\subsection{Cosmological constraints on neutrino masses}
\label{subsec:const_msum}

Cosmological constraints on the sum of each neutrino masses have been reported by DESI \cite{DESI:2024mwx}.

\medskip

Using only the Planck CMB and assuming flat $\Lambda$CDM, three degenerate neutrino masses, and $\sum m_\nu>0$, the sum of neutrino mass is constrained as
\begin{align}
\sum m_\nu < 0.21~{\rm eV}~(95\%~{\rm C.L.},~\text{CMB})~.\label{eq:const_CMB}
\end{align}
Including CMB lensing data from Planck and ACT together with the DESI Baryon Acoustic Oscillation (BAO) data, and assuming NO with the lower bound on the mass sum imposed, the constraint becomes
\begin{align}
\sum m_\nu < 0.113~{\rm eV}~(95\%~{\rm C.L.},~\text{DESI BAO + CMB},~\sum m_\nu > 0.059~{\rm eV})~,\label{eq:const_NO}
\end{align}
while for the IO, one finds
\begin{align}
\sum m_\nu < 0.145~{\rm eV}~(95\%~{\rm C.L.},~\text{DESI BAO + CMB},~\sum m_\nu > 0.10~{\rm eV})~.\label{eq:const_IO}
\end{align}

\subsection{Constraints from kinematics of weak decay}
\label{subsec:def_mnue}

The KATRIN experiment~\cite{KATRIN:2024cdt} sets a constraint on the effective mass of electron neutrino through measurements of $^3H$ beta decay.
The current limit is
\begin{align}
    0.45~{\eV} \geq m_{\nu_e}^{\rm eff} \equiv \sqrt{\sum_i m_i^2 \left| U_{e i} \right|^2}.
\end{align}

\subsection{Neutrinoless double-beta decay}
\label{subsec:def_mee}

If neutrinos are Majorana particles, neutrinoless double beta decay can occur. 
This process is characterized by the effective Majorana mass defined by
\begin{align}
\langle m_{ee}\rangle =  \left| \sum_{i=1}^3 m_i U_{e i}^2 \right|.
\end{align}
Neutrinoless double beta decay has not yet been observed 
despite ongoing experimental searches. 
Consequently, upper limits on the effective neutrino mass have been set by KamLAND-Zen~\cite{KamLAND-Zen:2022tow} and GERDA~\cite{GERDA:2020xhi} respectively as below;
\begin{align}
\langle m_{ee}\rangle < 36-156~ {\rm meV},\\
\langle m_{ee}\rangle < 79- 180 ~{\rm meV}.
\end{align}

\medskip

In the future, the nEXO experiment~\cite{nEXO:2021ujk} is expected to reach a sensitivity to the effective neutrino mass in the range: 
\begin{align}
\langle m_{ee}\rangle &> ~4.7 - 20.3~ {\rm meV}.
\end{align}

\subsection{Prediction from analysis}

In this subsection, we present the results of the analysis for each texture structure. 
The results are summarized in Table \ref{tab:two-zero-texture}, where $\bigcirc$ and $\times$ denote viable and non-viable structures, respectively.

\medskip

When applying the neutrino mass sum constraint in Eq.~\eqref{eq:const_CMB}, we find that six structures (namely $A_1$, $A_2$, $B_1$, $B_2$, $B_3$, and $B_4$) are viable for NO within the $3\sigma$ range of $\theta_{23}$. 
For IO, three structures ($B_1$, $B_3$, and $C$) are viable within the $3\sigma$ range of $\theta_{23}$. 
On the other hand, when the neutrino mass sum constraints in Eq.~\eqref{eq:const_NO} and Eq.~\eqref{eq:const_IO} are imposed, only two structures, $A_1$ and $A_2$, remain viable for NO.

\begin{table}[H]
    \centering
    \caption{Summary of the two-zero texture analysis.}
    \label{tab:two-zero-texture}

    \begin{tabular}{ll cccccccc}
        \toprule
        Structure & & $A_1$ & $A_2$ & $B_1$ & $B_2$ & $B_3$ & $B_4$ & $C$ & \\
        \midrule
        \multirow{2}{*}{CMB}
            & NO & \hyperref[fig:A1tex_NO]{$\bigcirc$}
                 & \hyperref[fig:A2tex_NO]{$\bigcirc$}
                 & \hyperref[fig:B1tex_NO]{$\bigcirc$}
                 & \hyperref[fig:B2tex_NO]{$\bigcirc$}
                 & \hyperref[fig:B3tex_NO]{$\bigcirc$}
                 & \hyperref[fig:B4tex_NO]{$\bigcirc$}
                 & $\times$
                 & \\
            & IO & $\times$
                 & $\times$
                 & \hyperref[fig:B1tex_IO]{$\bigcirc$}
                 & $\times$
                 & \hyperref[fig:B3tex_IO]{$\bigcirc$}
                 & $\times$
                 & \hyperref[fig:Ctex_IO]{$\bigcirc$}
                 & \\
        \midrule
        \multirow{2}{*}{CMB+BAO}
            & NO & \hyperref[fig:A1tex_NO]{$\bigcirc$}
                 & \hyperref[fig:A2tex_NO]{$\bigcirc$}
                 & $\times$
                 & $\times$
                 & $\times$
                 & $\times$
                 & $\times$
                 & \\
            & IO & $\times$
                 & $\times$
                 & $\times$
                 & $\times$
                 & $\times$
                 & $\times$
                 & $\times$
                 & \\

        \midrule
        \multicolumn{10}{c}{}
        \\[-0.8em]
        \midrule

        Structure & & $D_1$ & $D_2$ & $E_1$ & $E_2$ & $E_3$ & $F_1$ & $F_2$ & $F_3$ \\
        \midrule
        \multirow{2}{*}{CMB}
            & NO & $\times$ & $\times$ & $\times$ & $\times$ & $\times$ & $\times$ & $\times$ & $\times$ \\
            & IO & $\times$ & $\times$ & $\times$ & $\times$ & $\times$ & $\times$ & $\times$ & $\times$ \\
        \midrule
        \multirow{2}{*}{CMB+BAO}
            & NO & $\times$ & $\times$ & $\times$ & $\times$ & $\times$ & $\times$ & $\times$ & $\times$ \\
            & IO & $\times$ & $\times$ & $\times$ & $\times$ & $\times$ & $\times$ & $\times$ & $\times$ \\
        \bottomrule
    \end{tabular}
\end{table}

\begin{figure}[H]
  \centering
  \begin{subfigure}{0.49\textwidth}
    \centering
    \includegraphics[width=\linewidth]{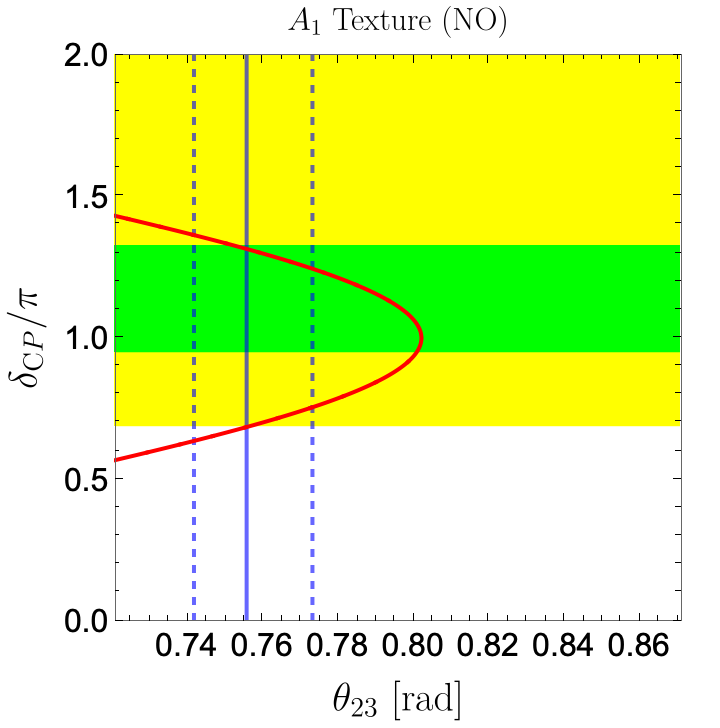}
    \label{fig:A1tex_NO_left}
  \end{subfigure}
  \hfill
  \begin{subfigure}{0.49\textwidth}
    \centering
    \includegraphics[width=\linewidth]{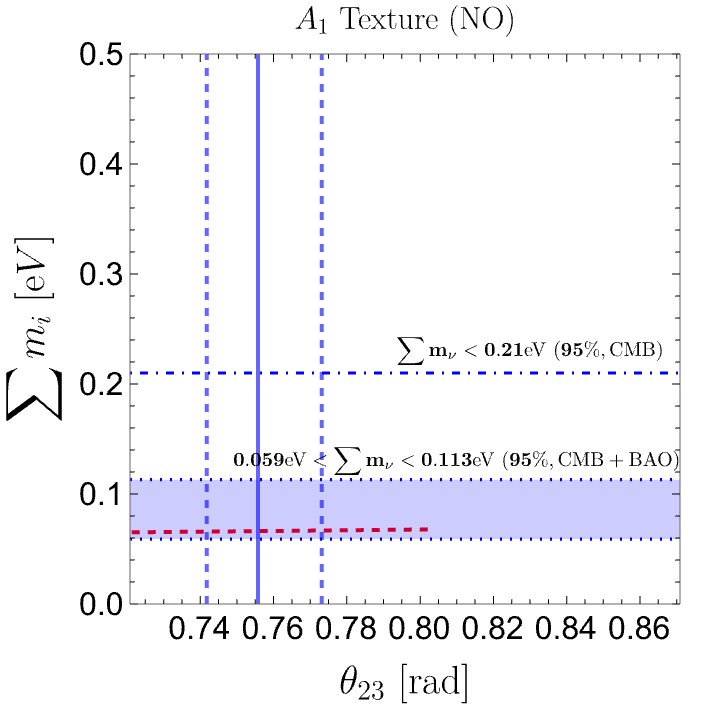}
    \label{fig:A1tex_NO_right}
  \end{subfigure}
  \caption{Distribution of observables for the $A_1$ structure with NO. The left and right panels show $\theta_{23}$ vs.~$\delta_{\mathrm{CP}}/\pi$ and $\theta_{23}$ vs.~$\Sigma\,m_{i}$, respectively.}
  \label{fig:A1tex_NO}
\end{figure}

\begin{figure}[H]
  \centering
  \begin{subfigure}{0.49\textwidth}
    \centering
    \includegraphics[width=\linewidth]{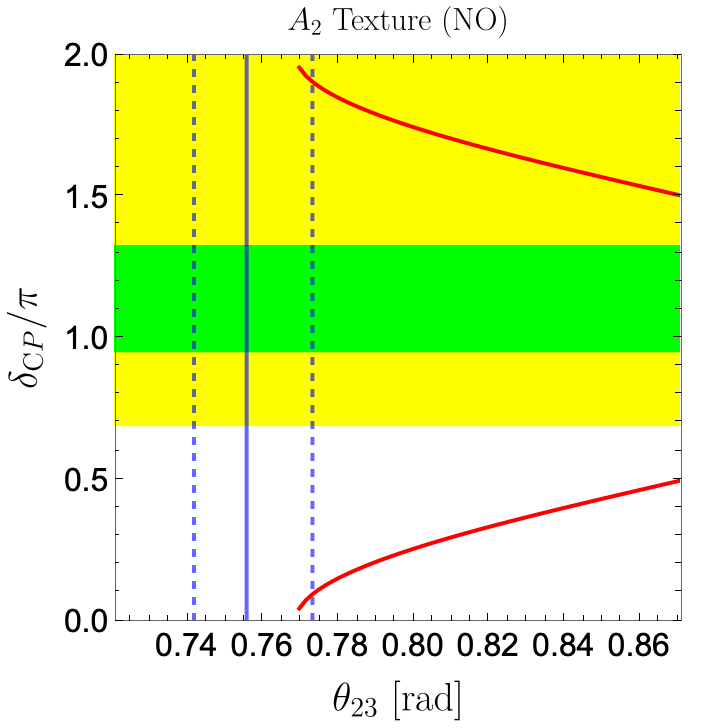}
    \label{fig:A2tex_NO_left}
  \end{subfigure}
  \hfill
  \begin{subfigure}{0.49\textwidth}
    \centering
    \includegraphics[width=\linewidth]{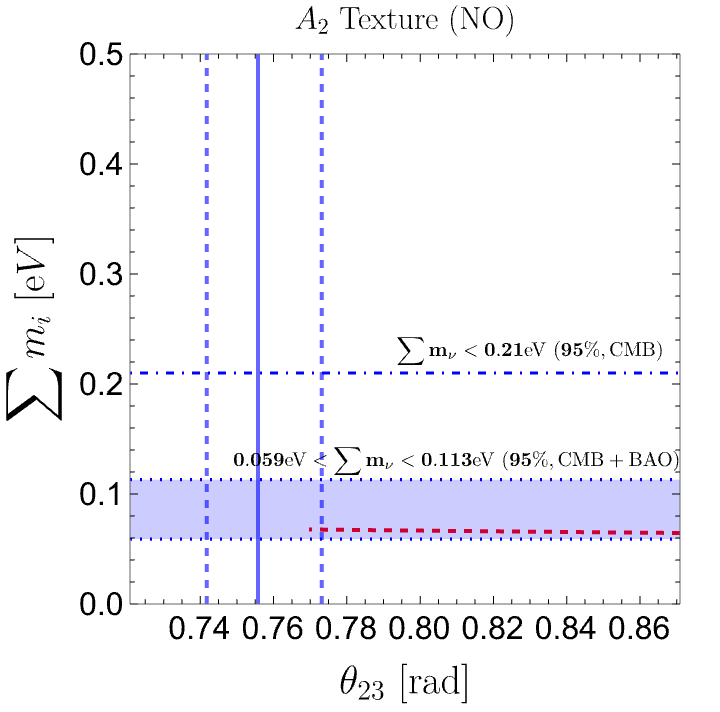}
    \label{fig:A2tex_NO_right}
  \end{subfigure}
  \caption{Distribution of observables for the $A_2$ structure with NO. The left and right panels show $\theta_{23}$ vs.~$\delta_{\mathrm{CP}}/\pi$ and $\theta_{23}$ vs.~$\Sigma\,m_{i}$, respectively.}
  \label{fig:A2tex_NO}
\end{figure}

\newpage
\noindent
As phenomenologically viable examples, we present the results for the $A_1$ and $A_2$ textures in NO in Figs.~\ref{fig:A1tex_NO} and~\ref{fig:A2tex_NO}. 
The left panel shows the $\theta_{23}$ dependence of $\delta_{\rm CP}$, where the green (yellow) band indicates the $1\sigma~(3\sigma)$ range of $\delta_{\rm CP}$ given in Table \ref{tab:NuFIT}. 
The red line represents the prediction obtained from our analysis. 
The blue solid (dashed) line indicates the best-fit $(1\sigma)$ value of $\theta_{23}$ from Table \ref{tab:NuFIT}. 
The horizontal axis range corresponds to the $3\sigma$ range of $\theta_{23}$ given in Table \ref{tab:NuFIT}. 

\medskip

The right panel shows the $\theta_{23}$ dependence of the sum of neutrino masses, where the region below the horizontal blue dash-dotted line is allowed by the neutrino mass sum constraint in Eq.~\eqref{eq:const_CMB}. 
The blue band represents the viable region under the neutrino mass sum constraints given in Eq.~\eqref{eq:const_NO} and Eq.~\eqref{eq:const_IO}. 
When calculating the sum of neutrino masses, we use values of $\delta_{\rm CP}$ within the $3\sigma$ range listed in Table \ref{tab:NuFIT}, obtained by solving Eq.~\eqref{eq:tex-NO_d-t23} and Eq.~\eqref{eq:tex-IO_d-t23}. 
The red dashed line indicates the prediction derived from the analysis. 
The blue solid and dashed lines respectively represent the best-fit value and $1\sigma$ range of $\theta_{23}$ given in Table \ref{tab:NuFIT}. 
As in the left panel, the horizontal axis range corresponds to the $3\sigma$ range of $\theta_{23}$ from Table \ref{tab:NuFIT}. 
Results for the other two-zero textures are summarized in Appendix \ref{app:two-zero}.

\medskip

Furthermore, the Majorana phases $\alpha_2, \alpha_3$ are expressed in terms of $\theta_{23}$ by substituting oscillation parameters $\theta_{12}, \theta_{13}$, together with the predicted parameters $\delta_{\rm CP}, m_1,m_2,m_3$ into Eqs.~\eqref{eq:R2_texture} and \eqref{eq:R3_texture}. 
Using these Majorana phases, the effective electron neutrino mass $m_{\nu_e}^{}$ and the effective neutrino mass $\langle m_{ee}\rangle$ can be evaluated. 
Predictions for viable structures are summarized in Table.~\ref{tab:prediction_mee_tex-NO} for NO and Table~\ref{tab:prediction_mee_tex-IO} for IO. 
These tables show the allowed regions of $\langle m_{ee}\rangle$, $m_{\nu_e}^{\rm eff}$, $\sum_i m_i$ and $\delta_{\rm CP}/\pi$ for each viable structure. 
We find characteristic predictions for $\delta_{\rm CP}$. 
In particular, the $B$-series structures predict values round $\delta_{\rm CP} \sim 1.5 \pi$ and $\delta_{\rm CP} \sim0.5\pi$.
$C$ structure predicts $1.3\pi \lesssim \delta_{\rm CP} \lesssim 1.6 \pi$ and $0.4 \pi\lesssim \delta_{\rm CP} \lesssim0.6\pi$. 
In addition, these structures will be tested by future experiments 
searching for neutrinoless double-beta decay such as nEXO. 

\medskip

When the minimum value of the sum of neutrino masses is considered, they suggest that even the surviving two-zero texture structures are driven into narrow corners of the allowed parameter space by neutrino oscillation data.
Thus, it is evident that these textures are now tightly constrained in light of improving experimental precision.

\newpage

\begin{table}[H]
    \caption{Summary of the effective neutrino mass for neutrinoless double beta decay $\langle m_{ee}\rangle$, the effective electron neutrino mass $m_{\nu_e}^{\rm eff}$, the sum of the predicted neutrino masses $\sum_i m_i$, and the Dirac CP phase $\delta_{\rm CP}/\pi$ for the viable two-zero textures in NO.}
    \label{tab:prediction_mee_tex-NO}
    \centering
    \begin{tabular}{|c||c|c|c|c|}\hline
         Structure& $\langle m_{ee}\rangle$ [eV] & $m_{\nu_e}^{\rm eff}$ [eV] & $\sum_i m_i$ [eV] & $\delta_{\rm CP}/\pi$  \\ \hline
         $A_1$ texture (NO)& $\sim 0$ & $\sim0.021$ & $0.065-0.068$ & $0.6 - 1.4$\\
         $A_2$ texture (NO)& $\sim 0$ & $\sim 0.021$ &$0.065-0.068$ & $0.0 - 0.5,~1.5 - 2.0$\\ \hline
         $B_1$ texture (NO)& $>0.058$& $>0.064$ & $>0.19$& $\sim 0.5,~\sim 1.5$\\
         $B_2$ texture (NO)& $>0.048$& $>0.054$ & $>0.16$ & $\sim 0.5,~\sim 1.5$\\
         $B_3$ texture (NO)& $>0.061$& $>0.067$ & $>0.20$ & $\sim 0.5,~\sim 1.5$\\
         $B_4$ texture (NO)& $>0.051$& $>0.057$ & $>0.17$ & $\sim 0.5,~\sim 1.5$\\ \hline
    \end{tabular}
\end{table}

\begin{table}[H]
    \caption{Summary of the effective neutrino mass for the neutrinoless double beta decay $\langle m_{ee}\rangle$, the effective electron neutrino mass $m_{\nu_e}^{\rm eff}$, the sum of the predicted neutrino masses $\sum_i m_i$, and the Dirac CP phase $\delta_{\rm CP}/\pi$ for the viable two-zero textures in IO.}
    \label{tab:prediction_mee_tex-IO}
    \centering
    \begin{tabular}{|c||c|c|c|c|}\hline
         Structure& $\langle m_{ee}\rangle$ [eV] & $m_{\nu_e}^{\rm eff}$ [eV] & $\sum_i m_i$ [eV] & $\delta_{\rm CP}/\pi$\\ \hline
         $B_1$ texture (IO)& $>0.068$& $>0.070$ & $>0.19$ & $\sim 0.5,~\sim 1.5$\\
         $B_3$ texture (IO)& $>0.070$& $>0.073$ & $>0.19$ & $\sim 0.5,~\sim 1.5$\\ \hline
         $C$ texture (IO)& $>0.044$& $>0.067$ & $>0.17$ & $0.4-0.6,~1.3 - 1.6$ \\ \hline
    \end{tabular}
\end{table}

\section{One-zero textures}
\label{sec:onezero}

In the previous section, we revisited the analysis of neutrino mass matrices with two-zero texture structures. 
However, as shown there, these structures are severely constrained. 
This motivates us to consider the next simplest class of structures, namely those with one-zero entry. 
Such structures are classified in Table \ref{tab:one-zero}.

\medskip

In this section, we discuss the general constraints on the neutrino mass matrix (or its inverse) with one-zero entry. 
We then examine the viable sets of neutrino parameters that satisfy these constraints using machine learning techniques.

\begin{table}[H]
    \caption{Classification of one-zero structures.}
    \label{tab:one-zero}
    \centering
    \begin{tabular}{|ccc|}\hline
         $G_1: \begin{pmatrix}  0 & * & * \\ * & * & * \\ * & * & * \end{pmatrix}$& 
         $G_2: \begin{pmatrix}  * & * & * \\ * & 0 & * \\ * & * & * \end{pmatrix}$ & 
         $G_3: \begin{pmatrix}  * & * & * \\ * & * & * \\ * & * & 0 \end{pmatrix}$ \\ \hline
         $H_1: \begin{pmatrix}  * & 0 & * \\ 0 & * & * \\ * & * & * \end{pmatrix}$&
         $H_2: \begin{pmatrix}  * & * & 0 \\ * & * & * \\ 0 & * & * \end{pmatrix}$& 
         $H_3: \begin{pmatrix}  * & * & * \\ * & * & 0\\ * & 0 & * \end{pmatrix}$ \\ \hline
    \end{tabular}
\end{table}

\subsection{Predictions from machine learning}

Recent applications of machine learning techniques to flavor physics (Refs.~\cite{Harvey:2021oue, Nishimura:2020nre, Matchev:2024ash, Kawai:2024pws, Nishimura:2024apb, Nishimura:2025rsk, Nishimura:2025knz, Baretz:2025zsv, Mendizabal:2025sbf, Abu-Ajamieh:2025mjk, Haba:2026uaf, Aranda:2026aaz}) have demonstrated the effectiveness of such approaches. 
In particular, generative artificial intelligence, i.e., generative AI, excels at producing new data based on a learned data distribution. 
For example, diffusion models have been employed to generate phenomenologically viable parameter points that satisfy experimental constraints, as shown in Refs.~\cite{Nishimura:2025rsk, Nishimura:2025knz}. 
Under these developments, we apply generative AI to generate data for the analysis of texture structures. 
Among various techniques, in this paper we adopt flow matching, which was proposed in Ref.~\cite{Lipman:2023hef} as a simulation-free framework for training continuous normalizing flows.

\medskip

Flow matching learns a vector field that transports samples from a simple source distribution, such as a Gaussian distribution, to the target data distribution. 
This vector field is conditioned on label information, enabling the generated samples to be guided toward desired physical observables. 
The implementation of flow matching consists of two phases: the training phase and the generation phase. 
During training, each data point $G$ with its label $L$ is connected to a randomly sampled point from the source distribution by a continuous path. 
The neural network is trained to learn the velocity field that moves points along these paths toward the data distribution. 
In the generation phase, a new point is first sampled from the noise distribution. 
This point is then transported by solving the learned flow equation under the desired label condition, producing a new sample that follows the learned data distribution.

\medskip

A useful feature of flow matching is the flexibility in choosing the probability path used for training. 
In particular, paths based on optimal transport displacement interpolation can produce straighter and simpler flows than standard diffusion-type paths. 
This can reduce the number of integration steps required during generation and improve sampling efficiency. 
Therefore, in this work, flow matching offers an efficient conditional generative method for producing data samples consistent with specified labels.

\subsubsection{Setup of flow matching}
\label{sec:flow_matching}

In this subsection, we describe the flow matching setup used in our numerical analysis. 
We employ flow matching as a conditional generative sampler for the parameter space of texture structures. 
The physical input to the neural network is prepared as a pair consisting of the generation target $G$ and the corresponding label $L$, where $G$ specifies the independent parameters of the mass matrix, while $L$ is calculated from the mass matrix.

\medskip

As a representative example, we explain the construction by using the one-zero texture $G_1$. 
Note that the same procedure applies to the other choices of the zero entry. 
For the $G_1$ texture, the mass matrix is written as
\begin{align}
    \mathcal{M}_{\nu}^{\rm flavor} = \Lambda_\nu
    \begin{pmatrix}
    0 & \alpha & \beta \\
    \alpha & \gamma & \delta \\
    \beta & \delta & \epsilon
    \end{pmatrix},
\end{align}
where $\{\alpha,\beta,\gamma,\delta,\epsilon\}$ are complex parameters and $\Lambda_\nu$ denotes the overall scale. 
We parameterize the latter by $s = \log_{10}\left(\Lambda_\nu/\meV\right)$.
Then, the generation target is defined as
\begin{align}
    G =
    \left\{
    \Re\,\alpha, \Im\,\alpha,
    \ldots,
    \Re\,\epsilon, \Im\,\epsilon,
    s
    \right\}.
\end{align}
Thus, $G$ has 11 real components. 
In preparing the training data, the real and imaginary parts of $\{\alpha,\beta,\gamma,\delta,\epsilon\}$ are sampled in the range
\begin{align}
    -1 \leq
    \left\{
    \Re\,\alpha, \Im\,\alpha,
    \ldots,
    \Re\,\epsilon, \Im\,\epsilon
    \right\}
    \leq 1,
\end{align}
while the scale parameter is taken as $-1 \leq s \leq 3$.
This range corresponds to $10^{-1}\,\meV \leq \Lambda_\nu \leq 10^{3}\,\meV$.

\medskip

For each sampled $G$, we construct $\mathcal{M}_{\nu}^{\rm flavor}$ and diagonalize it by the PMNS matrix as in Eq.~\eqref{eq:nu-mass_diag-flavor}. 
Then, we calculate the label $L$, which is used as the condition in flow matching. 
The label $L$ consists of two mass-squared differences and the absolute values of all PMNS matrix elements.
For the normal ordering, we use
\begin{align}
    L_{\rm NO}
    =
    \left\{
    \log_{10}\left(\Delta m_{21}^{2}/\meV^2\right),
    \log_{10}\left(\Delta m_{31}^{2}/\meV^2\right),
    V_{ij}
    \right\},
\end{align}
with $i=1,2,3$.
On the other hand, for the inverted ordering, the second mass-squared difference is replaced by $|\Delta m_{32}^{2}|$:
\begin{align}
    L_{\rm IO}
    =
    \left\{
    \log_{10}\left(\Delta m_{21}^{2}/\meV^2\right),
    \log_{10}\left(|\Delta m_{32}^{2}|/\meV^2\right),
    V_{ij}
    \right\}.
\end{align}
Therefore, $L$ has 11 real components in this setup. 
The logarithmic preprocessing is applied only to the mass-squared differences, whereas the absolute values of the PMNS matrix elements are used directly. 
This approach improves training efficiency for quantities that span several orders of magnitude.

\medskip

The simulated dataset consists of pairs $\left(G,L\right)$. 
This dataset is divided into training and validation sets with a ratio of 90\% and 10\%. 
For conditional flow matching, we use the simulation-based inference package \textit{sbi} introduced in Ref.~\cite{tejero-cantero2020sbi}. 
The flow-matching estimator is trained to learn the conditional distribution of $G$ given a label $L$. 
Regarding the training details, we adopt a transformer architecture for the flow-matching estimator, with a hidden dimension of 100 and 5 layers. 
The conditional labels $x=L$ are standardized using the $z$-score option ``independent'', while no $z$-score normalization is applied to the generated variables $\theta=G$. 
The estimator is trained by minimizing the flow-matching loss provided by \textit{sbi}. 
We use the Adam optimizer with a learning rate of $5.0\times10^{-4}$ and the OneCycleLR scheduler. 
The batch size is set to 256, and the network is trained for up to $5 \times 10^4$ optimization steps. 
The validation loss is evaluated every 1,000 steps, and early stopping with a patience of 20 is applied. 

\medskip

After training, we fix the labels to the experimental values of the neutrino mass-squared differences and mixing parameters, as specified in Table \ref{tab:NuFIT}. 
New candidates for $G$ are then generated from the learned conditional flow. 
These generated points are mapped back to the neutrino mass matrix, and the corresponding observables are recalculated to verify whether they satisfy the required phenomenological constraints. 
The CP phases are not imposed as direct labels in this setup; instead, they are evaluated from the generated mass matrices after the sampling procedure.

\medskip

The accuracy of the generated parameters is evaluated using the chi-squared value, defined as
\begin{align}
    \chi^{2} = \sum_{i} \left( \frac{P_{i}-\mu_{i}}{\sigma_{i}} \right)^{2}, \label{eq:chi_sq}
\end{align}
where $\{P_{i},\mu_{i},\sigma_{i}\}$ denotes the model-predicted value of a physical observable, its central value and the corresponding $1\sigma$ deviation, respectively. 
In our analysis, this quantity is calculated using the following five observables:
\begin{align}
\begin{split}
    P_{\mathrm{NO}}&=\left\{ \Delta m_{21}^{2},
    \Delta m_{31}^{2},\,
    \theta_{12},\,\theta_{23},\,\theta_{13} \right\},
    \\
    P_{\mathrm{IO}}&=\left\{ \Delta m_{21}^{2},
    |\Delta m_{32}^{2}|,\,
    \theta_{12},\,\theta_{23},\,\theta_{13} \right\}.
\end{split}
\label{eq:def_chisq}
\end{align}

\medskip

If the trained network does not achieve the required accuracy, several approaches can be considered to improve its performance. 
Fine-tuning is one such technique for enhancing the accuracy of generated samples, and its effectiveness has been demonstrated in Refs.~\cite{Nishimura:2025rsk,Nishimura:2025knz}. 
To obtain a sufficient number of viable parameter sets, we also apply fine-tuning to the flow matching estimator. 
In the actual analysis, fine-tuning is required for $G_2$ of NO and $\{G_2,G_3,H_1,H_2,H_3\}$ of IO to achieve a sufficiently accurate distribution of viable solutions. 
For these structures, we prepare 1,000,000 samples using a flow matching model that has been trained once, followed by the fine-tuning procedure.

\medskip

Only $G_2$ requires additional training in the case of NO, and the fine-tuning is conducted over 25 rounds. 
In the first round, we extract data with $\chi^2 \leq \chi^2_{\max}$ from the prepared 1,000,000 samples and retrain the neural network using this data. 
From the second round onward, 100,000 new samples are generated each round, and those satisfying $\chi^2 < \chi^2_{\max}$ are added to the accumulated training set. 
At each round, the estimator is initialized with the weights of the parent estimator, followed by fine-tuning using all accepted samples obtained up to that point. 
The $\chi^2$ threshold is gradually tightened during the procedure: $\chi^2_{\max} = 10{,}000$ for rounds 1--10, $\chi^2_{\max} = 5{,}000$ for rounds 11--20, and $\chi^2_{\max} = 1{,}000$ for rounds 21--25.

\medskip

On the other hand, the criterion for each round is organized as follows in the case of IO. 
We adopt $\chi^2_{\max}=5{,}000$ for the $G_3$ texture, while the $\{G_2,H_1,H_2,H_3\}$ textures are analyzed with $\chi^2_{\max}=10{,}000$. 
Then, 3, 14, 3, 5, and 12 rounds are conducted for $\{G_2,G_3,H_1,H_2,H_3\}$, respectively.

\subsubsection{Results from flow matching}
\label{sec:result_fm}

In this section, we present the numerical results for one-zero texture structures for NO and IO, following the method described in the previous section. 
The results are summarized in Table \ref{tab:one-zero-texture}. Compared to the two-zero texture case, one-zero textures have more degrees of freedom, and hence more structures remain viable. 
On the other hand, we find that some textures shown in Table~\ref{tab:one-zero-texture} are disfavored even in the one-zero texture case.
\begin{table}[H]
    \centering
    \caption{Summary of the one-zero texture analysis.}
    \label{tab:one-zero-texture}
    
    \begin{tabular}{ll cccccc}
        \toprule
        Structure & & $G_1$ & $G_2$ & $G_3$ & $H_1$ & $H_2$ & $H_3$ \\
        \midrule
        \multirow{2}{*}{CMB}     & NO & \hyperref[fig:G1tex_NO]{$\bigcirc$} & \hyperref[fig:G2tex_NO]{$\times$} & \hyperref[fig:G3tex_NO]{$\times$} & \hyperref[fig:H1tex_NO]{$\bigcirc$} & \hyperref[fig:H2tex_NO]{$\bigcirc$} & \hyperref[fig:H3tex_NO]{$\times$} \\
                                 & IO & \hyperref[fig:G1tex_IO]{$\times$}       & \hyperref[fig:G2tex_IO]{$\bigcirc$}       & \hyperref[fig:G3tex_IO]{$\bigcirc$} & \hyperref[fig:H1tex_IO]{$\bigcirc$}   & \hyperref[fig:H2tex_IO]{$\bigcirc$} & \hyperref[fig:H3tex_IO]{$\bigcirc$}    \\ 
        \midrule
        \multirow{2}{*}{CMB+BAO} & NO & \hyperref[fig:G1tex_NO]{$\bigcirc$} & \hyperref[fig:G2tex_NO]{$\times$} & \hyperref[fig:G3tex_NO]{$\times$}   & \hyperref[fig:H1tex_NO]{$\bigcirc$}   & \hyperref[fig:H2tex_NO]{$\bigcirc$}  & \hyperref[fig:H3tex_NO]{$\times$}   \\
                                 & IO & \hyperref[fig:G1tex_IO]{$\times$}       & \hyperref[fig:G2tex_IO]{$\bigcirc$}       & \hyperref[fig:G3tex_IO]{$\times$}   & \hyperref[fig:H1tex_IO]{$\bigcirc$}   & \hyperref[fig:H2tex_IO]{$\bigcirc$}   & \hyperref[fig:H3tex_IO]{$\times$}    \\
        \bottomrule
    \end{tabular}
\end{table}

\begin{figure}[H]
    \centering
    \includegraphics[width=107mm]{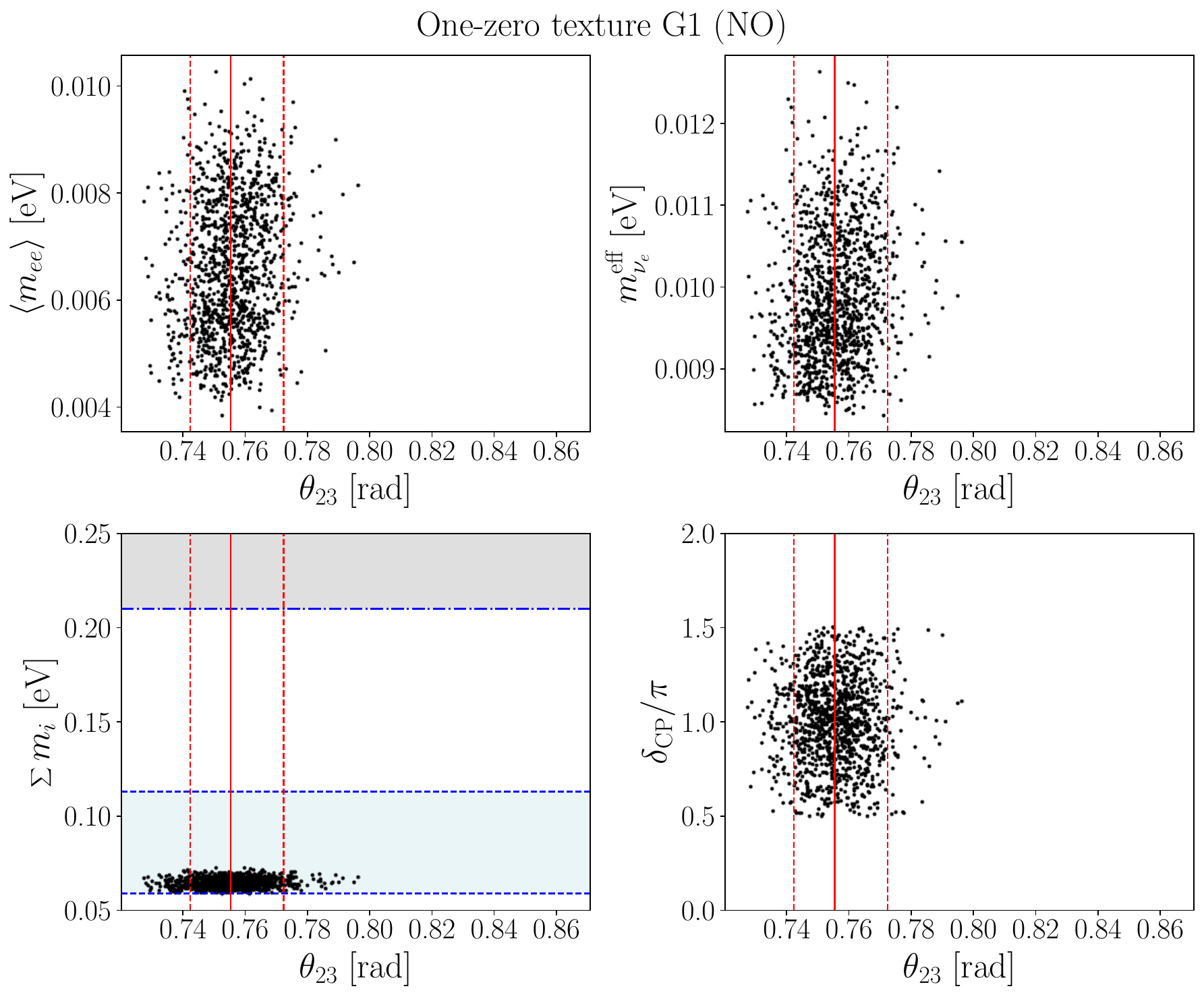}
    \caption{Distribution of observables for the $G_1$ structure with NO. In the upper row, the left (right) panel shows $\theta_{23}$ vs.~$\langle m_{ee}\rangle$ ($m_{\nu_e}^{\mathrm{eff}}$). In the lower row, the left (right) panel shows $\theta_{23}$ vs.~$\Sigma\,m_{i}$ ($\delta_{\mathrm{CP}}/\pi$). These 1,165 points satisfy the condition $\chi^2 < 45$.}
    \label{fig:ml_G1NO}
\end{figure}

\begin{figure}[H]
    \centering
    \includegraphics[width=107mm]{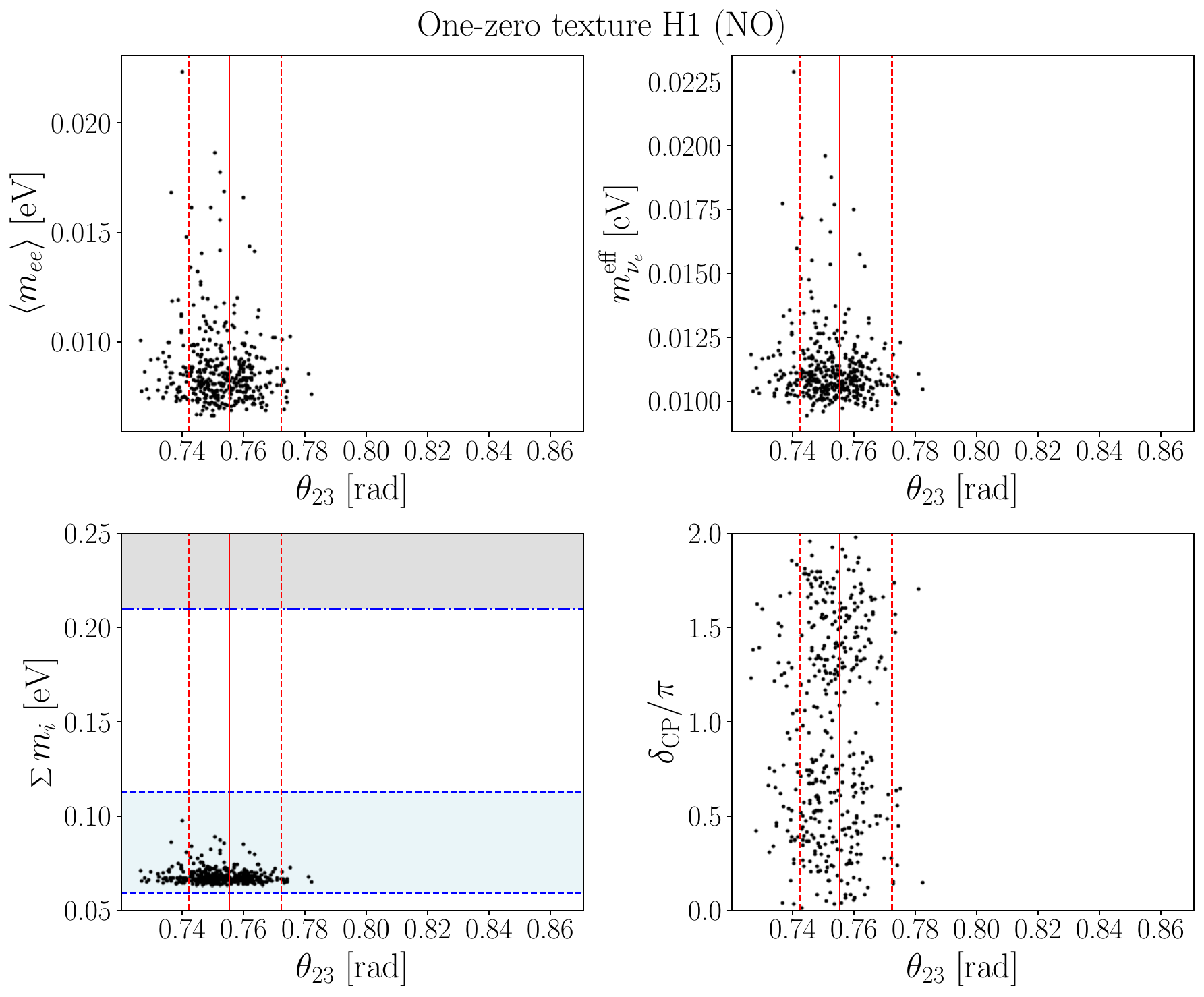}
    \caption{Distribution of observables for the $H_1$ structure with NO. The figure layout is the same as in Fig.~\ref{fig:ml_G1NO}. These 413 points satisfy the condition $\chi^2 < 45$.}
    \label{fig:ml_H1NO}
\end{figure}

\begin{figure}[H]
    \centering
    \includegraphics[width=107mm]{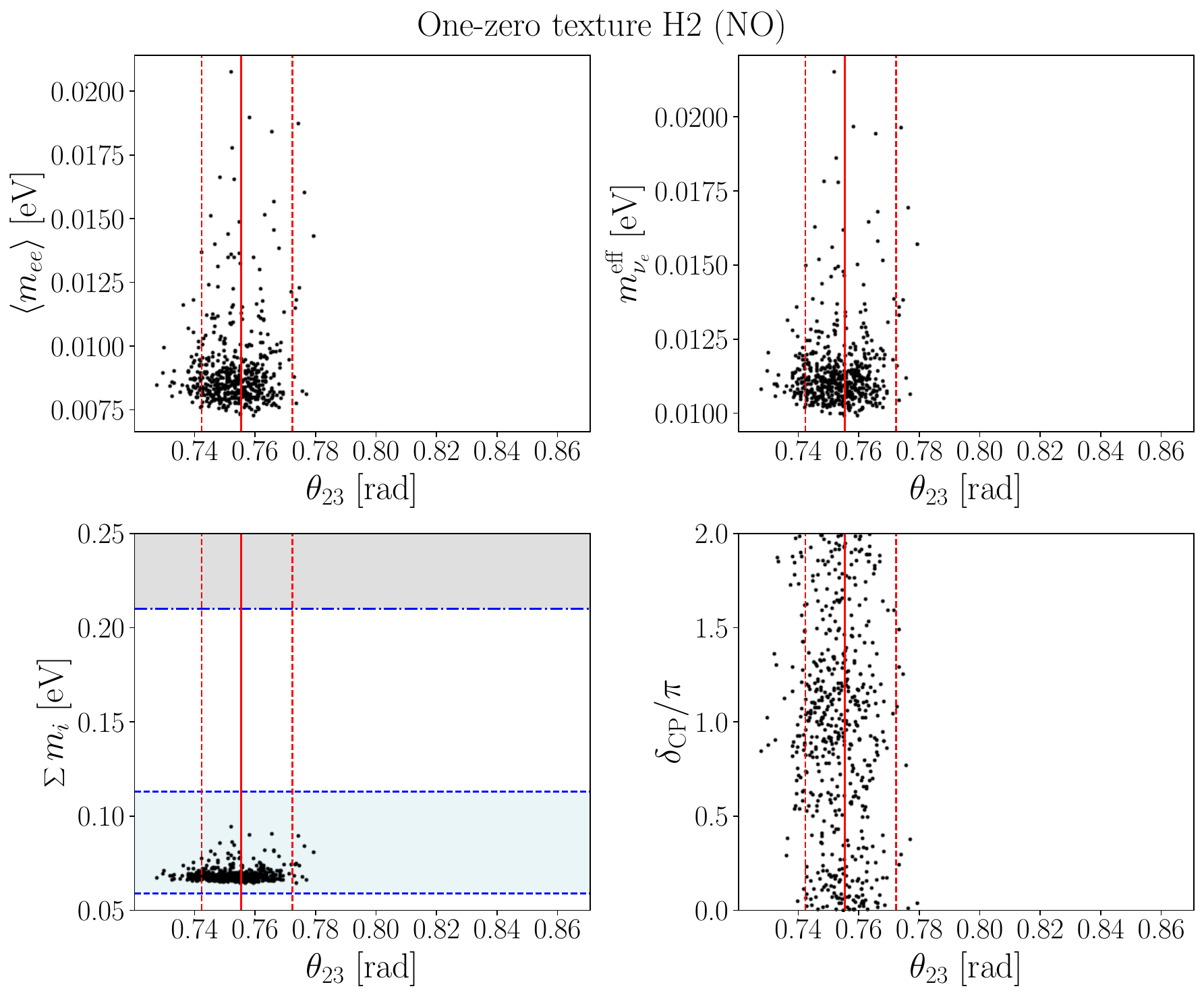}
    \caption{Distribution of observables for the $H_2$ structure with NO. The figure layout is the same as in Fig.~\ref{fig:ml_G1NO}. These 534 points satisfy the condition $\chi^2 < 45$.}
    \label{fig:ml_H2NO}
\end{figure}

\begin{figure}[H]
    \centering
    \includegraphics[width=107mm]{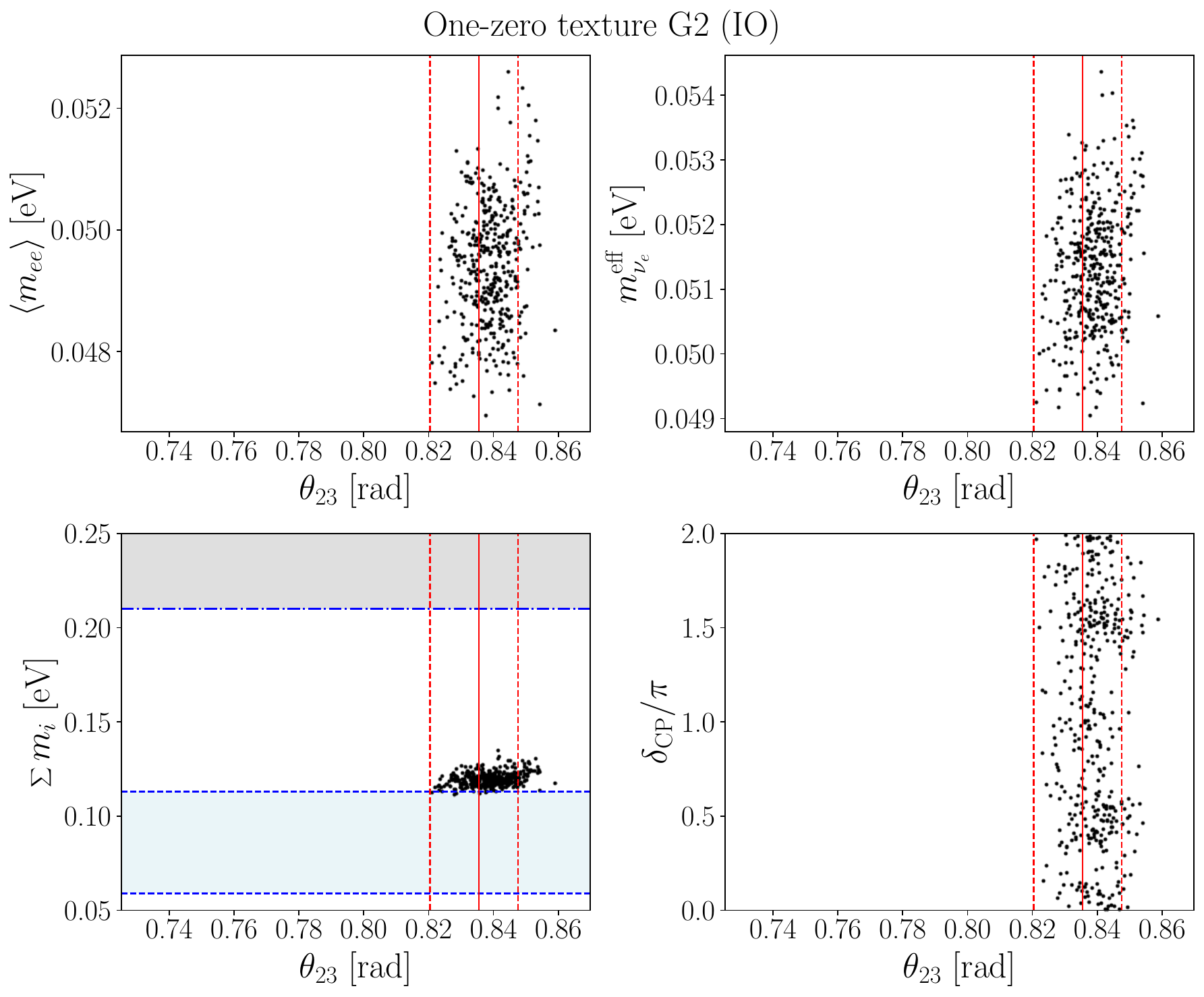}
    \caption{Distribution of observables for the $G_2$ structure with IO. The figure layout is the same as in Fig.~\ref{fig:ml_G1NO}. These 401 points satisfy the condition $\chi^2 < 45$.}
    \label{fig:ml_G2IO}
\end{figure}

\begin{figure}[H]
    \centering
    \includegraphics[width=107mm]{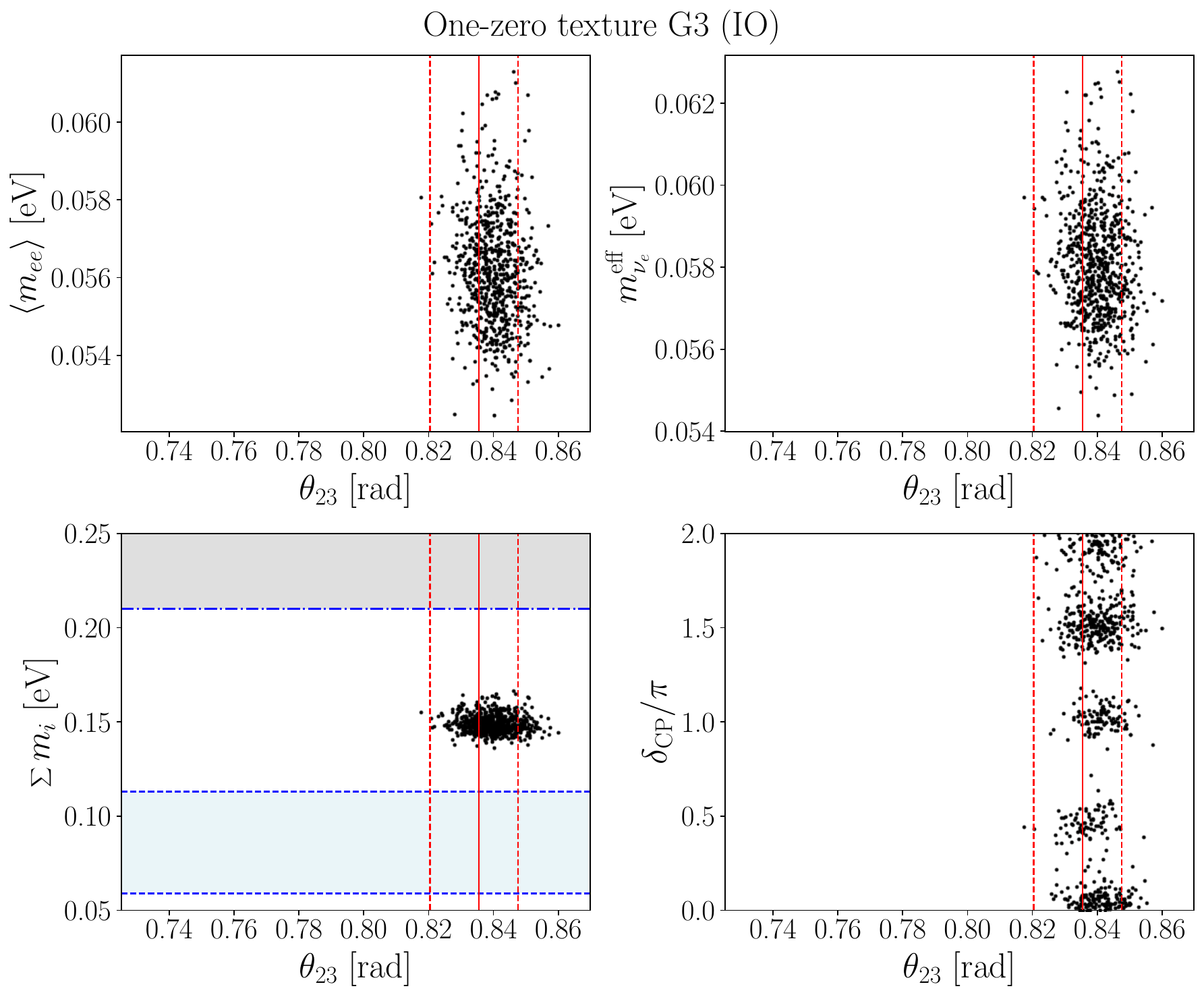}
    \caption{Distribution of observables for the $G_3$ structure with IO. The figure layout is the same as in Fig.~\ref{fig:ml_G1NO}. These 700 points satisfy the condition $\chi^2 < 45$.}
    \label{fig:ml_G3IO}
\end{figure}

\begin{figure}[H]
    \centering
    \includegraphics[width=107mm]{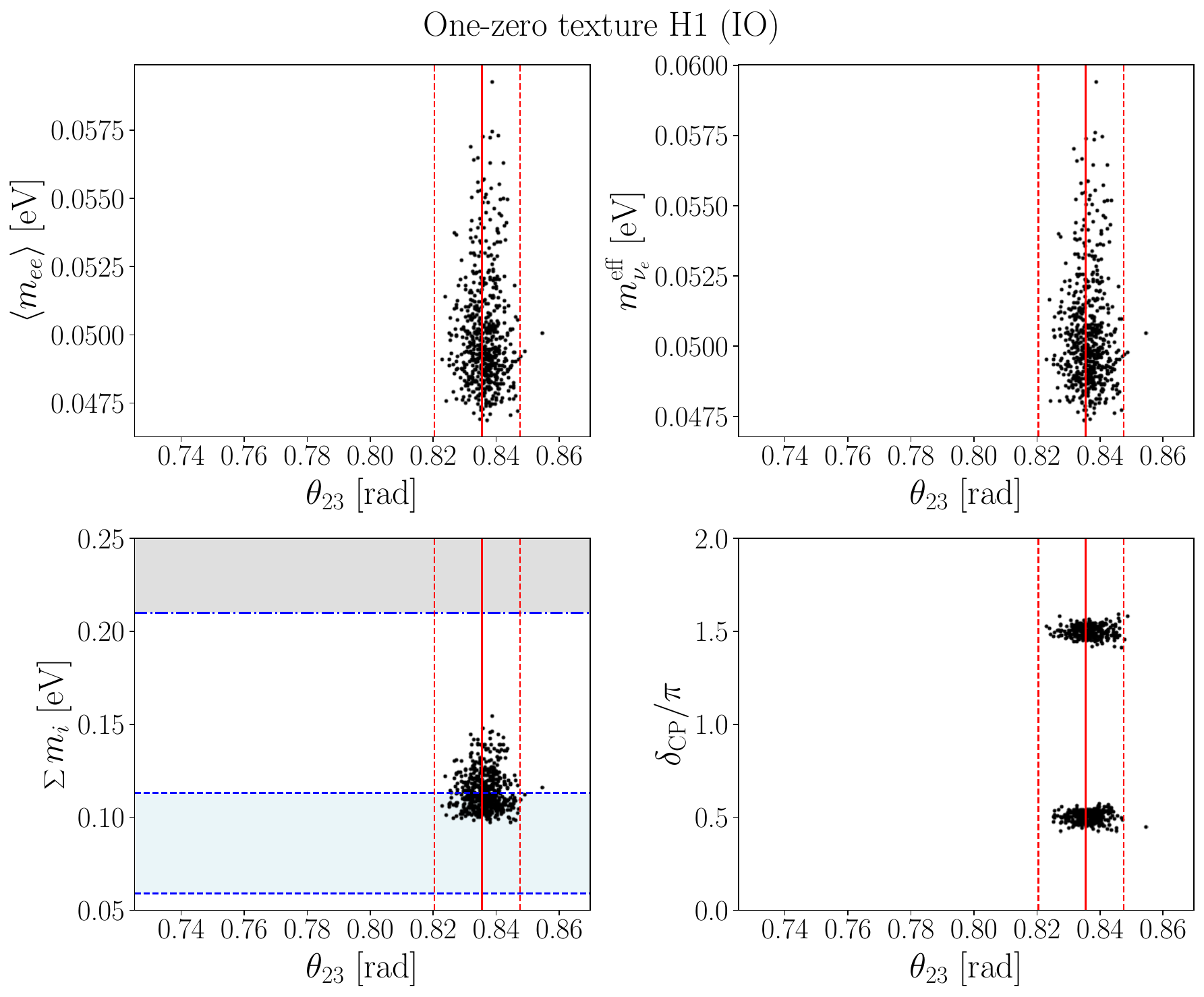}
    \caption{Distribution of observables for the $H_1$ structure with IO. The figure layout is the same as in Fig.~\ref{fig:ml_G1NO}. These 589 points satisfy the condition $\chi^2 < 45$.}
    \label{fig:ml_H1IO}
\end{figure}

\begin{figure}[H]
    \centering
    \includegraphics[width=107mm]{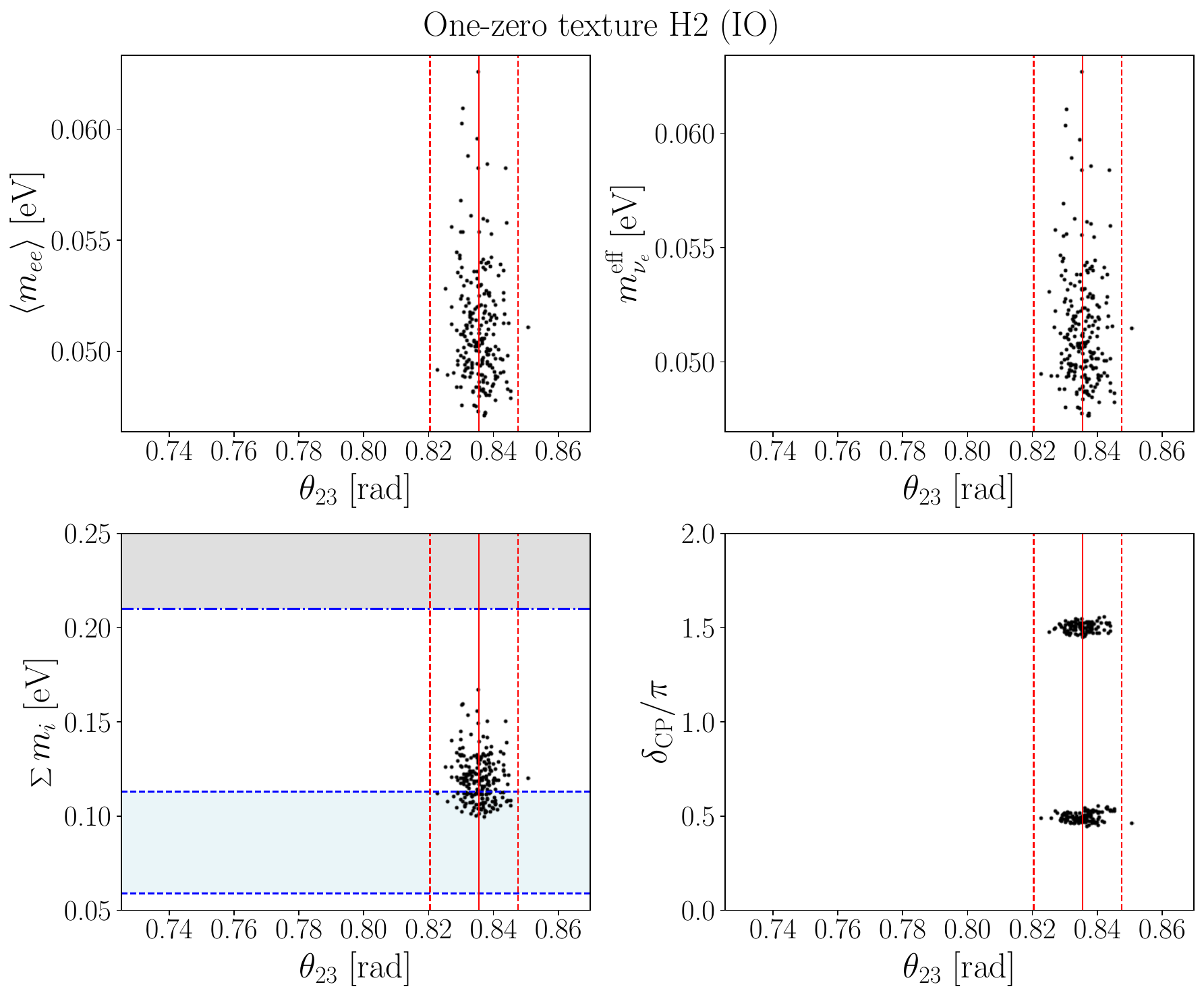}
    \caption{Distribution of observables for the $H_2$ structure with IO. The figure layout is the same as in Fig.~\ref{fig:ml_G1NO}. These 218 points satisfy the condition $\chi^2 < 45$.}
    \label{fig:ml_H2IO}
\end{figure}

\begin{figure}[H]
    \centering
    \includegraphics[width=107mm]{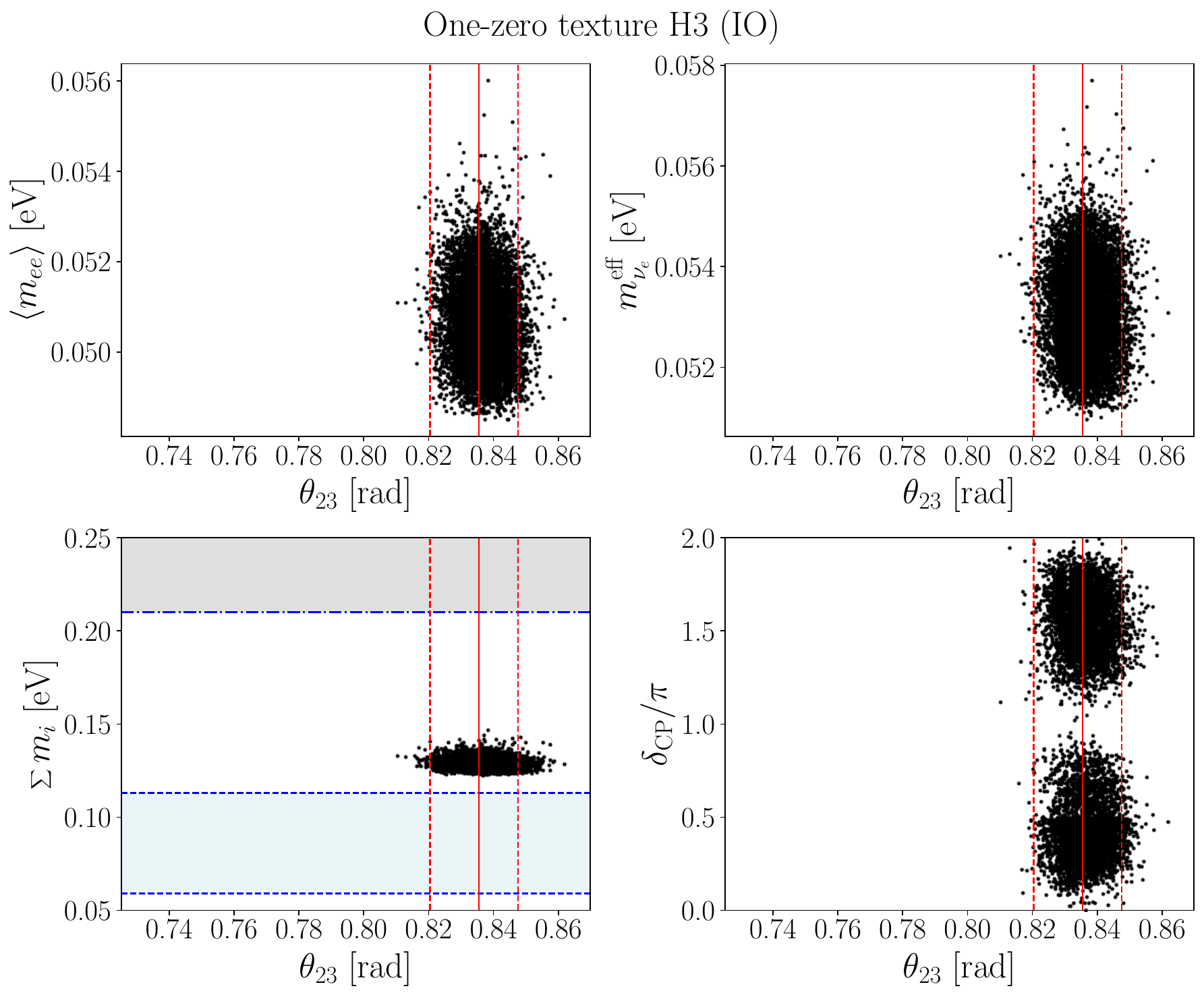}
    \caption{Distribution of observables for the $H_3$ structure with IO. The figure layout is the same as in Fig.~\ref{fig:ml_G1NO}. These 9,078 points satisfy the condition $\chi^2 < 45$.}
    \label{fig:ml_H3IO}
\end{figure}

We plot the distributions of observables generated by flow matching in Figs.~\ref{fig:ml_G1NO}-\ref{fig:ml_H2NO} for the $\{G_1,H_1,H_2\}$ textures with NO, and in Figs.~\ref{fig:ml_G2IO}-\ref{fig:ml_H3IO} for the $\{G_2,G_3,H_1,H_2,H_3\}$ textures with IO. 
In all figures, the red solid and dashed lines represent the experimental best-fit values and 1$\sigma$ ranges shown in Table \ref{tab:NuFIT}, respectively. 
Furthermore, the gray shaded regions are excluded by the CMB constraint, while the blue shaded regions are allowed by the combined CMB+BAO constraints for NO and IO, given in Eqs.~\eqref{eq:const_NO} and \eqref{eq:const_IO}, respectively. 
For structures other than those listed here, flow matching does not yield parameter points consistent with the observational constraints. 

\medskip

The predicted effective neutrino mass for neutrinoless double beta decay $\langle m_{ee}\rangle$, the effective electron neutrino mass $m_{\nu_e}^{\rm eff}$, the sum of the predicted neutrino masses $\sum_i m_i$, and the Dirac CP phase $\delta_{\rm CP}/\pi$ are summarized in Table~\ref{tab:prediction_mee_tex-NO-onezero} for NO and Table~\ref{tab:prediction_mee_tex-IO-onezero} for IO. 
For IO, the viable one-zero textures lead to sizable values of $\langle m_{ee}\rangle$, but their typical ranges are lower than those obtained in the two-zero texture case. 
A similar tendency is also seen in the predicted sum of neutrino masses, $\sum_i m_i$. Furthermore, $H_1$ and $H_2$ textures with IO prefer values of $\delta_{\rm CP}$ around $\pi/2$ and $3\pi/2$, which can be understood using the analytical methods introduced in the next section. 
Future cosmological observations and neutrino experiments will therefore be valuable not only for testing these predictions but also for discriminating between one-zero and two-zero texture scenarios. 

\begin{table}[H]
    \caption{Summary of the effective neutrino mass for neutrinoless double beta decay $\langle m_{ee}\rangle$, the effective electron neutrino mass $m_{\nu_e}^{\rm eff}$, the sum of the predicted neutrino masses $\sum_i m_i$, and the Dirac CP phase $\delta_{\rm CP}/\pi$ for the viable one-zero textures with NO.}
    \label{tab:prediction_mee_tex-NO-onezero}
    \centering
    \begin{tabular}{|c||c|c|c|c|}\hline
         Structure& $\langle m_{ee}\rangle$ [eV] & $m_{\nu_e}^{\rm eff}$ [eV] & $\sum_i m_i$ [eV] & $\delta_{\rm CP}/\pi$  \\ \hline
         $G_1$ texture (NO)& $0.004-0.010$ & $0.008-0.013$ & $0.0590-0.0721$ & $0.5 - 1.5$\\ \hline
         $H_1$ texture (NO)& $0.007-0.022$ & $0.009-0.023$ &$0.064-0.098$ & $0.0-2.0$\\
         $H_2$ texture (NO)& $0.007-0.021$& $0.010-0.022$ & $0.065-0.094$& $0.0-2.0$\\ \hline
    \end{tabular}
\end{table}

\newpage
\vspace*{\stretch{1}}

\begin{table}[H]
    \caption{Summary of the effective neutrino mass for the neutrinoless double beta decay $\langle m_{ee}\rangle$, the effective electron neutrino mass $m_{\nu_e}^{\rm eff}$, sum of the predicted neutrino masses $\sum_i m_i$, and the Dirac CP phase $\delta_{\rm CP}/\pi$ for the viable one-zero textures with IO.}
    \label{tab:prediction_mee_tex-IO-onezero}
    \centering
    \begin{tabular}{|c||c|c|c|c|}\hline
         Structure & $\langle m_{ee}\rangle$ [eV] & $m_{\nu_e}^{\rm eff}$ [eV] & $\sum_i m_i$ [eV] & $\delta_{\rm CP}/\pi$ \\ \hline

         $G_2$ texture (IO) & $0.047-0.053$ & $0.049-0.054$ & $0.111-0.135$ & $0.0-2.0$ \\

         $G_3$ texture (IO) & $0.052-0.061$ & $0.054-0.063$ & $0.136-0.166$ & $0.0 - 2.0$\\ \hline

         $H_1$ texture (IO) & $0.047 - 0.059$ & $0.047 - 0.059$ & $0.0975 - 0.1541$ &
         \makecell{
         $0.42 - 0.58$\\
         $1.4 - 1.6$
         } \\ \hline

         $H_2$ texture (IO) & $0.047 - 0.063$ & $0.048 - 0.063$ & $0.0997 - 0.1668$ &
         \makecell{
         $0.44 - 0.56$\\
         $1.4 - 1.6$
         } \\ \hline

         $H_3$ texture (IO) & $0.049 - 0.056$ & $0.051 - 0.058$ & $0.1228 - 0.1467$ & $0.0 - 2.0$ \\ \hline
    \end{tabular}
\end{table}

\vspace{\stretch{1}}

\vspace{\stretch{1}}
\newpage
\vspace*{\stretch{1}}

\vspace{\stretch{1}}
\newpage
\vspace*{\stretch{1}}

\vspace{\stretch{1}}
\newpage
\vspace*{\stretch{1}}

\vspace{\stretch{1}}
\newpage

\subsection{Analytical method for one-zero textures}

In this subsection, we introduce an analytical methodology used to verify the results obtained from flow matching. 
In the case of one-zero structures, we can use only a single equation: either one of Eqs.~\eqref{tex_component} is set to zero for the one-zero textures, or one of the corresponding minor conditions is set to zero for one-zero minors. 
However, this is not enough to obtain meaningful predictions due to the large number of unconstrained parameters. 
Even with the limited number of constraints, we can obtain the viable region by using the one-zero structure together with the cosmological constraints. 
We then consider how to constrain structures with one-zero components by using inequalities and the cosmological constraints on the sum of neutrino masses.

\medskip

First, we consider the texture case in which a $(i,j)$ component is equal to zero:
\begin{align}
    \left(\mathcal{M}_{\nu}^{\rm flavor}\right)^*_{ij}&= 0.
\end{align}
From Eqs.~\eqref{tex_component}, this complex equation can be written by
\begin{align}
    -V_{i1}V_{j1}m_1 &= e^{i\alpha_2}V_{i2}V_{j2}m_2+e^{i\alpha_3}V_{i3}V_{j3}m_3.
\end{align}
Then, applying the triangle inequality to the right-hand side of this equation and taking the absolute value of the left-hand side, we obtain
\begin{align}
    ||V_{i2}V_{j2}m_2|- |V_{i3}V_{j3}m_3|| \leq |V_{i1}V_{j1}m_1| \leq ||V_{i2}V_{j2}m_2|+ |V_{i3}V_{j3}m_3||.\label{eq:const_1zero_tex}
\end{align}
Here, we fix some parameters, $\theta_{ij}$ ($ij=13$ for $G_1$, and $ij=12, 13$ for others), $\Delta m^2_{21},\Delta m^2_{3\ell}$, to the best-fit value of the NuFIT 6.0 global fit~\cite{Esteban:2024eli}, as in the analysis in Sec.~\ref{sec:twozero}. 

\medskip

Since the parameter space remains too vast for an exhaustive analysis of the texture structures, we use the cosmological bounds discussed in Sec.~\ref{subsec:const_msum}. 
Fixing the sum of neutrino masses uniquely determines the individual active neutrino masses via the mass-squared differences $\Delta m^2_{ij}$ provided by the NuFIT 6.0 global fit \cite{Esteban:2024eli}. 
This enables us to scan the parameter space restricted to the cosmologically viable region.
Concretely, we explore the parameter space by varying the mass sum $\sum m_i$ in discrete steps within the viable region permitted by the cosmological constraints in Sec.~\ref{subsec:const_msum}. 
For each fixed value of the mass sum, the individual masses are sequentially evaluated using the $\Delta m^2_{ij}$ values provided by NuFIT 6.0 \cite{Esteban:2024eli}.

\medskip

Once the individual masses are fixed, Eqs.~\eqref{eq:const_1zero_tex} become an inequality involving $\theta_{23}$ ($\theta_{12}$ for $G_1$) and $\delta_{\rm CP}$ for each texture class. 
Then, by varying $\delta_{\rm CP}$ in discrete steps over the range of $[0,2\pi]$, the inequality finally gives the viable region of $\theta_{23}$ ($\theta_{12}$ for $G_1$) which is consistent with the constraints of the mass sum.
Combining these results, we derive constraints on the one-zero textures, as shown in the next subsection.

\subsection{Constraints on one-zero texture structures}

\begin{figure}[H] 
    \vspace{-10 mm}
  \centering
  \begin{minipage}{0.40\textwidth} 
    \centering
    \includegraphics[width=\linewidth]{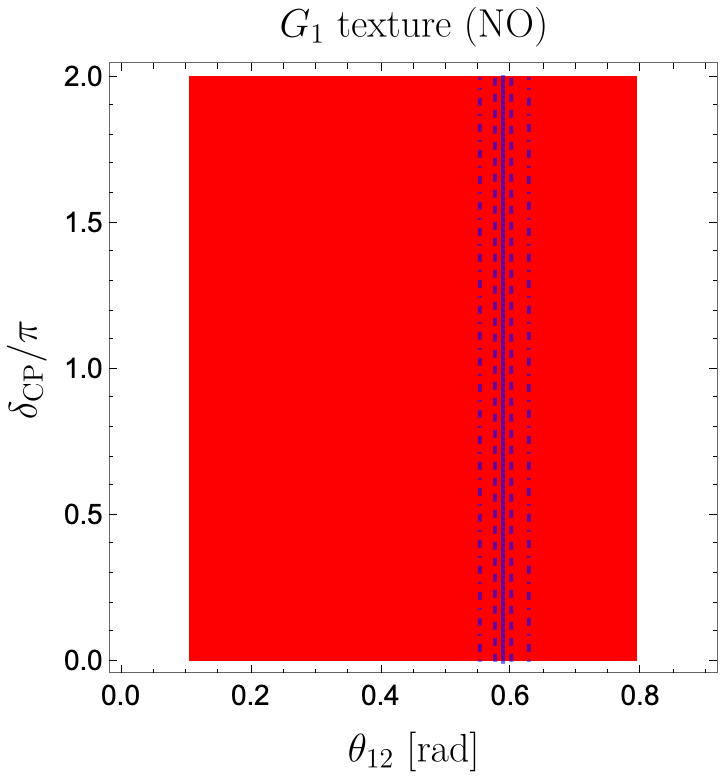}
    \label{fig:G1tex_NO}
  \end{minipage}\hfill 
  \begin{minipage}{0.4\textwidth}
    \centering
    \includegraphics[width=\linewidth]{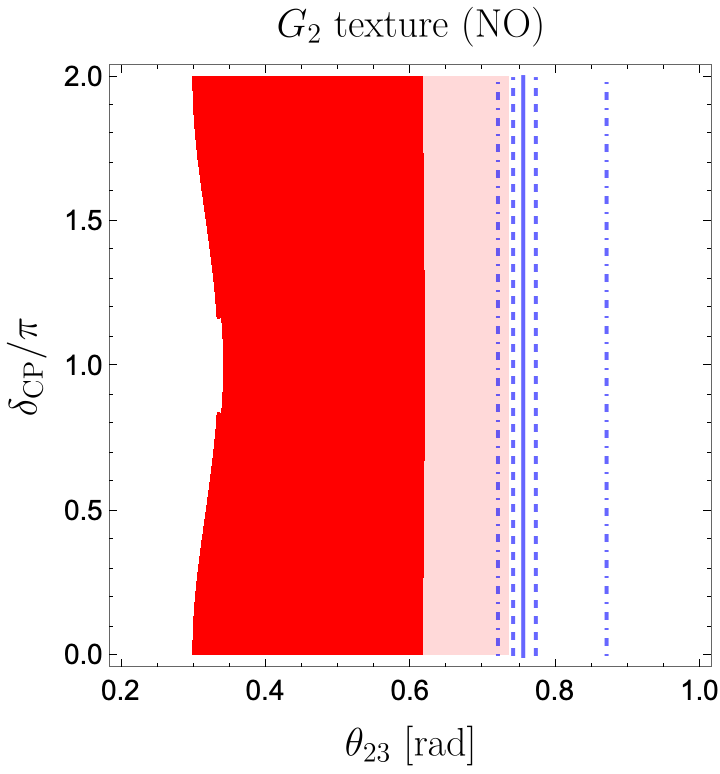}
    \label{fig:G2tex_NO}
  \end{minipage}

  \begin{minipage}{0.4\textwidth}
    \centering
    \includegraphics[width=\linewidth]{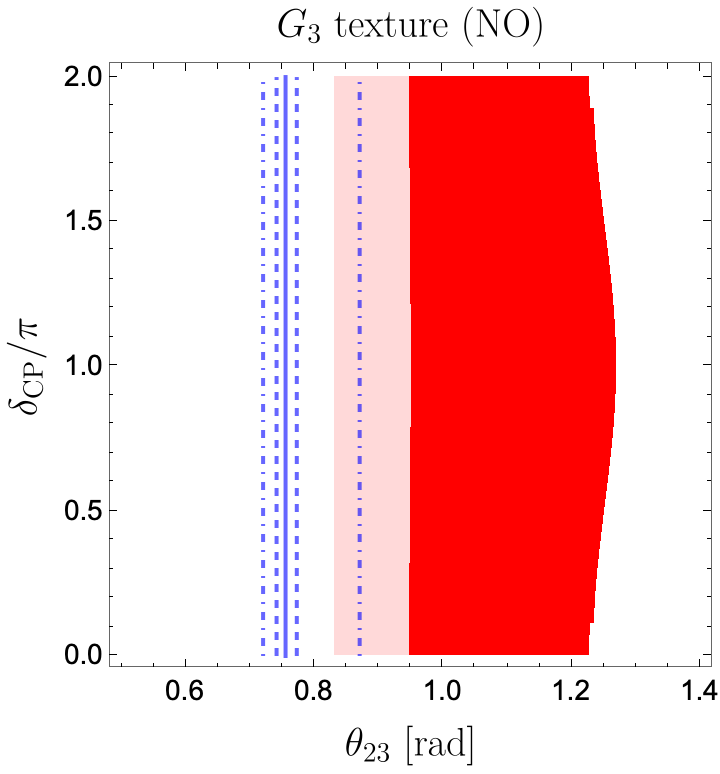}
    \label{fig:G3tex_NO}
  \end{minipage}\hfill
  \begin{minipage}{0.4\textwidth}
    \centering
    \includegraphics[width=\linewidth]{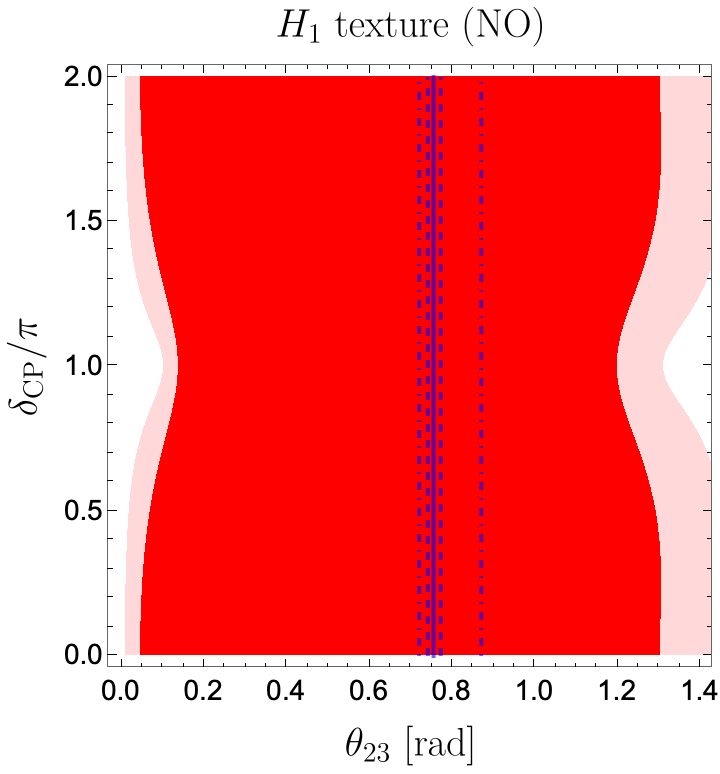}
    \label{fig:H1tex_NO}
  \end{minipage}
  

  \begin{minipage}{0.4\textwidth}
    \centering
    \includegraphics[width=\linewidth]{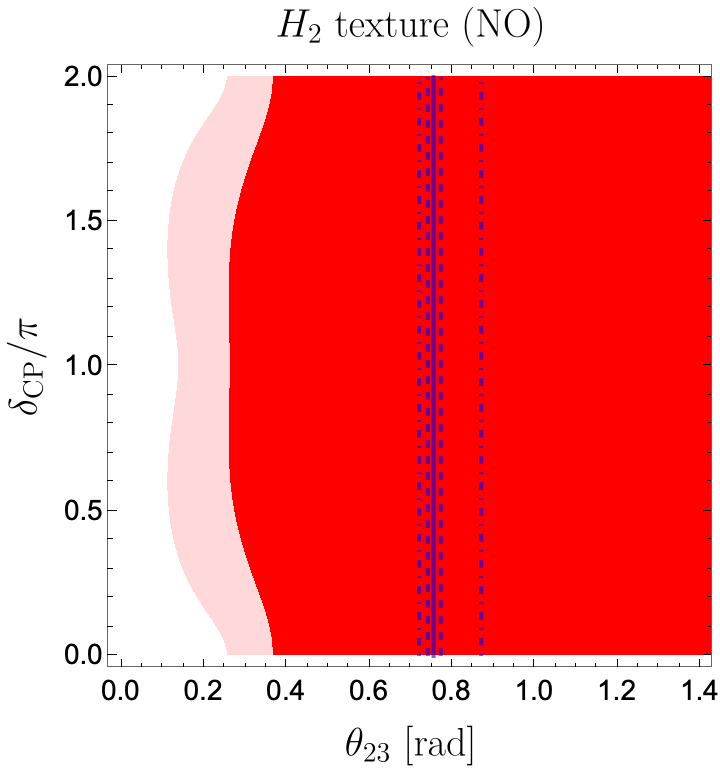}
    \label{fig:H2tex_NO}
  \end{minipage}\hfill
  \begin{minipage}{0.4\textwidth}
    \centering
    \includegraphics[width=\linewidth]{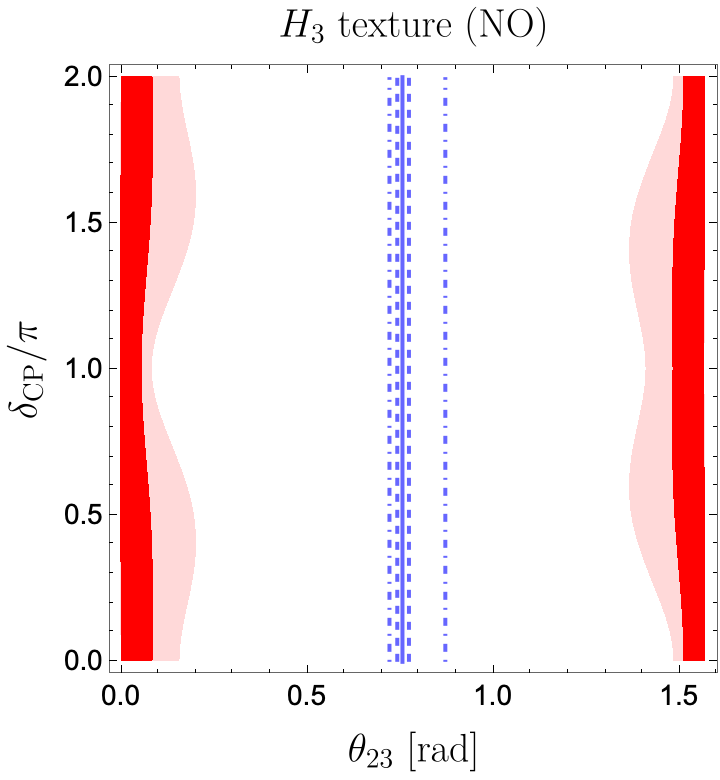}
    \label{fig:H3tex_NO}
  \end{minipage}
  
  \caption{Viable regions satisfying Eq.~\eqref{eq:const_1zero_tex} for NO. The vertical axis shows $\delta_{\rm CP}/\pi$ and the horizontal axis shows the $\theta_{23}$ ($\theta_{12}$ for $G_1$).
The blue solid, dashed and dashdotted lines show the best-fit value, $1\sigma$ range, $3\sigma$ range of $\theta_{23}$ in Table \ref{tab:NuFIT}, respectively. 
The light red region shows the viable region consistent with the constraint, $\sum m_\nu <0.21~{\rm eV}$, and the red region shows the viable region consistent with the constraint, $0.059 ~{\rm eV}< \sum m_\nu <0.113~{\rm eV}$. }
  \label{fig:one-zero_tex_NO}
\end{figure}

\begin{figure}[H] 
  \centering
  \vspace{-10 mm}
  \begin{minipage}{0.40\textwidth} 
    \centering
    \includegraphics[width=\linewidth]{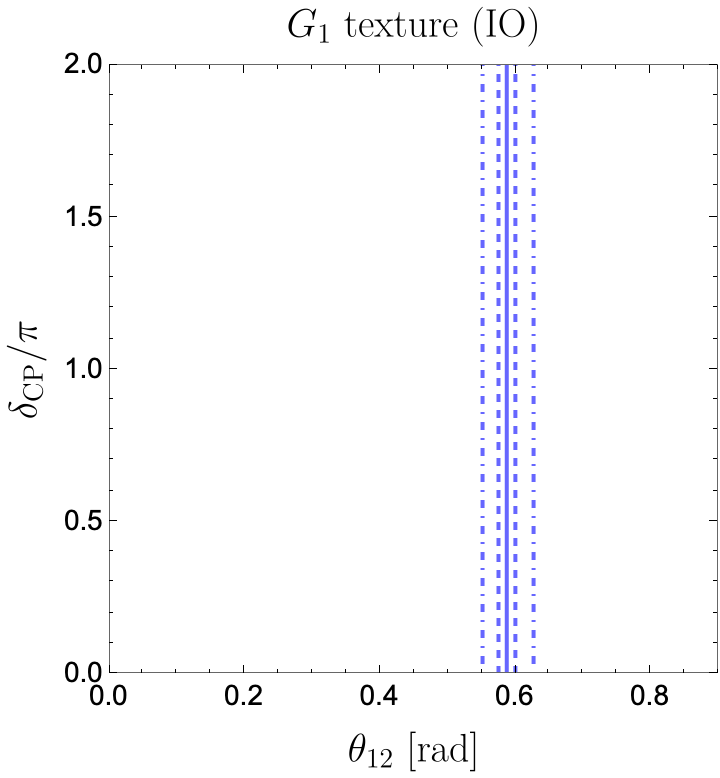}
    \label{fig:G1tex_IO}
  \end{minipage}\hfill 
  \begin{minipage}{0.4\textwidth}
    \centering
    \includegraphics[width=\linewidth]{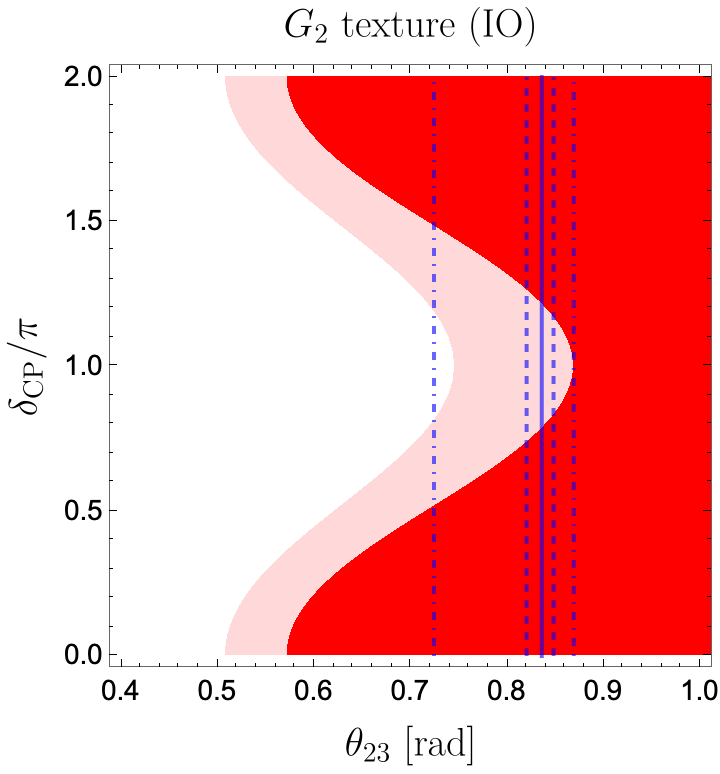}
    \label{fig:G2tex_IO}
  \end{minipage}
  
  \begin{minipage}{0.4\textwidth}
    \centering
    \includegraphics[width=\linewidth]{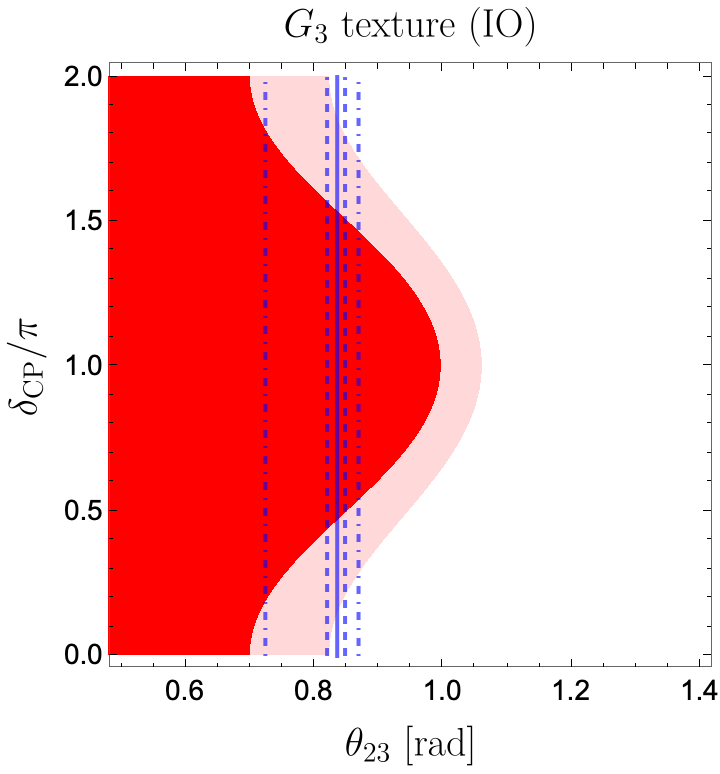}
    \label{fig:G3tex_IO}
  \end{minipage}\hfill
  \begin{minipage}{0.4\textwidth}
    \centering
    \includegraphics[width=\linewidth]{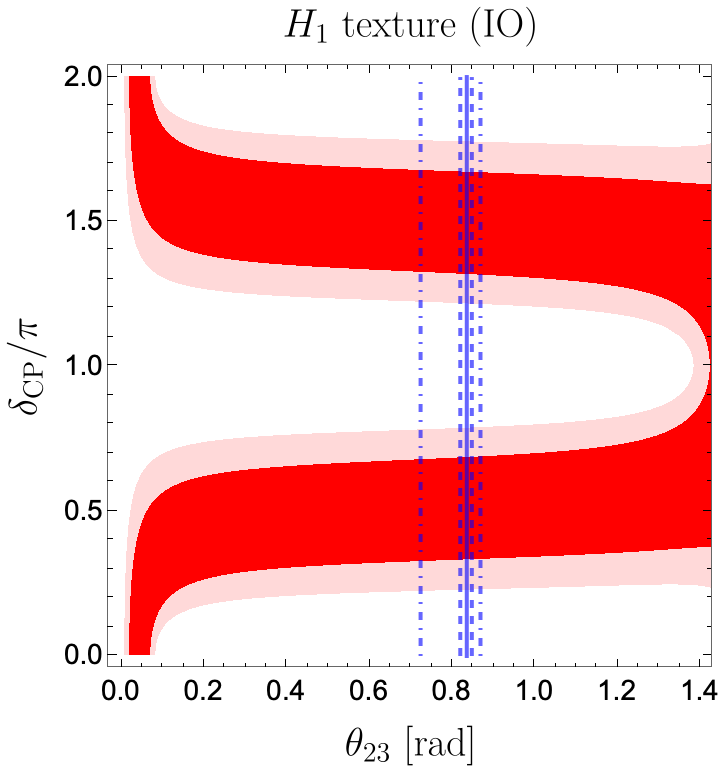}
    \label{fig:H1tex_IO}
  \end{minipage}
  
  \begin{minipage}{0.4\textwidth}
    \centering
    \includegraphics[width=\linewidth]{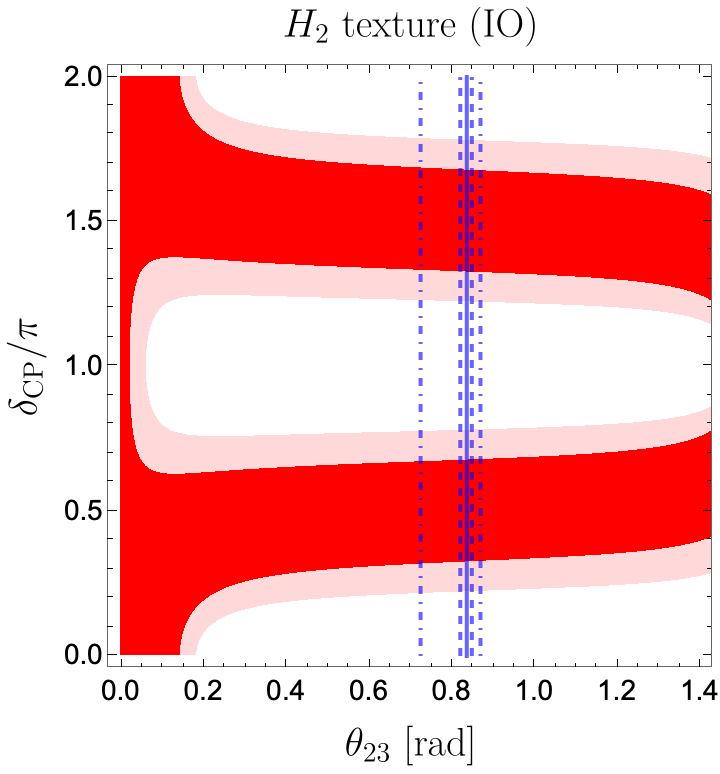}
    \label{fig:H2tex_IO}
  \end{minipage}\hfill
  \begin{minipage}{0.4\textwidth}
    \centering
    \includegraphics[width=\linewidth]{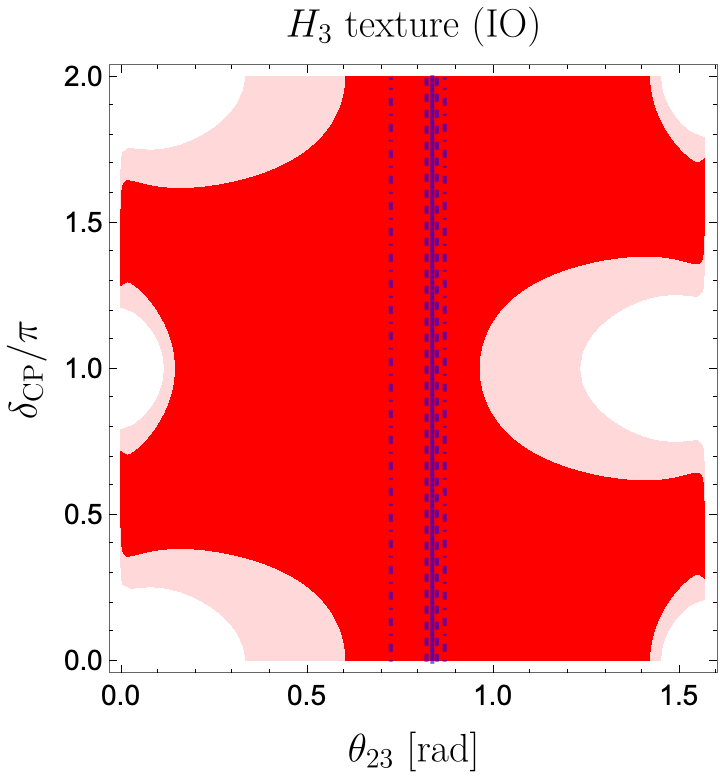}
    \label{fig:H3tex_IO}
  \end{minipage}
  
  \caption{Viable regions satisfying Eq.~\eqref{eq:const_1zero_tex} for IO. The vertical axis shows $\delta_{\rm CP}/\pi$ and the horizontal axis shows the $\theta_{23}$ ($\theta_{12}$ for $G_1$). 
The blue solid, dashed and dashdotted lines show the best-fit value, $1\sigma$ range, $3\sigma$ range of $\theta_{23}$ in Table \ref{tab:NuFIT}, respectively. 
The light red region shows the viable region consistent with the constraint, $\sum m_\nu <0.21~{\rm eV}$, and the red region shows the viable region consistent with the constraint, $0.10 ~{\rm eV}< \sum m_\nu <0.145~{\rm eV}$. }
  \label{fig:one-zero_tex_IO}
\end{figure}

The details of the results are shown in Fig.~\ref{fig:one-zero_tex_NO} and Fig.~\ref{fig:one-zero_tex_IO}. 
The vertical axis represents $\delta_{\rm CP}/\pi$ and the horizontal axis means $\theta_{23}$ ($\theta_{12}$ for $G_1$). 
The blue solid (dashed, dashdotted) line shows the best-fit $(1\sigma, 3\sigma)$ value of $\theta_{23}$ ($\theta_{12}$ for $G_1$) in Table \ref{tab:NuFIT}. 
The light red region denotes the viable region consistent with the constraint Eq.~\eqref{eq:const_CMB} from CMB, $\sum m_\nu <0.21~{\rm eV}$, and the red region indicates the viable region consistent with the constraints Eqs.~\eqref{eq:const_NO} and \eqref{eq:const_IO} from CMB+BAO, $0.059 ~{\rm eV}< \sum m_\nu <0.113~{\rm eV}$ for NO and $0.10 ~{\rm eV}< \sum m_\nu <0.145~{\rm eV}$ for IO.

\medskip

In the NO case, the $G_3$ texture is not viable with respect to Eq.~\eqref{eq:const_NO}, but it is viable under Eq.~\eqref{eq:const_CMB}. 
The $H_3$ texture with NO is not viable in either mass range. 
On the other hand, the other texture structures in NO remain viable under the cosmological mass sum constraints. 

\medskip

In the IO case, the $G_1$ structure has no viable region; however, the other structures can be realized consistently with the cosmological limits. 
In particular, the $H_1$ and $H_2$ textures predict that $\delta_{\rm CP}/\pi$ takes values around $0.5$ and $1.5$.

\medskip

Comparing these analytical results with those obtained from flow matching, we find that they are in good agreement. 
Indeed, the distribution of $\delta_{\mathrm{CP}}$ for the $H_1$ and $H_2$ textures with IO is consistent between the two approaches.

\newpage
\section{Realization of one-zero neutrino mass textures}
\label{sec:UV}

In this section, we propose a scenario to realize the one-zero textures of the neutrino mass matrix. 
Some neutrino masses discussed in the previous section cannot be realized based on group-theoretical symmetries. 

\medskip

To illustrate this point, let us discuss whether one can realize the $G_1$ texture based on a $U(1)$ symmetric model 
in which the neutrinos carry $U(1)$ charges $q_i$ with $i=1,2,3$. 
Note that the top-left $2 \times 2$ submatrix of $(m_\nu)_{ij}$ with $i,j=1,2$ has a specific structure in the $G_1$ texture, and it can only be achieved if the following charge conservation conditions hold:
\begin{align}
q_1 + q_1 \neq 0, \quad
q_2 + q_1 = 0, \quad
q_2 + q_2 = 0.
\end{align}
However, no viable charge assignments exist within $U(1)$ symmetric models that can successfully reproduce the $G_1$ texture. This constraint remains equally valid for non-Abelian symmetries. 
Here, we focus only on the neutrino mass matrix, but the same statement holds even when we consider the Weinberg operator generating the neutrino mass matrix. 
Furthermore, this statement applies to the other textures including $G_{2,3}$ and $H_{1,2,3}$ because they include the same submatrix discussed above. 

\subsection{Non-invertible selection rules}

To realize these neutrino mass textures, we deal with non-invertible selection rules realized by $\mathbb{Z}_2$ gauging of $\mathbb{Z}_N$~\cite{Kobayashi:2024yqq,Kobayashi:2024cvp}. 
Let $g$ denote a generator of the $\mathbb{Z}_N$ group, and let a $\mathbb{Z}_2$ outer automorphism $r$ be defined via the relations:
\begin{align}
r^2=e\,,\qquad r g^k r^{-1} = g^{-k}\, ,
\end{align}
where $e$ represents the group identity. Under this setup, the operator $r$ implements the outer automorphism by mapping each group element $g^k$ to $g^{-k}$, thereby allowing us to define the following equivalence classes:
\begin{align}
[g^k] = \{ h g^k h^{-1} | h = e, r \} = \{ g^k, g^{-k} \}\, ,
\label{eq:class}
\end{align}
where $k$ ranges over $0, 1, ..., \lfloor \frac{N}{2} \rfloor$ with $\lfloor \cdot \rfloor$ being the standard floor function.
These structures are identical to the $\mathbb{Z}_2$-invariant conjugacy classes of the Dihedral group $D_N \cong \mathbb{Z}_N \rtimes \mathbb{Z}_2$. The physical motivation for considering such classes comes from string compactifications, particularly type IIB string theory compactified on toroidal orbifolds with magnetic fluxes~\cite{Kobayashi:2024yqq}.\footnote{For alternative realizations of non-invertible selection rules, see also Refs.~\cite{Dong:2025pah,Dong:2025jra}.}
In the magnetized compactifications, the Kaluza-Klein reduction of a higher-dimensional Yang-Mills theory on a torus typically yields a $\mathbb{Z}_N$ flavor symmetry within the four-dimensional massless sector~\cite{Abe:2009vi,Berasaluce-Gonzalez:2012abm,Marchesano:2013ega}.
The introduction of a $\mathbb{Z}_2$ orbifold projection, however, explicitly breaks this flavor symmetry, leaving the $\mathbb{Z}_2$-even modes $\phi^k$ to be classified by the specific equivalence classes $[g^{k}]$ from Eq.~\eqref{eq:class}. These modes subsequently follow the fusion rule~\cite{Kobayashi:2024yqq}:
\begin{align}
[g^{k_1}] \cdot [g^{k_2}] = [g^{k_1+k_2}] + [g^{k_1-k_2}]\,.
\end{align}

\medskip

Since the fields are characterized by the classes rather than discrete individual group elements, the resulting selection rules depart significantly from conventional group-like symmetry patterns. More precisely, if $\tilde{g}^k$ represents an arbitrary element within the class $[g^k]$, a bare interaction term involving the 4D fields $\phi_{k_1} \cdots \phi_{k_n}$ (where each field carries a class $[g^{k_i}]$) is non-vanishing provided that at least one combination of representative elements $\tilde{g}^{k_i} \in [g^{k_i}]$ can be found to satisfy
\begin{align}
\label{eq:cond}
\tilde{g}^{k_1} \cdots \tilde{g}^{k_n} = e\, .
\end{align}
This mechanism contrasts with standard group-theoretic frameworks, where every field possesses a unique charge and a coupling is allowed if and only if the net charge vanishes. Under the non-invertible selection rule governed by Eq.~\eqref{eq:cond}, a single field is associated with a collective set of group elements. Consequently, an interaction is allowed as long as the conservation condition holds for any single combination of elements drawn from their corresponding classes. It is worth noting that a field and its charge conjugate fall into the same class $[g^{k_i}]$. Such non-invertible selection rules find a natural mathematical description within the framework of hypergroups and fusion algebras, which are well-established mathematical concepts known to be compatible with both quantum field theory and string theory (see, e.g., Ref.~\cite{Kaidi:2024wio}). 
In addition, phenomenological applications have been explored for realizing non-trivial mass matrices in the Standard Model~\cite{Kobayashi:2025znw,Kobayashi:2025ldi}, the minimal supersymmetric Standard Model~\cite{Nakai:2025thw} and grand unified theories~\cite{Kobayashi:2025thd,Kobayashi:2025rpx,Chen:2026mvi}.

\subsection{One-zero textures $H_{1,2,3}$}

Let us consider a case where the neutrino mass is generated from the Weinberg operator:
\begin{align}
    \frac{c_{ij}}{\Lambda}(L_i\tilde{H})(L_j\tilde{H})
\end{align}
with $i,j=1,2,3$, where $L_i$ and $\tilde{H}=i\sigma^2 H^\ast$ respectively denote the left-handed leptons and the Higgs field. 
In the case of $\mathbb{Z}_2$ gauging of $\mathbb{Z}_N$, both $L_iL_i$ and $HH$ include the identity class $[g^0]$. Since the selection rule associated with $\mathbb{Z}_2$ gauging of $\mathbb{Z}_N$ always allow the diagonal entries of $c_{ij}$, 
one cannot realize the textures $G_{1,2,3}$. 
Hence, we discuss the realization of the other textures $H_{1,2,3}$. 

\medskip

As discussed in Ref.~\cite{Kobayashi:2025ldi}, let us focus on $\mathbb{Z}_2$ gauging of $\mathbb{Z}_5$ which includes three distinct classes $\{[g^0],[g^1],[g^2]\}$. 
They obey the following commutative fusion rule:
\begin{align}
    [g^0]\cdot [g^m]&=[g^m]\,,\qquad
    [g^1]\cdot [g^1]= [g^0] + [g^2]\,,\qquad
    [g^2]\cdot [g^2]= [g^0] + [g^1]\,,
    \nonumber\\
    [g^1]\cdot [g^2] &= [g^1] + [g^2]\,.
\end{align}
with $m=0,1,2$. 
Under the non-invertible selection rule, one cannot find the $H_{1,2,3}$ textures within the Standard Model such that $c_{ij}$ has one of the $H_{1,2,3}$ textures and the Yukawa matrix of charged leptons is diagonal, i.e.,
\begin{align}
    Y_{ij} \Bar{L}_i H e_j,
\end{align}
with $Y_{ij}=\mathrm{diag}(\ast, \ast, \ast)$, where $\ast$ denotes the nonvanishing entry. 
Here, $e_j$ denotes the right-handed leptons. 
Similar things happen for the other cases such as $\mathbb{Z}_2$ gauging of $\mathbb{Z}_{3,4}$. 
Hence, we consider type II non-supersymmetric two Higgs doublet models or minimal supersymmetric Standard Model, where $H_u$ and $H_d$ appear in the Weinberg operators and Yukawa couplings of charged leptons, respectively. 
Then, the $H_{1,2,3}$ textures can be realized under the class assignments of matter fields shown in Table \ref{tab:H1H2H3}:
\begin{table}[H]
    \centering
    \caption{Class assignments of matter fields realizing the $H_1$, $H_2$ and $H_3$ textures.}
    \begin{tabular}{|c||c|c|c|}\hline
         &  $L_i = E_i$ & $H_u$ & $H_d$\\ \hline
         & $\{[g^0], [g^1], [g^2]\}$ & $[g^1]$ & $[g^0]$ \\
       $H_1$  & $\{[g^0], [g^2], [g^1]\}$ & $[g^2]$ & $[g^0]$ \\
         & $\{[g^1], [g^0], [g^2]\}$ & $[g^1]$ & $[g^0]$ \\ 
         & $\{[g^2], [g^0], [g^1]\}$ & $[g^2]$ & $[g^0]$ \\ \hline
         & $\{[g^0], [g^1], [g^2]\}$ & $[g^2]$ & $[g^0]$ \\
       $H_2$  & $\{[g^0], [g^2], [g^1]\}$ & $[g^1]$ & $[g^0]$ \\
         & $\{[g^1], [g^2], [g^0]\}$ & $[g^1]$ & $[g^0]$ \\ 
         & $\{[g^2], [g^1], [g^0]\}$ & $[g^2]$ & $[g^0]$ \\ \hline
         & $\{[g^1], [g^0], [g^2]\}$ & $[g^2]$ & $[g^0]$ \\
       $H_3$  & $\{[g^1], [g^2], [g^0]\}$ & $[g^2]$ & $[g^0]$ \\
         & $\{[g^2], [g^0], [g^1]\}$ & $[g^1]$ & $[g^0]$ \\ 
         & $\{[g^2], [g^1], [g^0]\}$ & $[g^1]$ & $[g^0]$ \\ \hline
    \end{tabular}
    \label{tab:H1H2H3}
\end{table}

\subsection{One-zero textures $G_{1,2,3}$}

As mentioned before, one cannot realize the $G_{1,2,3}$ textures under the 
non-invertible selection rule arising from $\mathbb{Z}_2$ gauging of $\mathbb{Z}_N$. 
Hence, we move to a different gauging scenario, specifically $\mathbb{Z}_3$ gauging of $\mathbb{Z}_7$~\cite{Dong:2025jra}.\footnote{For a realization of two-zero textures based on the $\mathbb{Z}_3$ gauging of $\mathbb{Z}_{13}$ and $\mathbb{Z}_{19}$, see Ref.~\cite{Qu:2026omn}.} 

\medskip

When we denote $a$ by the generator of $\mathbb{Z}_7$, the $\mathbb{Z}_3$ automorphism $b$ of the $\mathbb{Z}_7$ group acts as follows:
\begin{align}\label{eq:Z3-ZN}
      b^{-1}ab=a^2, \qquad b^3=e,\qquad a^7=e. 
\end{align}
Since $b$ acts on a generic element $a^k$ as
\begin{align}
    &b a^k b^{-1} = a^{4 k}, \qquad 
    b^2 a^k b^{-2} = a^{16 k},
\end{align}
one can define the following classes:
\begin{align}   
\label{eq:class-Z3} 
    &[a^k] \equiv \{a^{2^l k} | l = 0, 2, 4\} = \{a^k, a^{2k}, a^{4k}\}.
\end{align}
Specifically, for the $\mathbb{Z}_3$ gauging of $\mathbb{Z}_7$, there are three distinct classes: 
\begin{align}
\label{eq:class-Z3-7}
  &C^0 \equiv [a^0]=\{ e \},\quad
  C^1 \equiv [a^1]=[a^2]=\{ a,a^2,a^4 \},\quad
  C^2 \equiv [a^3]=\{ a^3,a^5,a^6 \},
\end{align}
which obey the following commutative fusion rules:
\begin{align}
&C^0\cdot C^m=C^m\,,
\qquad
C^1\cdot C^1 = C^1 + C^2\,,
\qquad
C^2 \cdot C^2 = C^1 + C^2\,,
\nonumber\\
&C^1 \cdot C^2 = C^0 + C^1 +C^2\,,
\end{align}
with $m=0,1,2$.\footnote{Here, we suppress the multiplicities for simplicity.} 
It includes a $S_2 \cong \mathbb{Z}_2$ permutation symmetry associated with $C^1 \leftrightarrow C^2$. 

\medskip

For the same reason as in the realization of the $H_{1,2,3}$ textures, we consider type II non-supersymmetric two Higgs doublet models or minimal supersymmetric Standard Model, where $H_u$ and $H_d$ appear in the Weinberg operators and Yukawa couplings of charged leptons, respectively. 
We find that the neutrino mass textures $G_{1,2,3}$ in the diagonal basis of charged leptons can be realized under the class assignments of matter fields shown in Table \ref{tab:G1G2G3}:
\begin{table}[H]
    \centering
    \caption{Class assignments of matter fields realizing the $G_1$, $G_2$ and $G_3$ textures.}
    \begin{tabular}{|c||c|c|c|c|}\hline
         &  $L_i$ &  $E_i$ & $H_u$ & $H_d$\\ \hline
         & $\{C^0, C^1, C^2\}$ & $\{C^0, C^2, C^1\}$ & $C^1$ & $C^0$ \\
       $G_1$  & $\{C^0, C^1, C^2\}$ & $\{C^0, C^2, C^1\}$ & $C^2$ & $C^0$ \\
         & $\{C^0, C^2, C^1\}$ & $\{C^0, C^1, C^2\}$ & $C^1$ & $C^0$ \\
         & $\{C^0, C^2, C^1\}$ & $\{C^0, C^1, C^2\}$ & $C^2$ & $C^0$ \\ \hline
         & $\{C^1, C^0, C^2\}$ & $\{C^2, C^0, C^1\}$ & $C^1$ & $C^0$ \\
       $G_2$  & $\{C^1, C^0, C^2\}$ & $\{C^2, C^0, C^1\}$ & $C^2$ & $C^0$ \\
         & $\{C^2, C^0, C^1\}$ & $\{C^1, C^0, C^2\}$ & $C^1$ & $C^0$ \\
         & $\{C^2, C^0, C^1\}$ & $\{C^1, C^0, C^2\}$ & $C^2$ & $C^0$ \\ \hline
         & $\{C^1, C^2, C^0\}$ & $\{C^2, C^1, C^0\}$ & $C^1$ & $C^0$ \\
       $G_3$  & $\{C^1, C^2, C^0\}$ & $\{C^2, C^1, C^0\}$ & $C^2$ & $C^0$ \\
         & $\{C^2, C^1, C^0\}$ & $\{C^1, C^2, C^0\}$ & $C^1$ & $C^0$ \\
         & $\{C^2, C^1, C^0\}$ & $\{C^1, C^2, C^0\}$ & $C^2$ & $C^0$ \\ \hline
    \end{tabular}
    \label{tab:G1G2G3}
\end{table}

\section{Conclusions}
\label{sec:con}

We have revisited one-zero and two-zero textures of the neutrino mass matrix using current experimental and cosmological constraints. We have examined which textures remain compatible with neutrino oscillation data, the cosmological bound on the sum of neutrino masses, the kinematic bound on the effective electron-neutrino mass, and limits from neutrinoless double-beta decay. 

\medskip

For two-zero textures, we updated the conventional analysis using recent oscillation data and cosmological bounds. 
If only the CMB bound on the sum of neutrino masses is imposed, several textures remain viable: $A_1$, $A_2$, and the $B$-series textures for NO, and $B_1$, $B_3$, and $C$ for IO. 
The results are summarized in Table~\ref{tab:two-zero-texture}. 
The $B$-series textures are particularly interesting, since they prefer $\delta_{\rm CP}$ around $\pi/2$ and $3\pi/2$ and predict relatively large values of $\langle m_{ee}\rangle$. 
Consequently, they are within the reach of future neutrinoless double-beta decay searches. 
Once the stronger CMB+BAO bound is included, however, the allowed possibilities are significantly reduced, leaving only the $A_1$ and $A_2$ textures for NO. 
These textures will also predict specific values of $\delta_{\mathrm{CP}}$ once the mixing angle $\theta_{23}$ is measured precisely, as shown in Figs.~\ref{fig:A1tex_NO} and~\ref{fig:A2tex_NO}.

\medskip

We have also conducted a comprehensive analysis of one-zero textures. 
Due to their greater number of free parameters, one-zero textures are less constrained than two-zero textures; however, not all remain viable. 
Utilizing flow matching as a conditional generative AI, we explored the parameter space and identified the structures capable of reproducing the observed neutrino oscillation parameters while satisfying experimental and cosmological constraints. 
When imposing only the CMB bound on the sum of neutrino masses, several textures remain viable: the $G_1$, $H_1$, and $H_2$ textures for NO, and $G_2$, $G_3$, and the $H$-series textures for IO. 
These results are summarized in Table~\ref{tab:one-zero-texture}. 
Remarkably, the viable IO textures predict relatively large values of $\langle m_{ee}\rangle$, although these values are relatively smaller than those predicted by the two-zero textures for IO. 
The same tendency is observed for the sum of neutrino masses $\sum_i m_i$. 
These predictions can be tested through future improvements in cosmological observations and neutrino experiments, potentially distinguishing one-zero textures from two-zero textures.

\medskip

We have mainly focused on the texture of the neutrino mass matrix, but one can also analyze one-zero and two-zero minor structures. 
The analytical methods and results are summarized in Appendix~\ref{app:two-zero_minor} for two-zero minors and Appendix~\ref{app:one-zero_minors} for one-zero minors. 
As shown in Table~\ref{tab:two-zero-minor}, most two-zero minor structures are disfavored once the CMB+BAO bound is imposed, whereas one-zero minors are still allowed by current cosmological constraints.

\medskip

Finally, we have discussed how one-zero texture structures can arise from non-invertible selection rules. 
This provides a possible theoretical origin for texture zeros and connects the phenomenological classification to an underlying structure in the lepton sector. Future data will further test these possibilities and may clarify which texture structures are realized in the neutrino mass matrix.

\acknowledgments

This work was supported in part by JSPS KAKENHI Grant Numbers JP26KJ0318 (C.M.), JP25KJ1927 (S.N.), JP25H01539 (H.O.) and JP26K07087 (H.O.).

\appendix
\section{Two-zero textures}
\label{app:two-zero}

In this appendix, we present the distributions of $\delta_{\mathrm{CP}}/\pi$ and $\Sigma\,m_{i}$ as functions of $\theta_{23}$ for the viable two-zero textures other than $A_1$ and $A_2$. 
The figure layout is the same as $A_1$ and $A_2$ 
shown in the main text. 

\subsection*{Normal ordering}

\begin{figure}[H]
  \centering
  \begin{subfigure}{0.49\textwidth}
    \centering
    \includegraphics[width=0.98\linewidth]{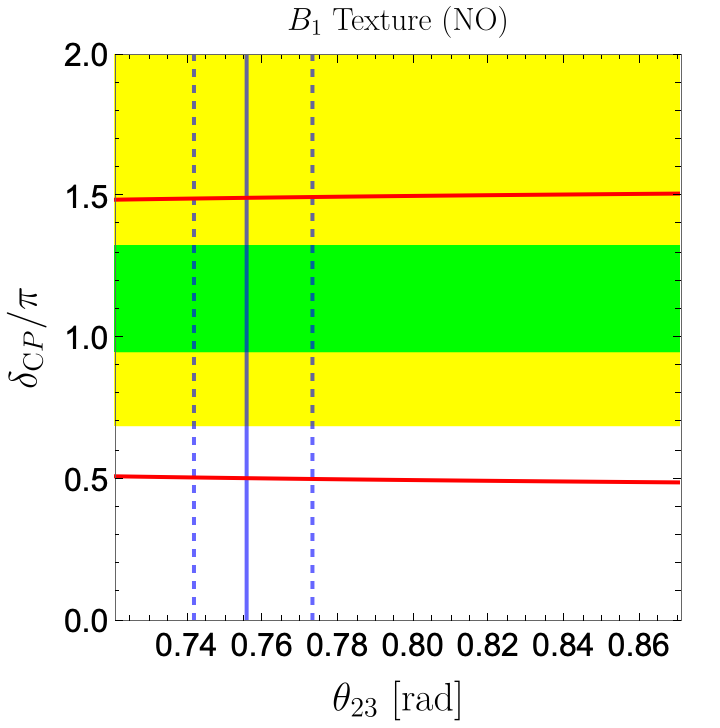}
    \label{fig:B1tex_NO_left}
  \end{subfigure}
  \hfill
  \begin{subfigure}{0.49\textwidth}
    \centering
    \includegraphics[width=0.98\linewidth]{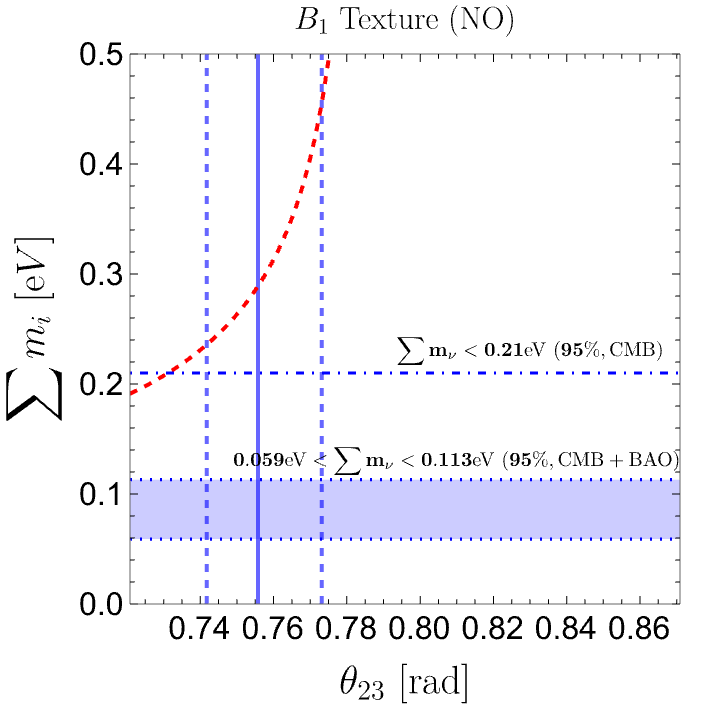}
    \label{fig:B1tex_NO_right}
  \end{subfigure}
  \caption{Distribution of observables for the $B_1$ structure with NO. The left and right panels show $\theta_{23}$ vs.~$\delta_{\mathrm{CP}}/\pi$ and $\theta_{23}$ vs.~$\Sigma\,m_{i}$, respectively.}
  \label{fig:B1tex_NO}
\end{figure}

\begin{figure}[H]
  \centering
  \begin{subfigure}{0.49\textwidth}
    \centering
    \includegraphics[width=0.98\linewidth]{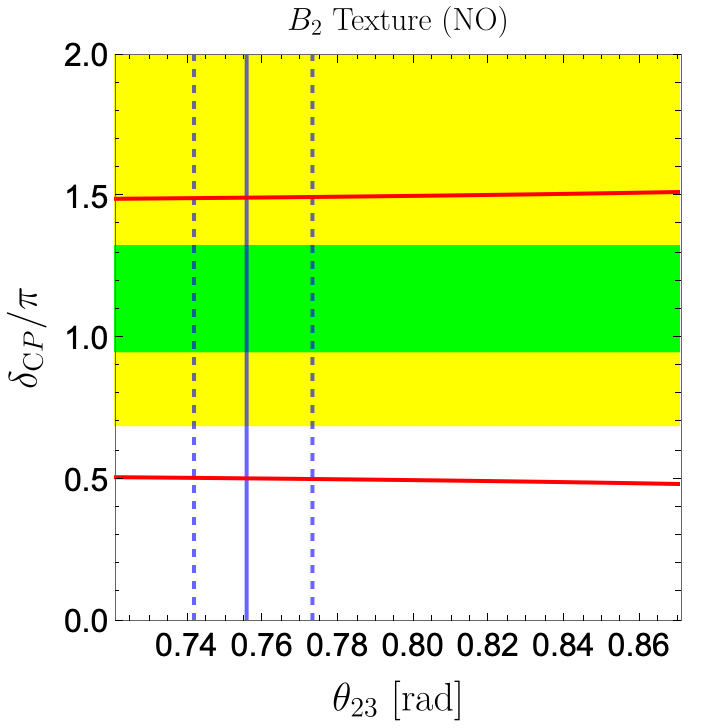}
    \label{fig:B2tex_NO_left}
  \end{subfigure}
  \hfill
  \begin{subfigure}{0.49\textwidth}
    \centering
    \includegraphics[width=0.98\linewidth]{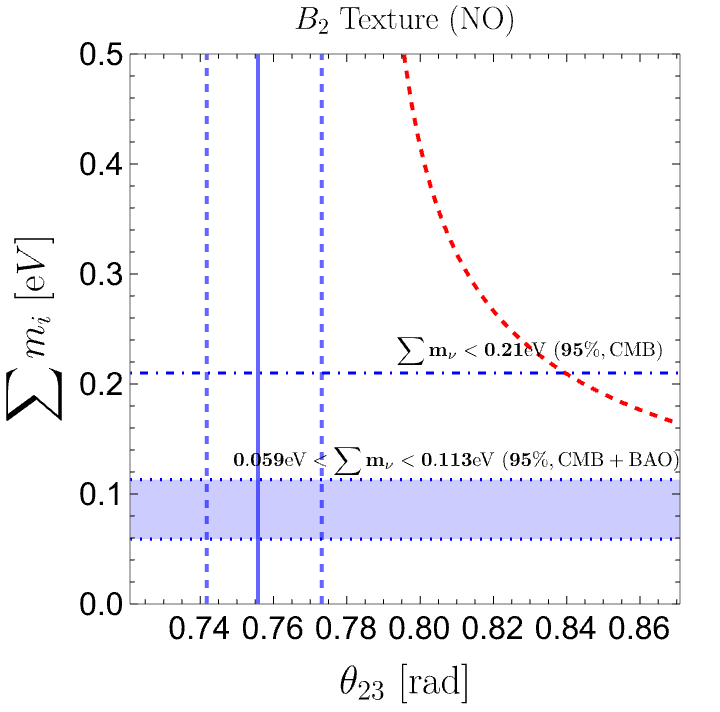}
    \label{fig:B2tex_NO_right}
  \end{subfigure}
  \caption{Distribution of observables for the $B_2$ structure with NO. The left and right panels show $\theta_{23}$ vs.~$\delta_{\mathrm{CP}}/\pi$ and $\theta_{23}$ vs.~$\Sigma\,m_{i}$, respectively.}
  \label{fig:B2tex_NO}
\end{figure}

\vspace{\stretch{1}}
\newpage
\vspace*{\stretch{1}}

\begin{figure}[H]
  \centering
  \begin{subfigure}{0.49\textwidth}
    \centering
    \includegraphics[width=0.98\linewidth]{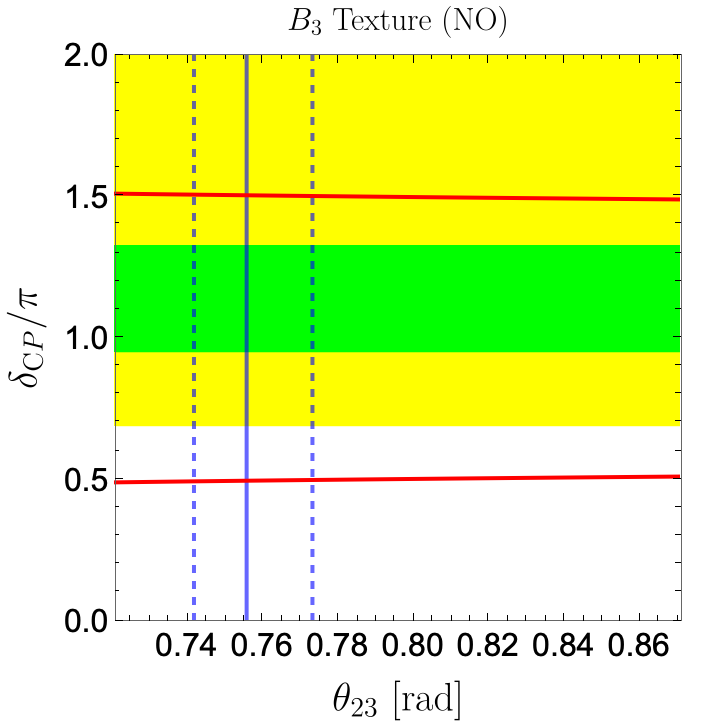}
    \label{fig:B3tex_NO_left}
  \end{subfigure}
  \hfill
  \begin{subfigure}{0.49\textwidth}
    \centering
    \includegraphics[width=0.98\linewidth]{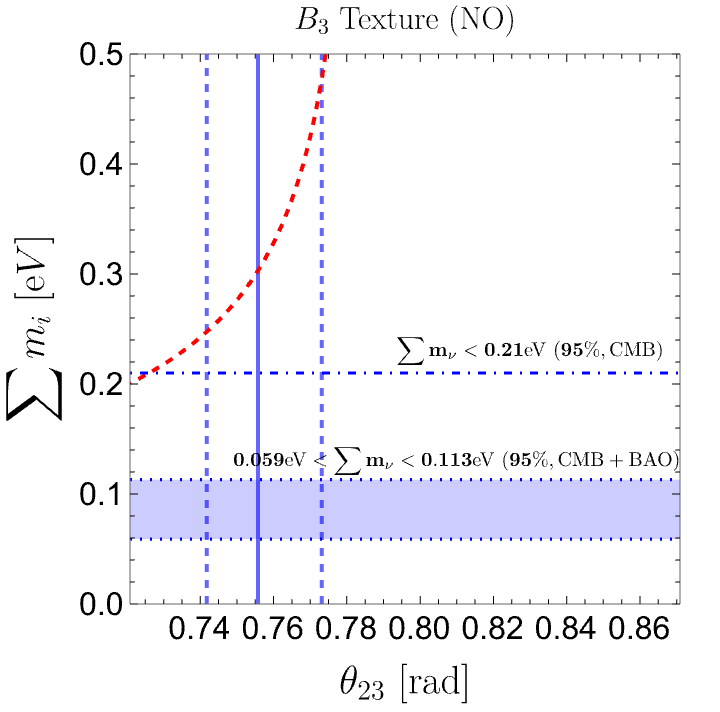}    
    \label{fig:B3tex_NO_right}
  \end{subfigure}
  \caption{Distribution of observables for the $B_3$ structure with NO. The left and right panels show $\theta_{23}$ vs.~$\delta_{\mathrm{CP}}/\pi$ and $\theta_{23}$ vs.~$\Sigma\,m_{i}$, respectively.}
  \label{fig:B3tex_NO}
\end{figure}

\begin{figure}[H]
  \centering
  \begin{subfigure}{0.49\textwidth}
    \centering
    \includegraphics[width=0.98\linewidth]{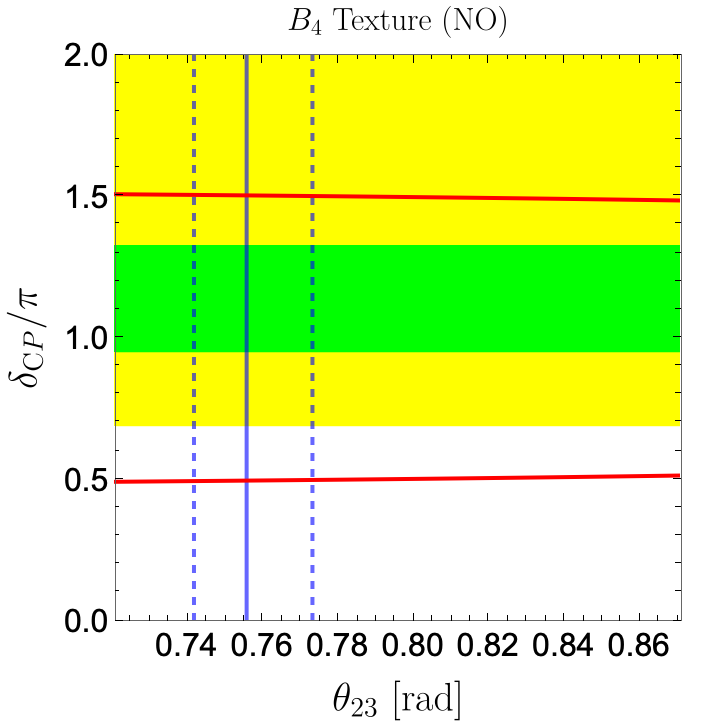}
    \label{fig:B4tex_NO_left}
  \end{subfigure}
  \hfill
  \begin{subfigure}{0.49\textwidth}
    \centering
    \includegraphics[width=0.98\linewidth]{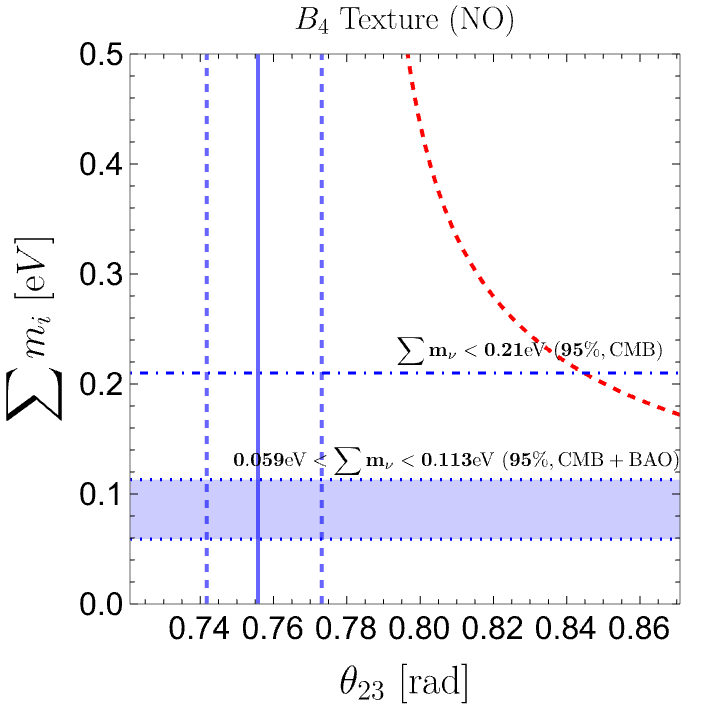}
    \label{fig:B4tex_NO_right}
  \end{subfigure}
  \caption{Distribution of observables for the $B_4$ structure with NO. The left and right panels show $\theta_{23}$ vs.~$\delta_{\mathrm{CP}}/\pi$ and $\theta_{23}$ vs.~$\Sigma\,m_{i}$, respectively.}
  \label{fig:B4tex_NO}
\end{figure}

\vspace{\stretch{1}}
\newpage

\subsection*{Inverted ordering}

\vspace{\stretch{1}}

\begin{figure}[H]
  \centering
  \begin{subfigure}{0.49\textwidth}
    \centering
    \includegraphics[width=0.98\linewidth]{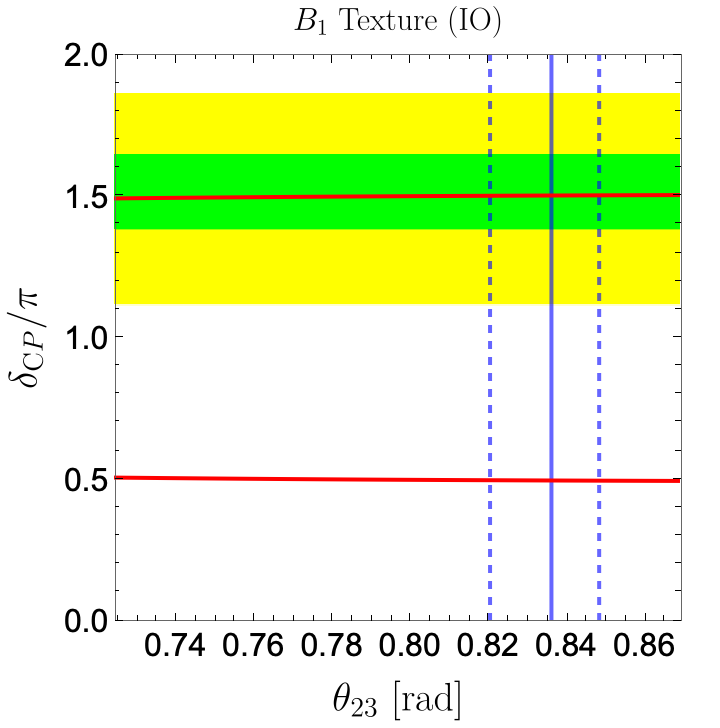}
    \label{fig:B1tex_IO_left}
  \end{subfigure}
  \hfill
  \begin{subfigure}{0.49\textwidth}
    \centering
    \includegraphics[width=0.98\linewidth]{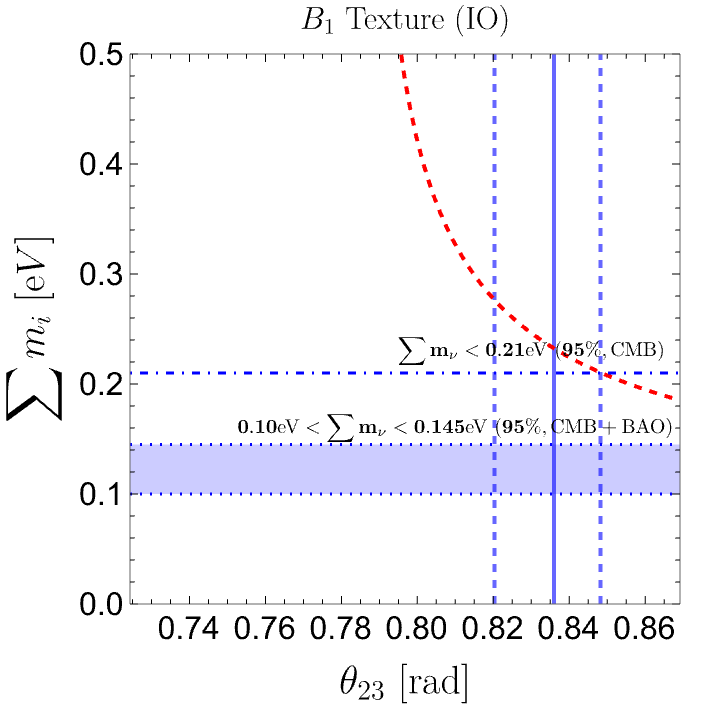}
    \label{fig:B1tex_IO_right}
  \end{subfigure}
  \caption{Distribution of observables for the $B_1$ structure with IO. The left and right panels show $\theta_{23}$ vs.~$\delta_{\mathrm{CP}}/\pi$ and $\theta_{23}$ vs.~$\Sigma\,m_{i}$, respectively.}
  \label{fig:B1tex_IO}
\end{figure}

\begin{figure}[H]
  \centering
  \begin{subfigure}{0.49\textwidth}
    \centering
    \includegraphics[width=0.98\linewidth]{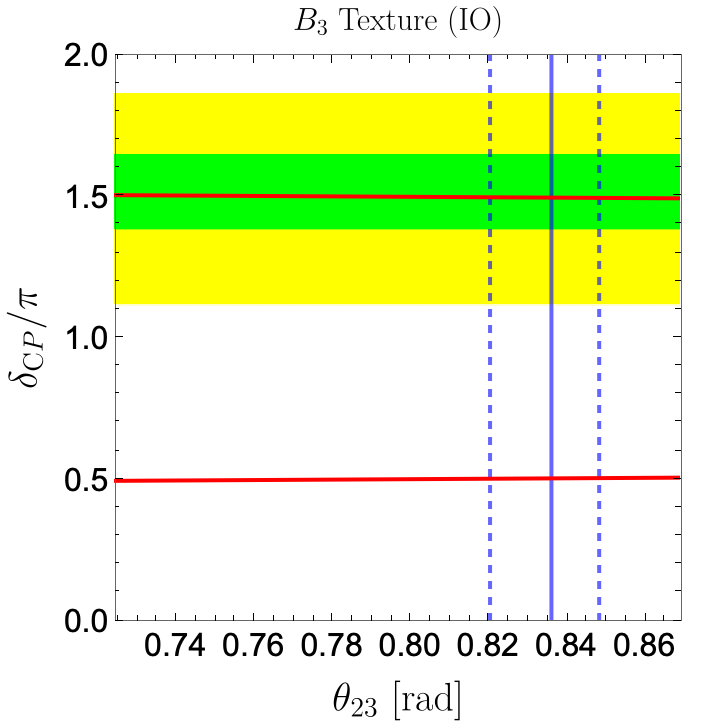}
    \label{fig:B3tex_IO_left}
  \end{subfigure}
  \hfill
  \begin{subfigure}{0.49\textwidth}
    \centering
    \includegraphics[width=0.98\linewidth]{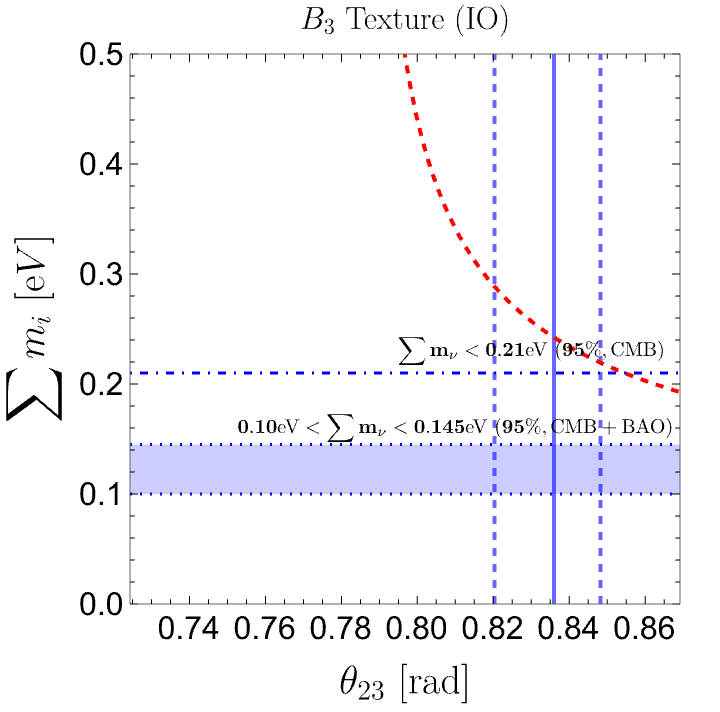}
    \label{fig:B3tex_IO_right}
  \end{subfigure}
  \caption{Distribution of observables for the $B_3$ structure with IO. The left and right panels show $\theta_{23}$ vs.~$\delta_{\mathrm{CP}}/\pi$ and $\theta_{23}$ vs.~$\Sigma\,m_{i}$, respectively.}
  \label{fig:B3tex_IO}
\end{figure}

\vspace{\stretch{1}}
\newpage

\begin{figure}[H]
  \centering
  \begin{subfigure}{0.49\textwidth}
    \centering
    \includegraphics[width=0.98\linewidth]{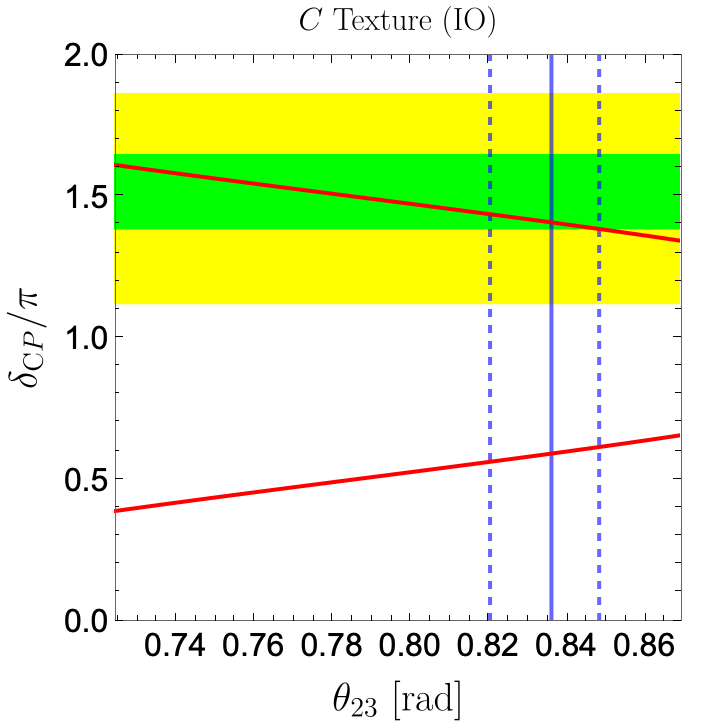}
    \label{fig:Ctex_IO_left}
  \end{subfigure}
  \hfill
  \begin{subfigure}{0.49\textwidth}
    \centering
    \includegraphics[width=0.98\linewidth]{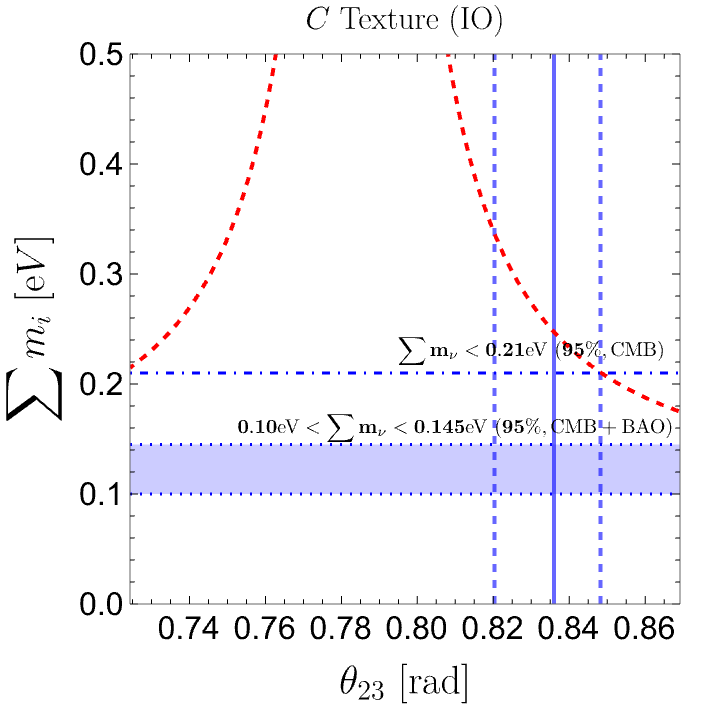}
    \label{fig:Ctex_IO_right}
  \end{subfigure}
  \caption{Distribution of observables for the $C$ structure with IO. The left and right panels show $\theta_{23}$ vs.~$\delta_{\mathrm{CP}}/\pi$ and $\theta_{23}$ vs.~$\Sigma\,m_{i}$, respectively.}
  \label{fig:Ctex_IO}
\end{figure}

\section{Two-zero minors}
\label{app:two-zero_minor}

In this appendix, we first present the analytical method for two-zero minors in Sec.~\ref{app:two_zero_minor_method}. 
In Sec.~\ref{app:two_zero_minor_results}, we present the distributions of $\delta_{\mathrm{CP}}/\pi$ and $\Sigma\,m_{i}$ as functions of $\theta_{23}$ for the viable two-zero minor structures listed in Table~\ref{tab:two-zero-minor}.

\subsection{Analytical method for two-zero minors}
\label{app:two_zero_minor_method}

The analysis of two-zero minor structures is similar to that of two-zero texture structures, except that the zero conditions are imposed on the inverse neutrino mass matrix. 
The components of the inverse mass matrix are given by
\begin{align}
        \left(\mathcal{M}_{\nu}^{\rm flavor}\right)^{-1}_{ij}&= \frac{V_{i1}V_{j1}}{m_1}+ e^{i\alpha_2} \frac{V_{i2}V_{j2}}{m_2}+ e^{i\alpha_3}\frac{V_{i3}V_{j3}}{m_3}.\label{min_component}
\end{align}
Imposing the two-zero minor conditions,
\begin{align}
\left(\mathcal{M}_{\nu}^{\rm flavor}\right)^{-1}_{ij} &=0,\qquad
\left(\mathcal{M}_{\nu}^{\rm flavor}\right)^{-1}_{\rho\sigma} =0,
\end{align}
with $(i,j)\ne(\rho,\sigma)$, we obtain
\begin{align}
e^{i\alpha_2} &= \frac{m_2}{m_1} \frac{-V_{i3}V_{j3}V_{\rho 1}V_{\sigma 1}+V_{i1}V_{j1}V_{\rho 3}V_{\sigma 3}}{V_{i3}V_{j3}V_{\rho 2}V_{\sigma 2}-V_{i 2}V_{j2}V_{\rho 3}V_{\sigma 3}} \equiv  \frac{m_2}{m_1} R_2 (\theta_{12},\theta_{13},\theta_{23},\delta_{\rm CP}), \\
e^{i\alpha_3} &= \frac{m_3}{m_1} \frac{V_{i2}V_{j2}V_{\rho1}V_{\sigma 1}-V_{i1}V_{j1}V_{\rho2}V_{\sigma2}}{V_{i3}V_{j3}V_{\rho2}V_{\sigma2}-V_{i2}V_{j2}V_{\rho3}V_{\sigma3}} \equiv  \frac{m_3}{m_1} R_3 (\theta_{12},\theta_{13},\theta_{23},\delta_{\rm CP}).
\end{align}
Since the magnitudes of $e^{i\alpha_2}$ and $e^{i\alpha_3}$ are unity, the mass ratios are given by
\begin{align}
    \frac{m_1}{m_2} &= |R_2 (\theta_{12},\theta_{13},\theta_{23},\delta_{\rm CP})|,\\
    \frac{m_1}{m_3} &= |R_3 (\theta_{12},\theta_{13},\theta_{23},\delta_{\rm CP})|.
\end{align}

\medskip

For NO, the mass-squared differences can be written as
\begin{align}
    \Delta m^2_{21} &\equiv m_2^2 -m_1^2 = m_1^2\left(\frac{1}{|R_2|^2}-1\right),\\
    \Delta m^2_{31} &\equiv m_3^2 -m_1^2 = m_1^2\left(\frac{1}{|R_3|^2}-1\right).
\end{align}
Combining these equations, we obtain
\begin{align}
    \frac{\Delta m^2_{21}}{\frac{1}{|R_2|^2}-1} = \frac{\Delta m^2_{31}}{\frac{1}{|R_3|^2}-1}.
\end{align}
By fixing $\theta_{12},\theta_{13},\Delta m^2_{21},\Delta m^2_{31}$ to the best-fit values in Table \ref{tab:NuFIT}, we can obtain predictions for the remaining neutrino parameters, including the individual neutrino masses:
\begin{align}
    m_1&=\sqrt{\frac{\Delta m_{21}^2}{\frac{1}{|R_2|^2}-1}} = \sqrt{\frac{\Delta m_{31}^2}{\frac{1}{|R_3|^2}-1}}, \label{eq:nu-minor_m1_NO}\\
    m_2&=\sqrt{\frac{\Delta m_{21}^2}{1-|R_2|^2}}, \label{eq:nu-minor_m2_NO}\\
    m_3&= \sqrt{\frac{\Delta m_{31}^2 }{1-|R_3|^2}}. \label{eq:nu-minor_m3_NO}
\end{align}

\medskip

On the other hand, for IO, the mass-squared differences are given by
\begin{align}
    \Delta m^2_{21} &\equiv m_2^2 -m_1^2 = m_1^2(\frac{1}{|R_2|^2}-1),\\
    \Delta m^2_{32} &\equiv m_3^2 -m_2^2 = m_1^2(\frac{1}{|R_3|^2}-\frac{1}{|R_2|^2}).
\end{align}
Then the equation can be written by
\begin{align}
    \frac{\Delta m^2_{21}}{\frac{1}{|R_2|^2}-1} = \frac{\Delta m^2_{32}}{\frac{1}{|R_3|^2}-\frac{1}{|R_2|^2}}.
\end{align}
As in the two-zero texture case, by fixing $\theta_{12},\theta_{13},\Delta m^2_{21},\Delta m^2_{32}$ to the best-fit values in Table \ref{tab:NuFIT}, we obtain the predictions.
The concrete form of the each neutrino mass for the two-zero minor structures in IO are given by
\begin{align}
    m_1&=\sqrt{\frac{\Delta m_{21}^2}{\frac{1}{|R_2|^2}-1}} = \sqrt{\frac{\Delta m_{32}^2}{\frac{1}{|R_3|^2}-\frac{1}{|R_2|^2}}}, \label{eq:nu-minor_m1_IO}\\
    m_2&=\sqrt{\frac{\Delta m_{21}^2}{1-|R_2|^2}}, \label{eq:nu-minor_m2_IO}\\
    m_3&= \sqrt{\frac{\Delta m_{32}^2+\Delta m_{21}^2 }{1-|R_3|^2}}. \label{eq:nu-minor_m3_IO}
\end{align}

Taking into account the neutrino mass sum constraint in Sec.~\ref{subsec:const_msum}, the results of the analysis for each minor structure are summarized in Table ~\ref{tab:two-zero-minor}, where $\bigcirc$ and and $\times$ denote viable and non-viable structures, respectively.

\begin{table}[H]
    \centering
    \caption{Summary of the two-zero minor analysis.}
    \label{tab:two-zero-minor}

    \begin{tabular}{ll cccccccc}
        \toprule
        Structure & & $A_1$ & $A_2$ & $B_1$ & $B_2$ & $B_3$ & $B_4$ & $C$ & \\
        \midrule
        \multirow{2}{*}{CMB}
            & NO & $\times$
                 & $\times$
                 & \hyperref[fig:B1min_NO]{$\bigcirc$}
                 & \hyperref[fig:B2min_NO]{$\bigcirc$}
                 & \hyperref[fig:B3min_NO]{$\bigcirc$}
                 & \hyperref[fig:B4min_NO]{$\bigcirc$}
                 & \hyperref[fig:Cmin_NO]{$\bigcirc$}
                 & \\
            & IO & $\times$
                 & $\times$
                 & $\times$
                 & \hyperref[fig:B2min_IO]{$\bigcirc$}
                 & $\times$
                 & \hyperref[fig:B4min_IO]{$\bigcirc$}
                 & $\times$
                 & \\
        \midrule
        \multirow{2}{*}{CMB+BAO}
            & NO & $\times$
                 & $\times$
                 & $\times$
                 & $\times$
                 & $\times$
                 & $\times$
                 & $\times$
                 & \\
            & IO & $\times$
                 & $\times$
                 & $\times$
                 & $\times$
                 & $\times$
                 & $\times$
                 & $\times$
                 & \\

        \midrule
        \multicolumn{10}{c}{}
        \\[-0.8em]
        \midrule

        Structure & & $D_1$ & $D_2$ & $E_1$ & $E_2$ & $E_3$ & $F_1$ & $F_2$ & $F_3$ \\
        \midrule
        \multirow{2}{*}{CMB}
            & NO & \hyperref[fig:D1min_NO]{$\bigcirc$}
                 & \hyperref[fig:D2min_NO]{$\bigcirc$}
                 & $\times$
                 & $\times$
                 & $\times$
                 & $\times$
                 & $\times$
                 & $\times$ \\
            & IO & $\times$
                 & $\times$
                 & $\times$
                 & $\times$
                 & $\times$
                 & $\times$
                 & $\times$
                 & $\times$ \\
        \midrule
        \multirow{2}{*}{CMB+BAO}
            & NO & \hyperref[fig:D1min_NO]{$\bigcirc$}
                 & \hyperref[fig:D2min_NO]{$\bigcirc$}
                 & $\times$
                 & $\times$
                 & $\times$
                 & $\times$
                 & $\times$
                 & $\times$ \\
            & IO & $\times$
                 & $\times$
                 & $\times$
                 & $\times$
                 & $\times$
                 & $\times$
                 & $\times$
                 & $\times$ \\
        \bottomrule
    \end{tabular}
\end{table}

\subsection{Results}
\label{app:two_zero_minor_results}

In this section, we present the analysis results for two-zero minor structures. 
Predictions for viable structures are summarized in Table \ref{tab:prediction_mee_min-NO} for NO and Table \ref{tab:prediction_mee_min-IO} for IO. 
These tables show the allowed regions of $\langle m_{ee}\rangle$, $m_{\nu_e}^{\rm eff}$, $\sum_i m_i$ and $\delta_{\rm CP}/\pi$ for each viable structure. 
We find characteristic predictions for $\delta_{\rm CP}$. 
In particular, the $B$-series structures predict values around $\delta_{\rm CP} \sim 1.5 \pi$ and $\delta_{\rm CP} \sim0.5\pi$. 
$C$ structure predicts $1.3 \pi\lesssim \delta_{\rm CP} \lesssim 1.6 \pi$ and $0.4 \pi\lesssim \delta_{\rm CP} \lesssim0.7\pi$. 

\medskip

Furthermore, we present the distributions of $\delta_{\mathrm{CP}}/\pi$ and $\Sigma\,m_{i}$ as functions of $\theta_{23}$ for the viable two-zero minors, using the same figure layout as that employed for the two-zero textures.

\begin{table}[H]
    \caption{Summary of the effective neutrino mass for the neutrinoless double beta decay $\langle m_{ee}\rangle$, the effective electron neutrino mass $m_{\nu_e}^{\rm eff}$, the sum of the predicted neutrino masses $\sum_i m_i$, and the Dirac CP phase $\delta_{\rm CP}/\pi$ for the viable two-zero minors with NO.}
    \label{tab:prediction_mee_min-NO}
    \centering
    \begin{tabular}{|c||c|c|c|c|}\hline
         Structure& $\langle m_{ee}\rangle$ [eV] & $m_{\nu_e}^{\rm eff}$ [eV] & $\sum_i m_i$ [eV] & $\delta_{\rm CP}/\pi$  \\ \hline
         $B_1$ minor (NO)& $>0.048$& $>0.054$ & $>0.16$& $\sim 0.5,~\sim 1.5$\\
         $B_2$ minor (NO)& $>0.058$& $>0.064$ & $>0.19$ & $\sim 0.5,~\sim 1.5$\\
         $B_3$ minor (NO)& $>0.051$& $>0.057$ & $>0.17$ & $\sim 0.5,~\sim 1.5$\\
         $B_4$ minor (NO)& $>0.061$& $>0.067$ & $>0.20$ & $\sim 0.5,~\sim 1.5$\\ \hline
         $C$ minor (NO)& $>0.029$ & $>0.049$ &$ > 0.15$ & $0.4 - 0.7,~1.3 - 1.6$\\ \hline
         $D_1$ minor (NO)& $\sim 0$ & $\sim0.020$ & $0.065-0.068$ & $0.0 - 0.5,~1.5 - 2.0$\\
         $D_2$ minor (NO)& $\sim 0$ & $\sim 0.021$ &$0.065-0.068$ & $0.6 - 1.4$\\ \hline
    \end{tabular}
\end{table}

\begin{table}[H]
    \caption{Summary of the effective neutrino mass for the neutrinoless double beta decay $\langle m_{ee}\rangle$, the effective electron neutrino mass $m_{\nu_e}^{\rm eff}$, the sum of the predicted neutrino masses $\sum_i m_i$, and the Dirac CP phase $\delta_{\rm CP}/\pi$ for the viable two-zero minors with IO.}
    \label{tab:prediction_mee_min-IO}
    \centering
    \begin{tabular}{|c||c|c|c|c|}\hline
         Structure& $\langle m_{ee}\rangle$ [eV] & $m_{\nu_e}^{\rm eff}$ [eV] & $\sum_i m_i$ [eV] & $\delta_{\rm CP}/\pi$\\ \hline
         $B_2$ texture (IO)& $>0.068$& $>0.071$ & $>0.19$ & $\sim 0.5,~\sim 1.5$\\
         $B_4$ texture (IO)& $>0.070$& $>0.073$ & $>0.19$ & $\sim 0.5,~\sim 1.5$\\ \hline
    \end{tabular}
\end{table}

\subsubsection*{Normal ordering}

\begin{figure}[H]
  \centering
  \begin{subfigure}{0.49\textwidth}
    \centering
    \includegraphics[width=0.98\linewidth]{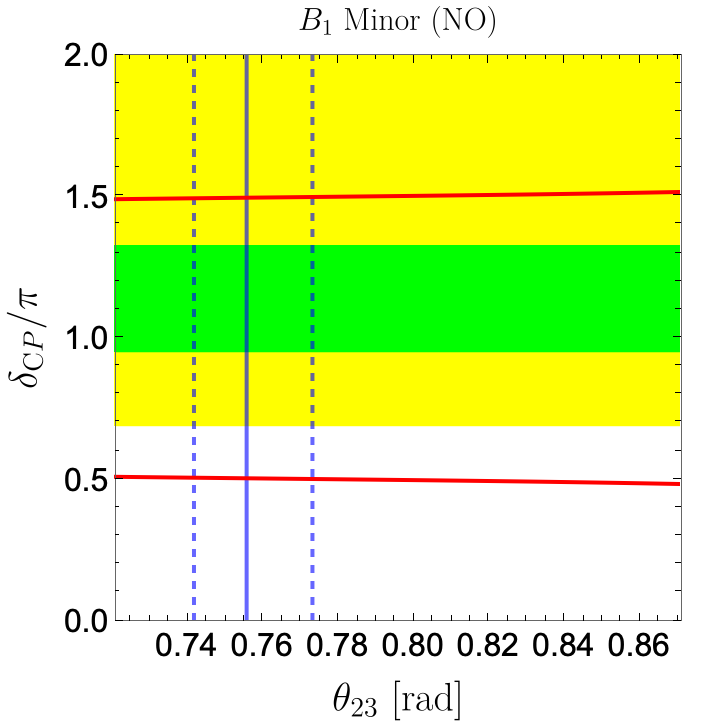}
    \label{fig:B1min_NO_left}
  \end{subfigure}
  \hfill
  \begin{subfigure}{0.49\textwidth}
    \centering
    \includegraphics[width=0.98\linewidth]{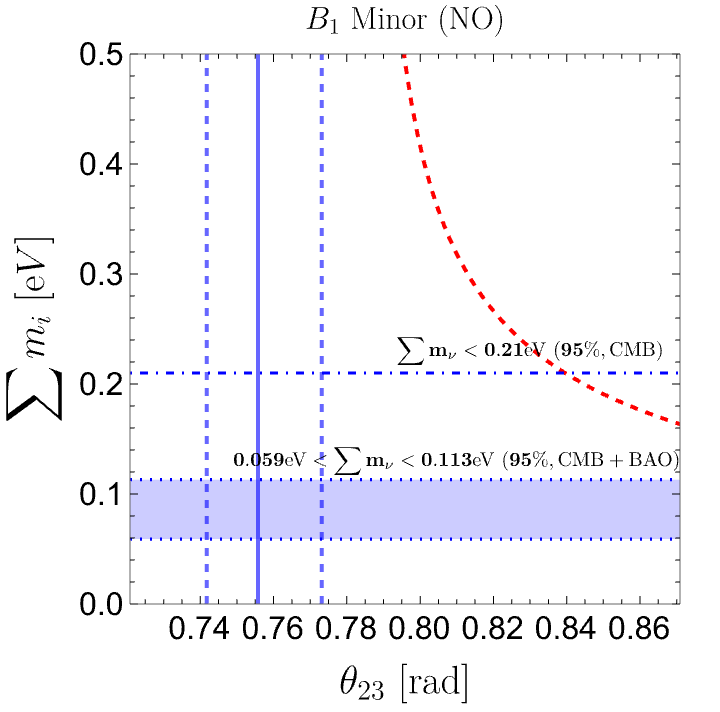}
    \label{fig:B1min_NO_right}
  \end{subfigure}
  \caption{Distribution of observables for the $B_1$ minor with NO. The left and right panels show $\theta_{23}$ vs.~$\delta_{\mathrm{CP}}/\pi$ and $\theta_{23}$ vs.~$\Sigma\,m_{i}$, respectively.}
  \label{fig:B1min_NO}
\end{figure}

\newpage
\vspace*{\stretch{1}}

\begin{figure}[H]
  \centering
  \begin{subfigure}{0.49\textwidth}
    \centering
    \includegraphics[width=0.98\linewidth]{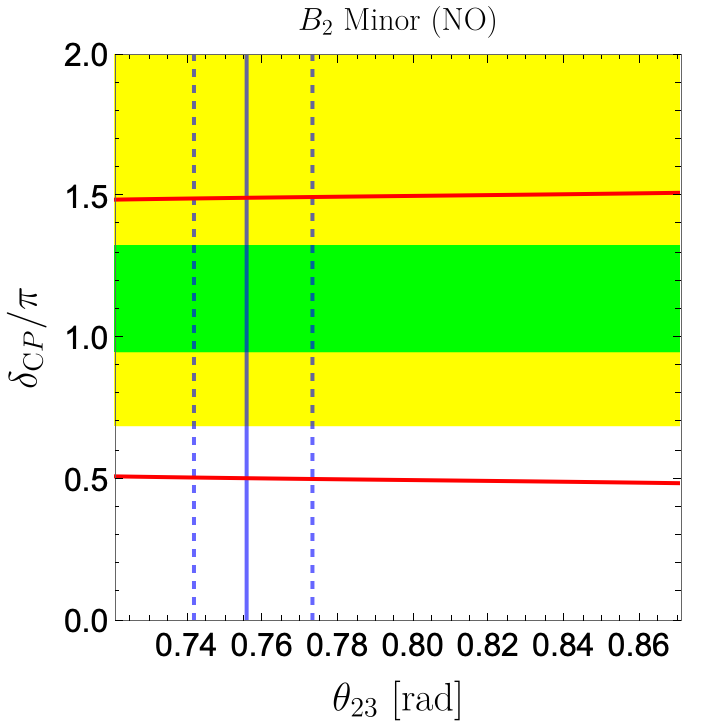}
    \label{fig:B2min_NO_left}
  \end{subfigure}
  \hfill
  \begin{subfigure}{0.49\textwidth}
    \centering
    \includegraphics[width=0.98\linewidth]{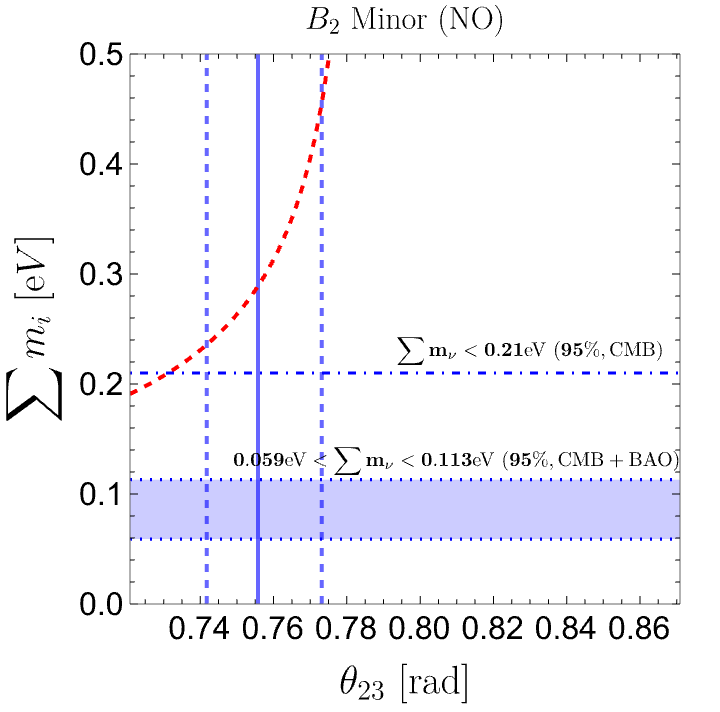}
    \label{fig:B2min_NO_right}
  \end{subfigure}
  \caption{Distribution of observables for the $B_2$ minor with NO. The left and right panels show $\theta_{23}$ vs.~$\delta_{\mathrm{CP}}/\pi$ and $\theta_{23}$ vs.~$\Sigma\,m_{i}$, respectively.}
  \label{fig:B2min_NO}
\end{figure}

\begin{figure}[H]
  \centering
  \begin{subfigure}{0.49\textwidth}
    \centering
    \includegraphics[width=0.98\linewidth]{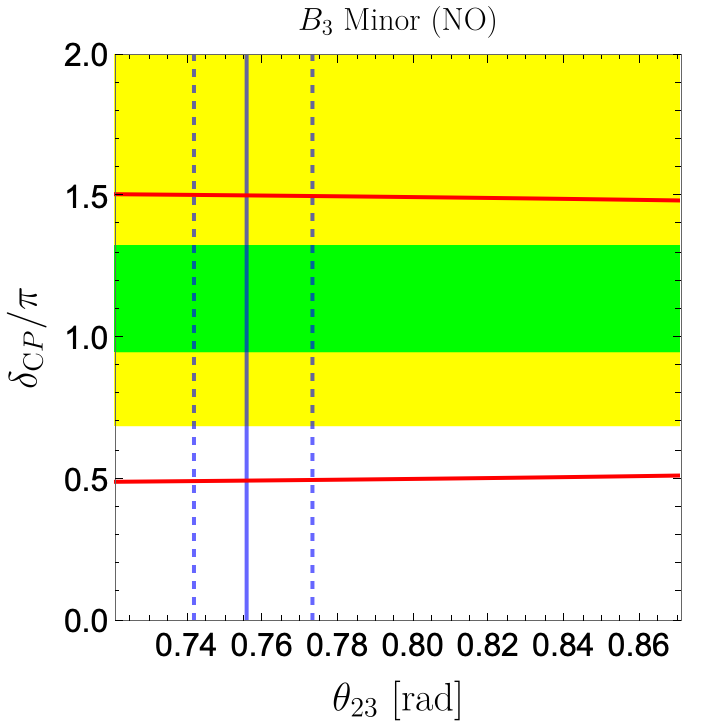}
    \label{fig:B3min_NO_left}
  \end{subfigure}
  \hfill
  \begin{subfigure}{0.49\textwidth}
    \centering
    \includegraphics[width=0.98\linewidth]{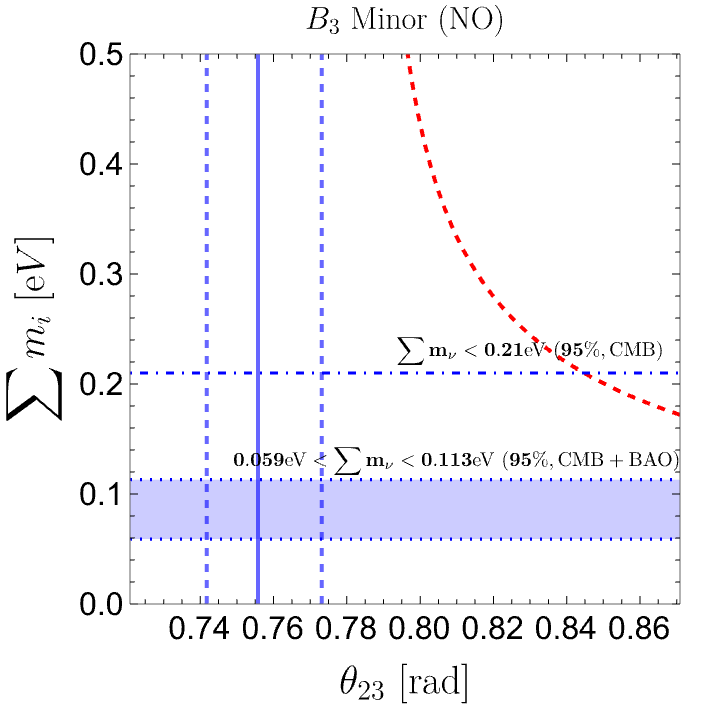}
    \label{fig:B3min_NO_right}
  \end{subfigure}
  \caption{Distribution of observables for the $B_3$ minor with NO. The left and right panels show $\theta_{23}$ vs.~$\delta_{\mathrm{CP}}/\pi$ and $\theta_{23}$ vs.~$\Sigma\,m_{i}$, respectively.}
  \label{fig:B3min_NO}
\end{figure}

\vspace{\stretch{1}}
\newpage
\vspace*{\stretch{1}}

\begin{figure}[H]
  \centering
  \begin{subfigure}{0.49\textwidth}
    \centering
    \includegraphics[width=0.98\linewidth]{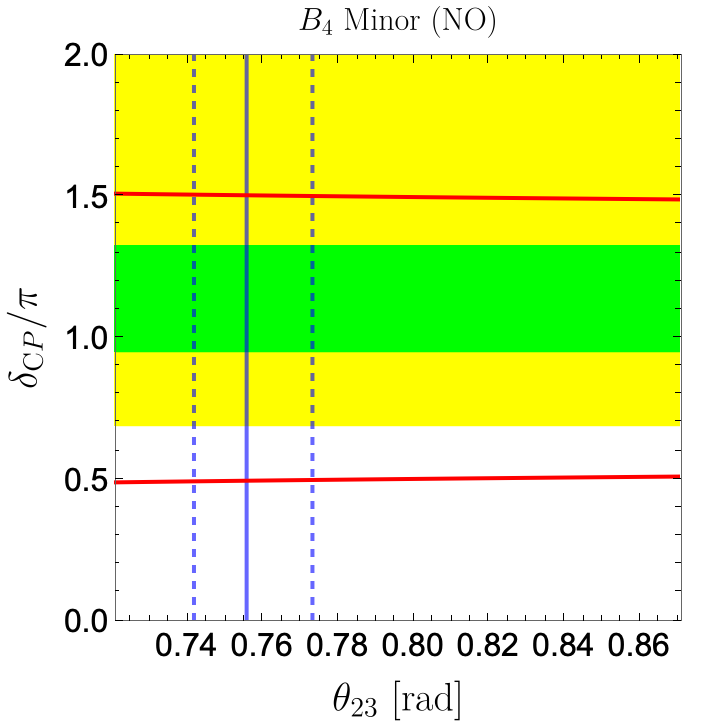}
    \label{fig:B4min_NO_left}
  \end{subfigure}
  \hfill
  \begin{subfigure}{0.49\textwidth}
    \centering
    \includegraphics[width=0.98\linewidth]{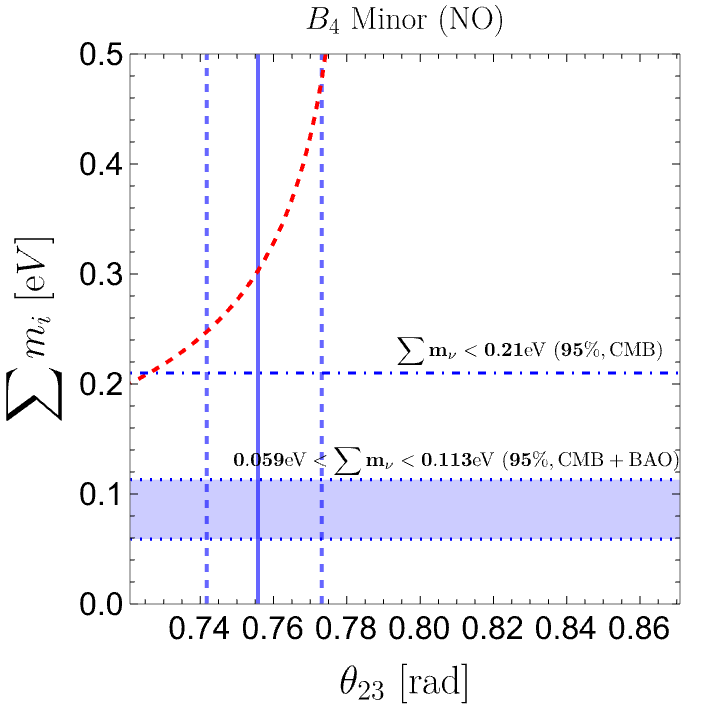}
    \label{fig:B4min_NO_right}
  \end{subfigure}
  \caption{Distribution of observables for the $B_4$ minor with NO. The left and right panels show $\theta_{23}$ vs.~$\delta_{\mathrm{CP}}/\pi$ and $\theta_{23}$ vs.~$\Sigma\,m_{i}$, respectively.}
  \label{fig:B4min_NO}
\end{figure}

\begin{figure}[H]
  \centering
  \begin{subfigure}{0.49\textwidth}
    \centering
    \includegraphics[width=0.98\linewidth]{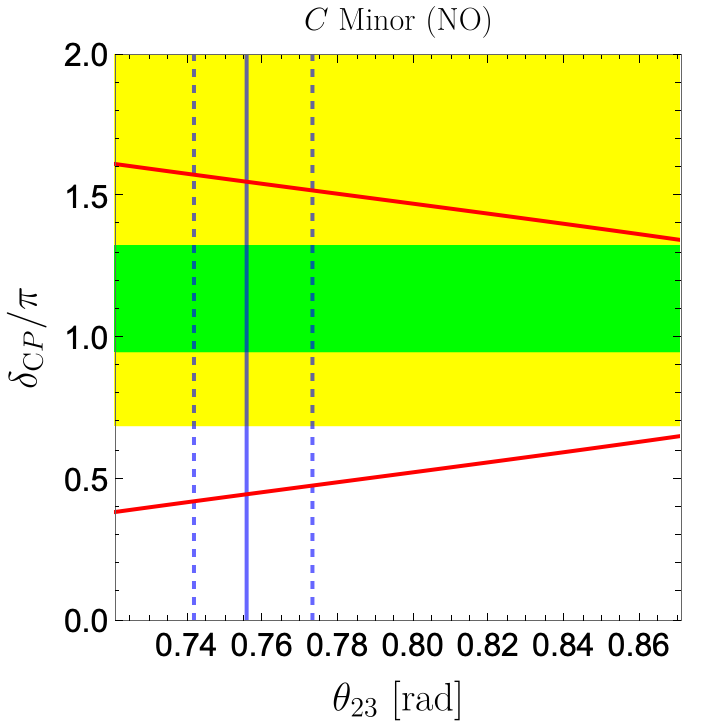}
    \label{fig:Cmin_NO_left}
  \end{subfigure}
  \hfill
  \begin{subfigure}{0.49\textwidth}
    \centering
    \includegraphics[width=0.98\linewidth]{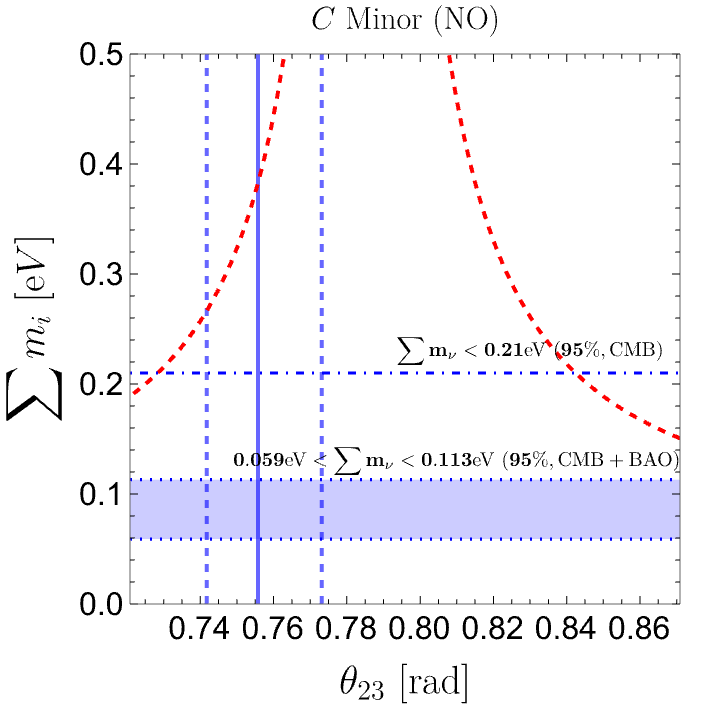}
    \label{fig:Cmin_NO_right}
  \end{subfigure}
  \caption{Distribution of observables for the $C$ minor with NO. The left and right panels show $\theta_{23}$ vs.~$\delta_{\mathrm{CP}}/\pi$ and $\theta_{23}$ vs.~$\Sigma\,m_{i}$, respectively.}
  \label{fig:Cmin_NO}
\end{figure}

\vspace{\stretch{1}}
\newpage
\vspace*{\stretch{1}}

\begin{figure}[H]
  \centering
  \begin{subfigure}{0.49\textwidth}
    \centering
    \includegraphics[width=0.98\linewidth]{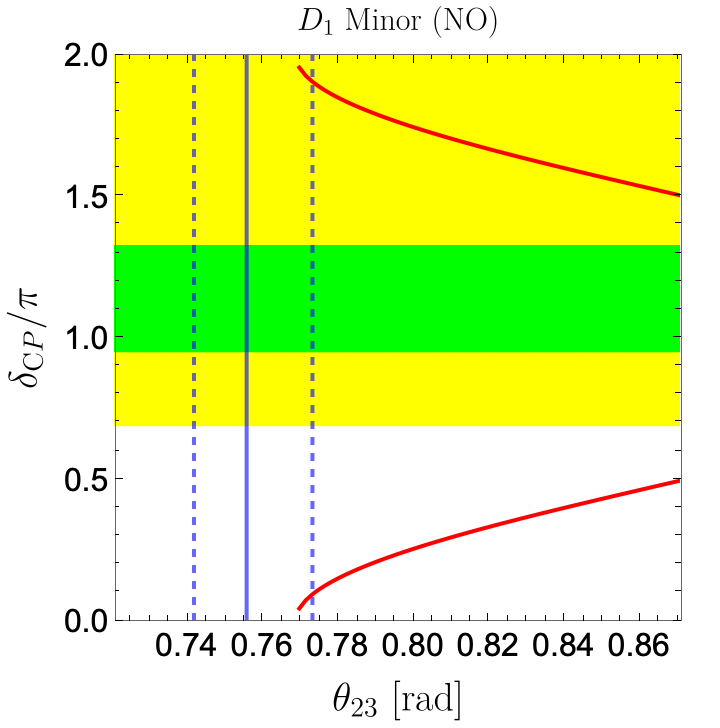}
    \label{fig:D1min_NO_left}
  \end{subfigure}
  \hfill
  \begin{subfigure}{0.49\textwidth}
    \centering
    \includegraphics[width=0.98\linewidth]{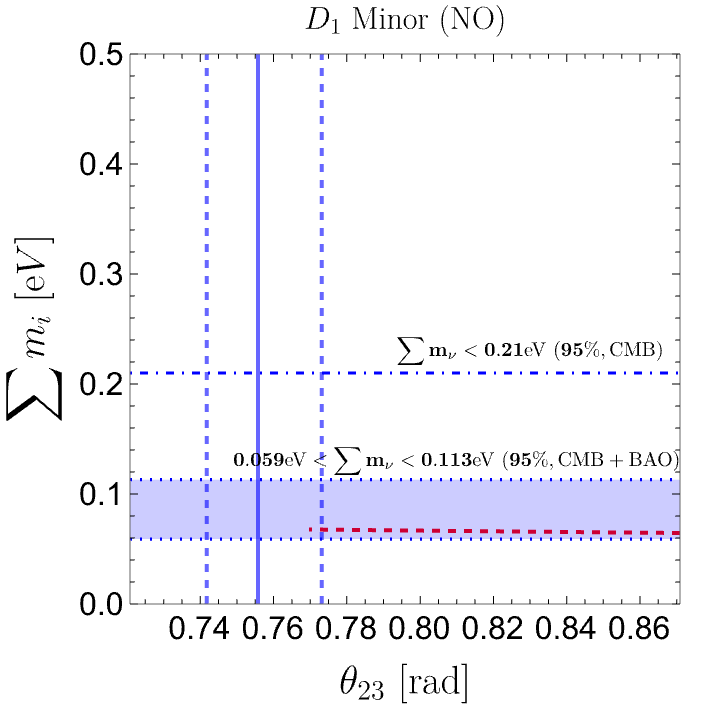}
    \label{fig:D1min_NO_right}
  \end{subfigure}
  \caption{Distribution of observables for the $D_1$ minor with NO. The left and right panels show $\theta_{23}$ vs.~$\delta_{\mathrm{CP}}/\pi$ and $\theta_{23}$ vs.~$\Sigma\,m_{i}$, respectively.}
  \label{fig:D1min_NO}
\end{figure}

\begin{figure}[H]
  \centering
  \begin{subfigure}{0.49\textwidth}
    \centering
    \includegraphics[width=0.98\linewidth]{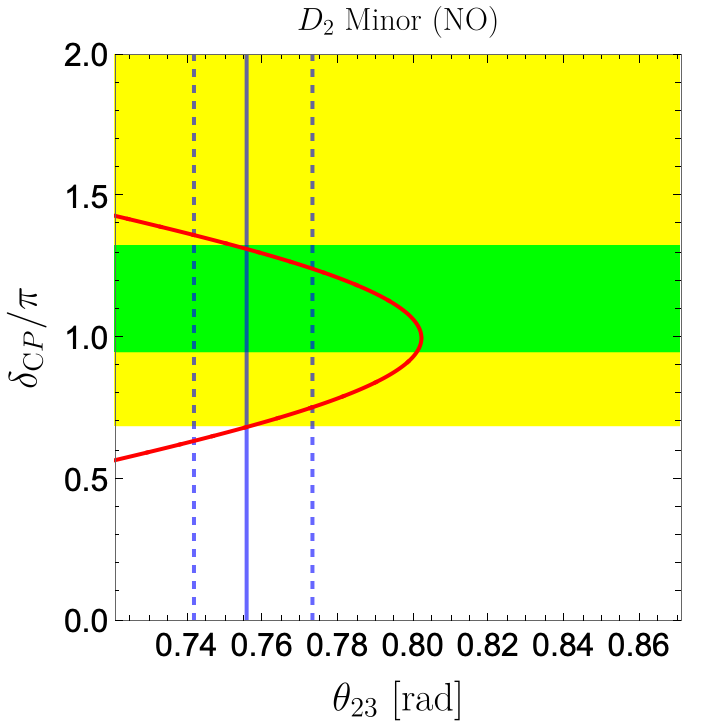}
    \label{fig:D2min_NO_left}
  \end{subfigure}
  \hfill
  \begin{subfigure}{0.49\textwidth}
    \centering
    \includegraphics[width=0.98\linewidth]{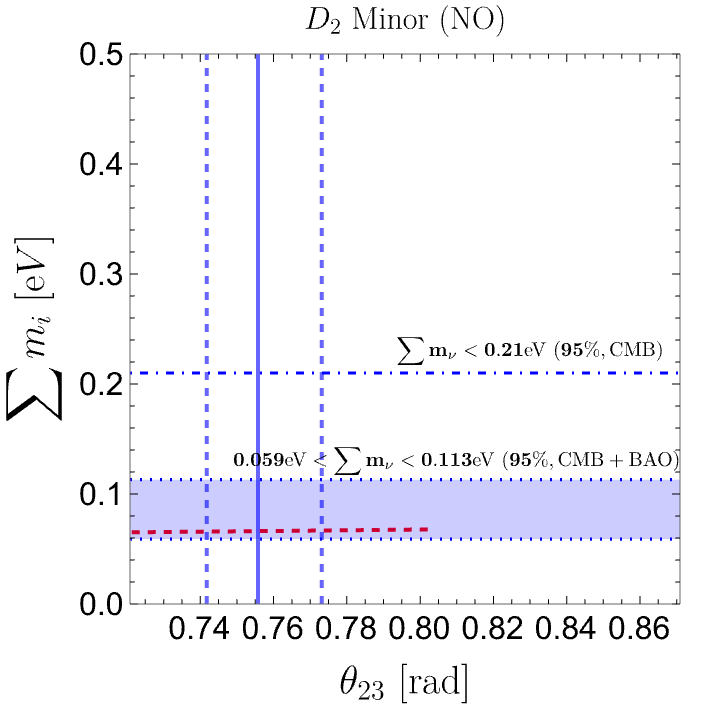}
    \label{fig:D2min_NO_right}
  \end{subfigure}
  \caption{Distribution of observables for the $D_2$ minor with NO. The left and right panels show $\theta_{23}$ vs.~$\delta_{\mathrm{CP}}/\pi$ and $\theta_{23}$ vs.~$\Sigma\,m_{i}$, respectively.}
  \label{fig:D2min_NO}
\end{figure}

\vspace{\stretch{1}}
\newpage

\subsubsection*{Inverted ordering}

\vspace{\stretch{1}}

\begin{figure}[H]
  \centering
  \begin{subfigure}{0.49\textwidth}
    \centering
    \includegraphics[width=0.98\linewidth]{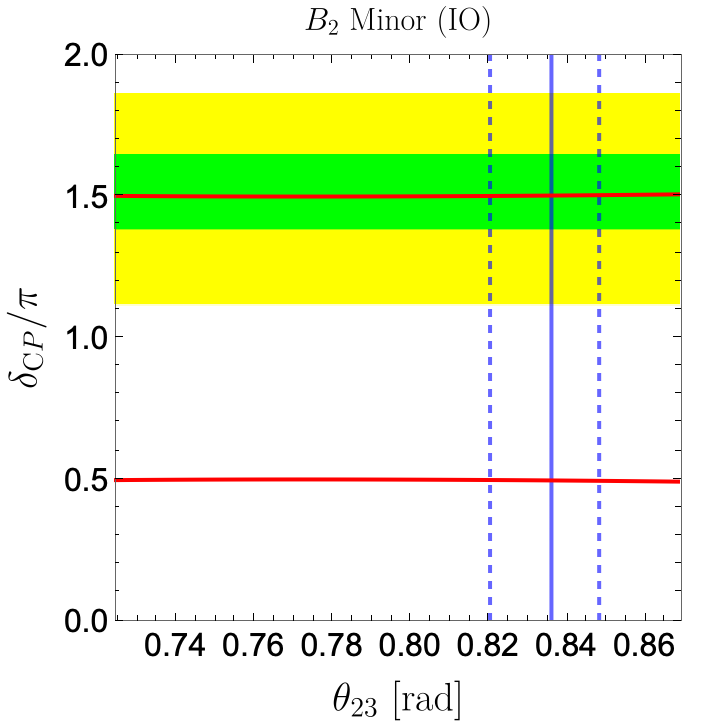}
    \label{fig:B2min_IO_left}
  \end{subfigure}
  \hfill
  \begin{subfigure}{0.49\textwidth}
    \centering
    \includegraphics[width=0.98\linewidth]{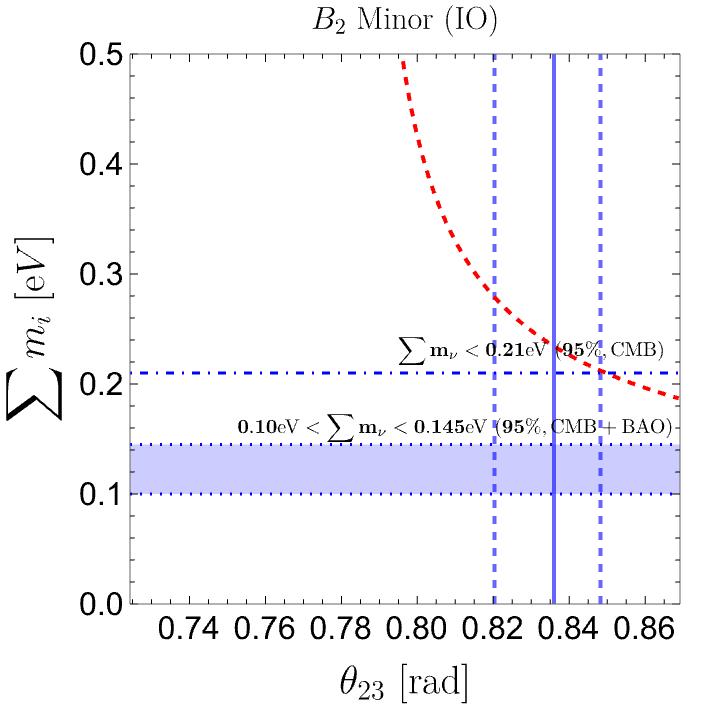}
    \label{fig:B2min_IO_right}
  \end{subfigure}
  \caption{Distribution of observables for the $B_2$ minor with IO. The left and right panels show $\theta_{23}$ vs.~$\delta_{\mathrm{CP}}/\pi$ and $\theta_{23}$ vs.~$\Sigma\,m_{i}$, respectively.}
  \label{fig:B2min_IO}
\end{figure}

\begin{figure}[H]
  \centering
  \begin{subfigure}{0.49\textwidth}
    \centering
    \includegraphics[width=0.98\linewidth]{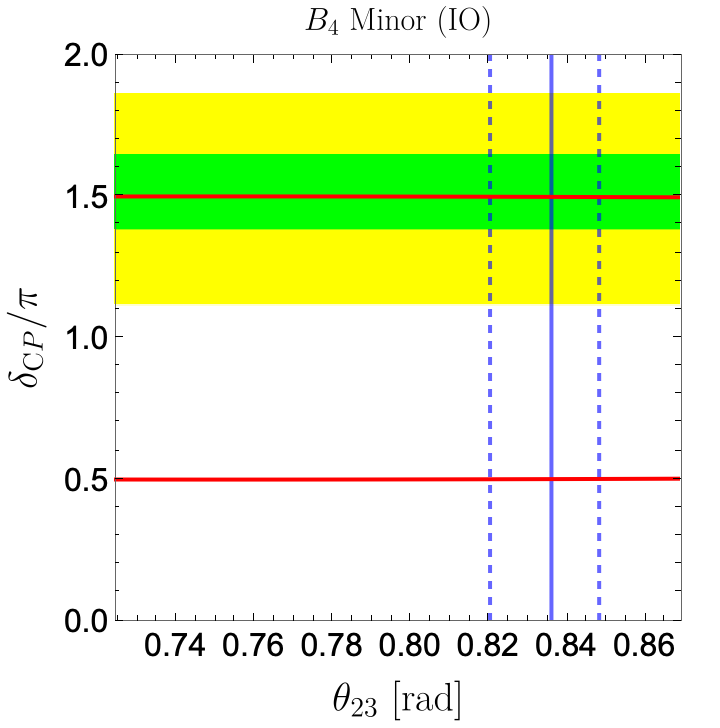}
    \label{fig:B4min_IO_left}
  \end{subfigure}
  \hfill
  \begin{subfigure}{0.49\textwidth}
    \centering
    \includegraphics[width=0.98\linewidth]{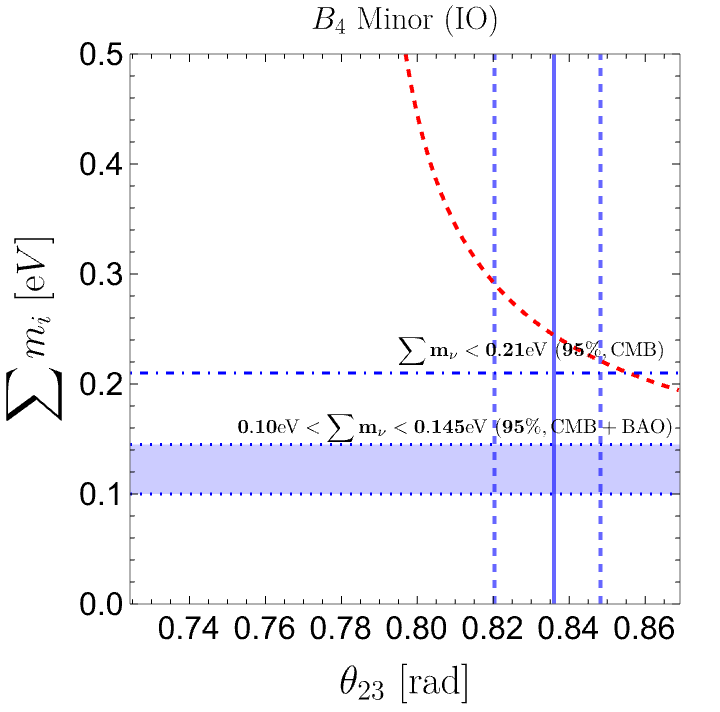}
    \label{fig:B4min_IO_right}
  \end{subfigure}
  \caption{Distribution of observables for the $B_4$ minor with IO. The left and right panels show $\theta_{23}$ vs.~$\delta_{\mathrm{CP}}/\pi$ and $\theta_{23}$ vs.~$\Sigma\,m_{i}$, respectively.}
  \label{fig:B4min_IO}
\end{figure}

\vspace{\stretch{1}}
\newpage

\section{One-zero minors}
\label{app:one-zero_minors}

In this appendix, we present the analytical method for one-zero minors. 
The analysis of one-zero minor structures is similar to that of one-zero texture structures, except that the zero condition is imposed on the inverse neutrino mass matrix.
Here, we consider the case in which the $(i,j)$ component of the inverse mass matrix vanishes
\begin{align}
    \left(\mathcal{M}_{\nu}^{\rm flavor}\right)^{-1}_{ij}&= 0.
\end{align}
The analysis proceeds in the same way as in the one-zero texture case. 
From Eqs.~\eqref{tex_component}, we obtain
\begin{align}
    -\frac{V_{i1}V_{j1}}{m_1} = e^{i\alpha_2} \frac{V_{i2}V_{j2}}{m_2}+ e^{i\alpha_3}\frac{V_{i3}V_{j3}}{m_3}.
\end{align}
This leads to the following inequality:
\begin{align}
    \Bigg| \Bigg|\frac{V_{i2}V_{j2}}{m_2}\Bigg|- \Bigg|\frac{V_{i3}V_{j3}}{m_3}\Bigg|\Bigg| \leq \Bigg|\frac{V_{i1}V_{j1}}{m_1}\Bigg| \leq \Bigg|\Bigg| \frac{V_{i2}V_{j2}}{m_2}\Bigg|+ \Bigg|\frac{V_{i3}V_{j3}}{m_3}\Bigg| \Bigg|.
    \label{eq:const_1zero_minor}
\end{align}
From this inequality, we can obtain the viable region of $\theta_{23}$ for each value of $\delta_{\rm CP}$ by fixing $\theta_{12},\theta_{13},\Delta m^2_{21},\Delta m^2_{3\ell}$ and by imposing the cosmological constraints on the sum of neutrino masses.

\subsection*{Results}

We next present the analysis results for one-zero minor structures for NO and IO.
The results are summarized in Table \ref{tab:one-zero-minor}. 
We find that some structures are excluded even in the one-zero minor case: $G_1$ for NO and $G_2$, $G_3$, $H_3$ for IO.

\medskip

\begin{figure}[H] 
  \centering
\vspace{-10 mm}  
  \begin{minipage}{0.4\textwidth}
    \centering
    \includegraphics[width=\linewidth]{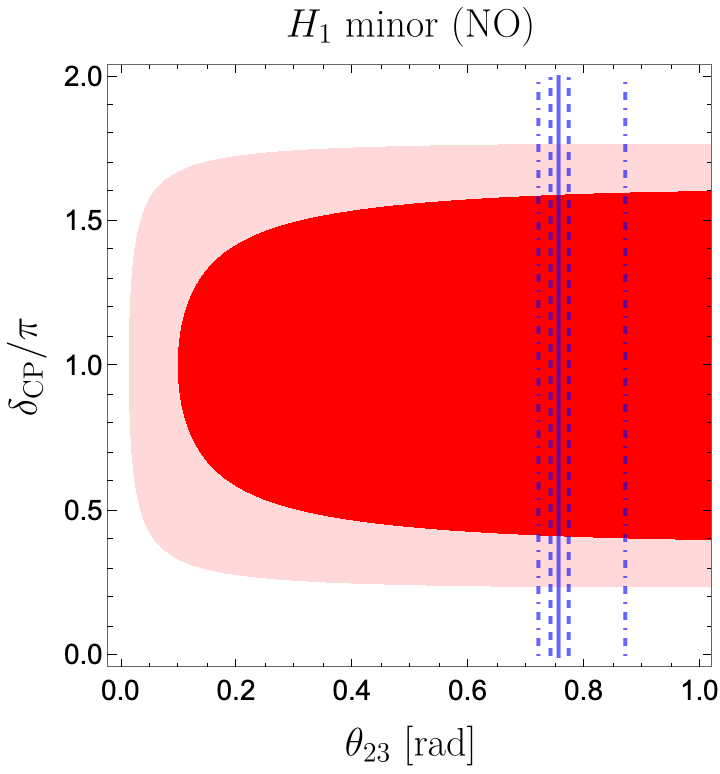}
    \label{fig:H1min_NO}
  \end{minipage}\hfill
  \begin{minipage}{0.4\textwidth}
    \centering
    \includegraphics[width=\linewidth]{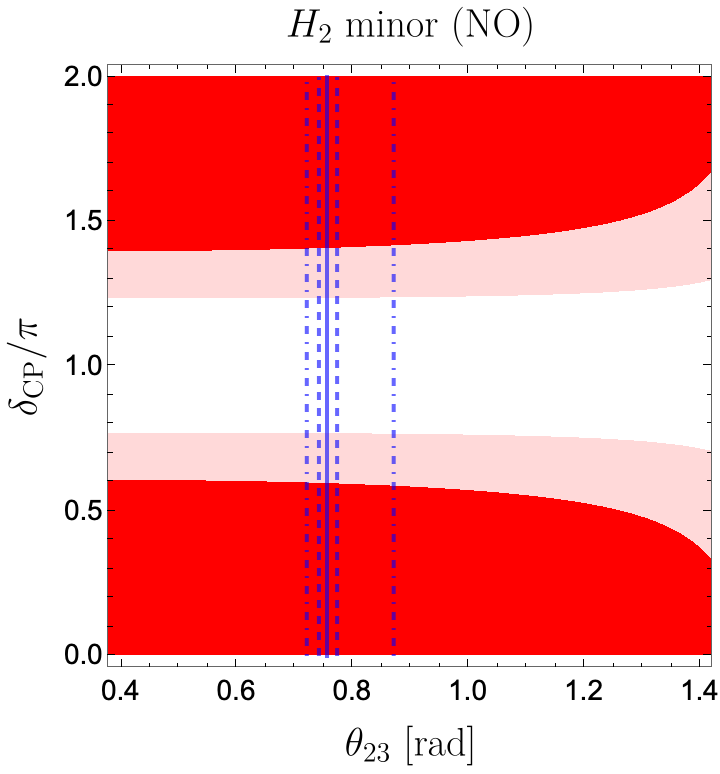}
    \label{fig:H2min_NO}
  \end{minipage}
  \caption{Viable regions in Eq.~\eqref{eq:const_1zero_minor} for NO. The vertical axis shows $\delta_{\rm CP}/\pi$, and the horizontal axis shows $\theta_{23}$, except for $G_1$, where $\theta_{12}$ is shown instead. 
The blue solid, dashed, and dash-dotted lines indicate the best-fit value, the $1\sigma$ range, and the $3\sigma$ range of $\theta_{23}$ given in Table \ref{tab:NuFIT}.
The light-red regions show the viable regions consistent with the CMB constraint, $\sum m_\nu <0.21~{\rm eV}$, while the red regions indicate those consistent with the CMB+BAO constraint, $0.059 ~{\rm eV}< \sum m_\nu <0.113~{\rm eV}$.}
  \label{fig:one-zero_min_NO}
\end{figure}

The characteristic results are shown in Fig.~\ref{fig:one-zero_min_NO}, where the figure layout is the same as Figs.~\ref{fig:one-zero_tex_NO} and \ref{fig:one-zero_tex_IO}.
We find that $H_1$ minor in the NO case predicts $0.25\pi \lesssim \delta_{\rm CP} \lesssim 1.75 \pi$ and $H_2$ minor in the NO case predicts $0 \lesssim \delta_{\rm CP} \lesssim 0.75 \pi$, $1.3\pi \lesssim \delta_{\rm CP} \lesssim 2.0 \pi$. We cannot find a peculiar pattern for the distribution of $\delta_{\mathrm{CP}}$ for the other cases.

\begin{table}[H]
    \centering
    \caption{Summary of the one-zero minor analysis.}
    \label{tab:one-zero-minor}
    
    \begin{tabular}{ll cccccc}
        \toprule
        Structure & & $G_1$ & $G_2$ & $G_3$ & $H_1$ & $H_2$ & $H_3$ \\
        \midrule
        \multirow{2}{*}{CMB}     & NO & \hyperref[fig:G1min_NO]{$\times$} & \hyperref[fig:G2min_NO]{$\bigcirc$} & \hyperref[fig:G3min_NO]{$\bigcirc$} & \hyperref[fig:H1min_NO]{$\bigcirc$} & \hyperref[fig:H2min_NO]{$\bigcirc$} & \hyperref[fig:H3min_NO]{$\bigcirc$} \\
                                 & IO & \hyperref[fig:G1min_IO]{$\bigcirc$}       & \hyperref[fig:G2min_IO]{$\bigcirc$}       & \hyperref[fig:G3min_IO]{$\bigcirc$} & \hyperref[fig:H1min_IO]{$\bigcirc$}   & \hyperref[fig:H2min_IO]{$\bigcirc$} & \hyperref[fig:H3min_IO]{$\times$}    \\ 
        \midrule
        \multirow{2}{*}{CMB+BAO} & NO & \hyperref[fig:G1min_NO]{$\times$} & \hyperref[fig:G2min_NO]{$\bigcirc$} & \hyperref[fig:G3min_NO]{$\bigcirc$}   & \hyperref[fig:H1min_NO]{$\bigcirc$}   & \hyperref[fig:H2min_NO]{$\bigcirc$}  & \hyperref[fig:H3min_NO]{$\bigcirc$}   \\
                                 & IO & \hyperref[fig:G1min_IO]{$\bigcirc$}       & \hyperref[fig:G2min_IO]{$\times$}       & \hyperref[fig:G3min_IO]{$\times$}   & \hyperref[fig:H1min_IO]{$\bigcirc$}   & \hyperref[fig:H2min_IO]{$\bigcirc$}   & \hyperref[fig:H3min_IO]{$\times$}   \\
        \bottomrule
    \end{tabular}
\end{table}

\bibliography{references}{}
\bibliographystyle{JHEP} 

\end{document}